\documentclass[a4paper,11pt]{article}
\pdfoutput=1 
\usepackage[english]{babel}
\usepackage{microtype,xspace}
\usepackage[utf8]{inputenc}	

\usepackage{jheppub} 
\usepackage{bm,amssymb,amsmath,slashed,graphicx,multirow,soul,mathtools,xspace,array}
\usepackage{siunitx}
\usepackage{bm} 
\usepackage{float}
\usepackage{feynmf}
\allowdisplaybreaks
\usepackage{ bbold }
\usepackage{subfigure}
\usepackage[normalem]{ulem} 

\usepackage{feynmf}

\newcommand{\g}{\gamma}

\newcommand{\fbar}{\bar{f}}
\newcommand{\tbar}{\bar{t}}

\newcommand{\nbar}{\bar{\nu}}
\newcommand{\gt}{\gamma_T}
\newcommand{\zg}{Z/\gamma}
\newcommand{\zgt}{Z/\gamma_T}
\newcommand{\zgp}{Z/\gamma_+}
\newcommand{\zgm}{Z/\gamma_-}
\newcommand{\gm}{\gamma_-}
\newcommand{\gp}{\gamma_+}
\newcommand{\wt}{W_T}
\newcommand{\zt}{Z_T}
\newcommand{\Wp}{W_+}
\newcommand{\Wpp}{W_+^+}
\newcommand{\Wpm}{W_-^+}
\newcommand{\Wpl}{W_L^+}
\newcommand{\zp}{Z_+}
\newcommand{\wm}{W_-}
\newcommand{\Wmp}{W_+^-}
\newcommand{\Wmm}{W_-^-}
\newcommand{\Wml}{W_L^-}
\newcommand{\zm}{Z_-}
\newcommand{\gtm}{\gamma_{-T}}

\newcommand{\wtm}{W_{-T}}
\newcommand{\ztm}{Z_{-T}}
\newcommand{\wl}{W_L}
\newcommand{\zl}{Z_L}
\newcommand{\fl}{f_L}
\newcommand{\fr}{f_R}
\newcommand{\flb}{\bar{f}_L}
\newcommand{\frb}{\bar{f}_R}
\newcommand{\tl}{t_L}
\newcommand{\bl}{b_L}
\newcommand{\tr}{t_R}

\newcommand{\tlb}{\bar{t}_L}
\newcommand{\blb}{\bar{b}_L}
\newcommand{\trb}{\bar{t}_R}

\newcommand{\ulbar}{\bar{u}_L}
\newcommand{\urbar}{\bar{u}_R}
\newcommand{\dlbar}{\bar{d}_L}
\newcommand{\drbar}{\bar{d}_R}

\newcommand{\zmax}{z_{\rm max}^{ABC}}
\newcommand{\Nmax}{N_{\rm max}^{ABC}}

\newcommand{\TeV}{\text{TeV}}
\newcommand{\GeV}{\text{GeV}}

\newcommand{\beq}{\begin{equation} }
\newcommand{\eeq}{\end{equation}}
\newcommand{\be}{\begin{equation} }
\newcommand{\ee}{\end{equation}}
\newcommand{\bi}{\begin{itemize} }
\newcommand{\ei}{\end{itemize} }

\newcommand{\zb}{\bar{z}}

\definecolor{Red}{rgb}{1.,0.,0.}
\definecolor{Grn}{rgb}{0.,0.75,0.}
\definecolor{Blu}{rgb}{0.,0.,1.}

\definecolor{red}{rgb}{0.6,.0706,.1373}
\definecolor{blue}{rgb}{0,0.396,0.741}

\usepackage[T1]{fontenc} 

\usepackage{amsmath,amssymb,epsfig,color,slashed}

\allowdisplaybreaks

\title{\boldmath LePDF: Standard Model PDFs for High-Energy Lepton Colliders}

\author[a,b]{Francesco Garosi,}
\author[b]{David Marzocca,}
\author[c]{Sokratis Trifinopoulos}

\affiliation[a]{SISSA International School for Advanced Studies, Via Bonomea 265, 34136, Trieste, Italy}
\affiliation[b]{INFN, Sezione di Trieste, SISSA, Via Bonomea 265, 34136, Trieste, Italy}
\affiliation[c]{Center  for  Theoretical  Physics,  Massachusetts  Institute  of  Technology,  Cambridge,  MA  02139,  USA}

\abstract{The emission of collinear radiation off an elementary lepton can be factorised from the hard scattering process by introducing Parton Distribution Functions of a Lepton (LePDF), which, contrary to protons, can be derived from first principles. 
In case of multi-TeV lepton colliders, such as the muon colliders currently being proposed, the complete structure of Standard Model interactions must be taken into account.
In this work we solve numerically the corresponding DGLAP equations at the double-log order and provide public files with LePDFs for both muons and electrons, including polarisation effects.
We discuss several interesting aspects of the resulting PDFs and compare them with the Effective Vector Approximation, showing that the latter fails to describe well the vector bosons PDFs at small momentum fractions, unless it is extended to higher orders.
}

\begin{document}

\maketitle

\section{Introduction}
\label{sec:intro}

At present, the Large Hadron Collider at CERN is our only tool for direct exploration of physics at the electroweak scale and above and the high-luminosity phase is planned to last until the early 2040s. It proved to be a formidable machine for both searches of new heavy particles as well as precision studies at the electroweak scale, of which the Higgs discovery and the precision study of its coupling is a prime example.
Nevertheless, the next large-scale experiment in high-energy physics is likely to be a lepton collider.
The proposed options include circular or linear electron-positron colliders (FCC-ee \cite{FCC:2018evy} and CEPC \cite{CEPCStudyGroup:2018ghi} for the former, ILC \cite{Behnke:2013xla}, CLIC \cite{CLIC:2018fvx} and CCC \cite{Bai:2021rdg} for the latter) as well as $\mu^+ \mu^-$ colliders \cite{Stratakis:2022zsk,Jindariani:2022gxj,Aime:2022flm,DeBlas:2022wxr,Accettura:2023ked}. Among these, the linear $e^+ e^-$ colliders and the muon colliders could achieve multi-TeV center-of-mass energies.

While leptons are elementary particles in the Standard Model (SM), the process of collinear emission of initial state radiation (ISR), with transverse momentum $p_T$ much smaller than the energy of the hard scattering process, can be factorized and a description in terms of parton distribution functions (PDFs) can be introduced, similarly to what is done in case of proton colliders and the parton content of a proton.
The case of collinear photon emission from an electron is known since almost a century and at leading order can be described using the effective photon approximation (EPA) \cite{Fermi:1924tc,vonWeizsacker:1934nji,Williams:1934ad,Landau:1934zj}
\be
    f_{\gamma}^{\rm EPA}(x) = \frac{\alpha_\gamma}{2\pi} P_{\gamma e}(x) \log \frac{E^2}{m_e^2}~,
    \label{eq:epa}
\ee
where $E$ is the energy of the initial electron and $P_{\gamma e}(x) = \frac{1+(1-x)^2}{x}$ is the splitting function, that describes the probability of an electron to emit a photon with a fraction $x$ of its energy and virtuality $- p_T^2 / (1-x)$.

\begin{figure}[t]
\centering
\includegraphics[width=6cm]{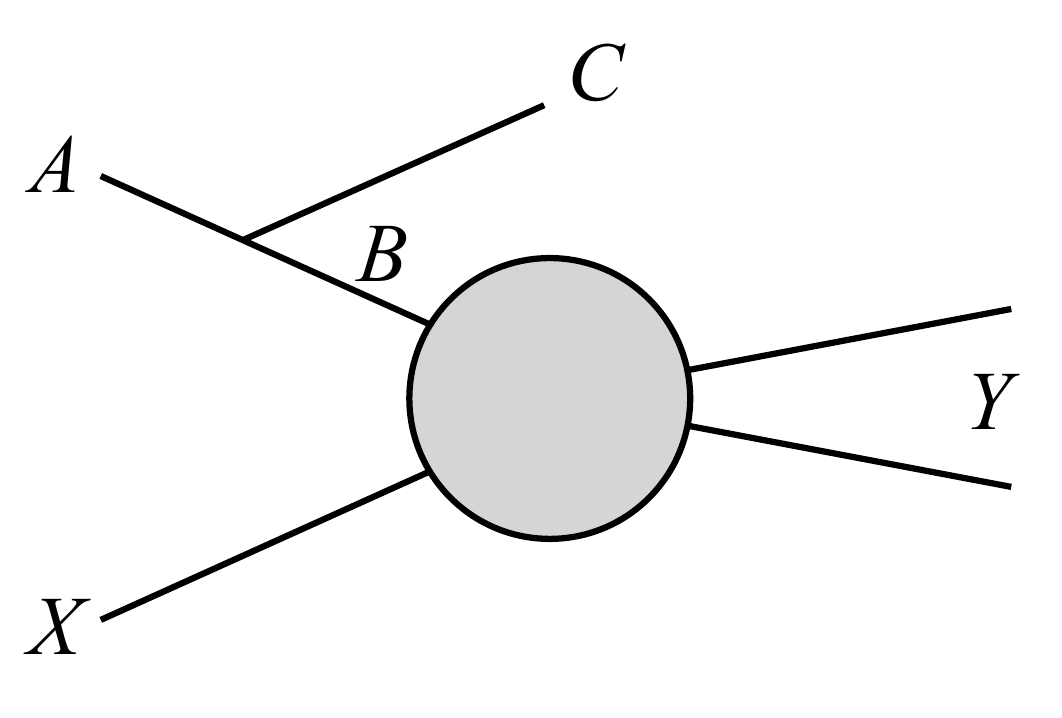} 
\caption{\label{fig:AX>CY} Diagram of a process $A X \to C Y$ with an initial-state splitting $A \to B C$.}
\end{figure}

When $E \gg m_e$ the large logarithms can be resummed in order to improve the perturbative expansion, and the factorization scale $Q$ is introduced.
Invariance of the physics under the factorization scale leads to the DGLAP equations \cite{Gribov:1972ri,Dokshitzer:1977sg,Altarelli:1977zs}.
For a generic splitting of massless partons $A \to B + C$ (as in Fig.~\ref{fig:AX>CY}), choosing $p_T$ as factorization scale and working at the leading logarithm (LL) order one has
\begin{equation}
\begin{split}
    Q^2 \frac{df_B(x,Q^2)}{dQ^2} &= P_B^v \, f_B(x,Q^2)+\sum_{A,C}\frac{\alpha_{ABC}(Q)}{2\pi}\int_x^1 \frac{dz}{z}P_{BA}^C\left(z \right)f_A\left(\frac{x}{z},Q^2\right)~,
    \label{eq:DGLAP_massless}
\end{split}
\end{equation}
where $P_{BA}^C(z)$ are the leading order (LO) splitting functions (listed in App.~\ref{app:splitting}), $\alpha_{ABC}(Q)$ the corresponding running coupling, and the term $P^v_B f_B$ describes virtual corrections (see App.~\ref{app:radiative} for details).
In case of a proton, due to its non-perturbative nature, the initial conditions for this system must be fitted from collider data.
For a lepton, instead, the initial condition can be computed perturbatively and the system can be solved from first principles. The initial condition is 
\be
	f_{\ell}(x, m_\ell^2) = \delta(1-x) + \mathcal{O}(\alpha)~,
 \label{eq:initial_condition}
\ee
for initially unpolarised beams\footnote{Here we assume unpolarised beams, so that polarisation effects are only due to the DGLAP evolution. If needed, it will be straightforward to implement a given intrinsic beam polarisation in future versions of the code.}, while all other PDFs vanish for $Q^2 = m_\ell^2$ at this order. Next-to-leading order (NLO) corrections to the initial conditions have also been computed \cite{Frixione:2019lga} and become relevant when next-to-leading log (NLL) evolution is considered \cite{Bertone:2019hks,Bertone:2022ktl}.
In this work we limit ourselves to LL evolution and LO initial conditions.

In case of multi-TeV lepton colliders one can be interested in factorization scales much higher than the electroweak scale. In this case all SM interactions and fields should be considered \cite{Ciafaloni:2005fm}.
In this aspect, lepton colliders differ qualitatively from hadron colliders. For the latter, QCD interactions are the dominant contributions to the DGLAP evolution in the whole energy range (see however e.g. Refs.~\cite{Bertone:2015lqa,Manohar:2016nzj,Manohar:2017eqh,Fornal:2018znf,Buonocore:2020nai} about the photon and lepton content of the proton). In the case of lepton collider, instead, QED and electroweak (EW) interactions are the leading ones, with QCD playing an important but not dominant role.

The facts that SM gauge group is non abelian, that it is spontaneously broken at the electroweak scale, and that interactions are chiral have several crucial implications for the evaluation of collinear radiation in this regime.
The non abelian nature of EW interactions imply a lack of cancellation of infrared (IR) divergencies between virtual corrections and real emission, which generates Sudakov double logarithms \cite{Ciafaloni:1998xg,Ciafaloni:2000df,Ciafaloni:2001vt}.
Electroweak symmetry breaking effects have been shown to provide important contributions and to be the dominant ones in case of longitudinal polarisations of electroweak gauge bosons \cite{Chen:2016wkt}.
Since the SM interactions are chiral, PDFs become polarized above the EW scale \cite{Bauer:2018arx}.

The goal of this work is to numerically solve the DGLAP equations from the initial condition at $Q = m_\ell$ up to multi-TeV scales, taking into account all SM interactions (including the effects mentioned above), and to provide public results with the complete LePDFs.
For concreteness, in the following we assume the initial lepton to be a muon, since muon colliders could achieve higher energies, for which our discussion is more relevant. However, all results can be equally applied to $e^+ e^-$ colliders with suitable substitutions thus we provide numerical results for both.

A preliminary study of PDFs for muon colliders by us, mainly focusing on the fermionic degrees of freedom, was included in Ref.~\cite{Azatov:2022itm}.
In this work we extend it by including all SM interactions, Sudakov double logs, polarisation, and EW symmetry breaking effects.
One similar study has already been performed in the literature, specifically in Refs.~\cite{Han:2020uid,Han:2021kes}, and we compare our results with the plots presented in those works. We also compare against approximate solutions obtained by solving iteratively, at fixed-order, the DGLAP equations, which provide the analogous of the Effective Vector Boson Approximation (EVA) \cite{Costantini:2020stv,AlAli:2021let,Ruiz:2021tdt}. A noteworthy result of this comparison is the realisation that the LO EVA for transverse EW gauge bosons PDFs is insufficient for correctly describing the full result, which is instead well approximated by including contributions up to $\mathcal{O}(\alpha^2)$. Crucially, Sudakov double logarithms appear at this order  due to the gauge boson splitting $V \to V V$ and the virtual corrections to the muon PDF.

In Section \ref{sec:QED} we present the first part of the evolution below the EW scale, where only QED and QCD interactions are relevant. In Section~\ref{sec:SM} we discuss the main aspects of the evolution above the EW scale. Our results are collected in Section~\ref{sec:results}, where several notable features of LePDFs are showed, and a comparison with EVA is presented. We conclude in Section~\ref{sec:conclusions}, while many details of our computations, the numerical implementation and the formatting of our LePDF files are collected in several Appendices. The numerical results for the LePDFs can be downloaded from GitHub at the link:  {\tt \href{https://github.com/DavidMarzocca/LePDF}{https://github.com/DavidMarzocca/LePDF}}.

\section{QED and QCD evolution}
\label{sec:QED}

For factorization scales below the EW scale the relevant degrees of freedom are light quarks and charged leptons, with vectorlike QED and QCD gauge interactions.
Neutrinos, while having negligible masses, become relevant only above the EW scale where the $W$ boson can go on-shell.\footnote{A possible effect of neutrinos even below the EW scale is due to neutrinos from muon decay $\mu^- \to e^- \bar{\nu}_e \nu_\mu$ which present IR singularities in the physical region when scattering with the incoming $\mu^+$ (or viceversa) \cite{Coleman:1965xm}. In case of muon colliders such singularities are cutoff by the finite width of the muon beam \cite{Ginzburg:1995bc,Melnikov:1996na}. We neglect such effects with our PDF formalism, assuming that it can be described independently of PDFs.}
Because the initial condition and the evolution equations are vectorlike, in this regime no polarisation effects are induced, i.e. PDFs will be the same for both fermion chiralities or gauge boson polarisation.

In the DGLAP evolution from the muon mass up to the EW scale one encounters several mass thresholds for each fermion species as well as at the QCD scale $Q_{\rm QCD}$. At each threshold a matching should be performed. For our purposes, we take all fermions except bottom and top quarks to be massless.\footnote{In future versions we plan to add also the $\tau$ and charm quark mass thresholds.}
The $Q_{\rm QCD}$ scale sets the onset of QCD interactions, which become relevant after the $\gamma \to q \bar{q}$ splitting. This can be interpreted as the QCD structure of a photon, and can be divided into a perturbative and a non-perturbative component, mainly due to the photon mixing with QCD vector mesons \cite{Borzumati:1992za,Aurenche:1993pk,Drees:1994eu,Schuler:1995fk,Schuler:1996fc}. The latter one will be power-suppressed at the large scales  we are eventually interested in, so we neglect it. The choice of $Q_{\rm QCD}$ depends on how many resonances are included in the non-perturbative component and a value close to $m_\rho$ has been argued to provide a good benchmark \cite{Drees:1994eu,Schuler:1995fk,Schuler:1996fc}. In practice, we follow the prescription of Ref.~\cite{Drees:1994eu,Han:2021kes} with  $Q_{\rm QCD} = 0.7 ~\GeV$.\footnote{We study the dependence of our results on $Q_{\rm QCD}$ by running the evolution also for values $Q_{\rm QCD} = 0.52 (1.0)~\GeV$ and interpreting the differences as theory uncertainty on our final results due to non-perturbative QCD dynamics, see dedicated discussion in Sec.~\ref{sec:uncertainties}.}

Therefore, from $m_\mu$ to $Q_{\rm QCD}$ we consider only QED interactions, including all charged leptons ad well as the light quarks ($u$, $d$, $s$, $c$).
At $Q_{\rm QCD}$ we match the PDFs and continue the evolution up to the bottom mass scale adding also QCD interactions and setting the initial condition for the gluon PDF as $f_g(x, Q_{\rm QCD}) = 0$. At $Q = m_b$ we perform another matching and continue the evolution up to the EW scale $Q_{\rm EW}$ including also the bottom quark, setting $f_b(x, m_b) = f_{\bar{b}}(x, m_b) = 0$ as its initial conditions.

Given the $C$ and $P$ symmetries of QED and QCD, and the fact that all fermions except for bottom and top quarks are taken massless, below $Q_{\rm EW}$ several PDFs are related:
\be\begin{split}
	f_{\ell_{\rm sea}} &= f_e = f_\tau = f_{\bar{e}} = f_{\bar{\mu}} = f_{\bar{\tau}} ~, \\
	f_{q^u} &= f_u = f_{\bar{u}} = f_c = f_{\bar{c}} ~, \\
	f_{q^d} &= f_d = f_{\bar{d}} = f_s = f_{\bar{s}} ~, \\
	f_b &= f_{\bar{b}} ~.
 \label{eq:PDFs_QED_QCD}
\end{split}\ee
The DGLAP equations, according to Eq.~\eqref{eq:DGLAP_massless}, are then given by
\be
\begin{split}
\frac{df_\ell}{dt} =& P^v_\ell f_\ell + \frac{\alpha_\gamma(t)}{2\pi}\left[P_{ff}^V\otimes f_\ell + P_{fV}^f\otimes f_{\gamma}\right]~, \\
\frac{df_{q^u}}{dt} =& P^v_{q^u} f_{q^u} + \frac{\alpha_\gamma(t)}{2\pi}Q_{u}^2\left[P_{ff}^V\otimes f_{q^u} + N_c P_{fV}^f\otimes f_{\gamma}\right]\\
&+\frac{\alpha_{3}(t)}{2\pi}\left[C_F P_{ff}^V\otimes f_{q^u} + T_F P_{fV}^f\otimes f_{g}\right]~,\\ 
\frac{df_{q^d,b}}{dt} =& P^v_{q^d} f_{q^d} + \frac{\alpha_\gamma(t)}{2\pi}Q_{d}^2\left[P_{ff}^V\otimes f_{q^d,b} + N_c P_{fV}^f\otimes f_{\gamma}\right]\\
&+\frac{\alpha_{3}(t)}{2\pi}\left[C_F P_{ff}^V\otimes f_{q^d,b} + T_F P_{fV}^f\otimes f_{g}\right]~,\\
\frac{df_\gamma}{dt} =& P^v_{\gamma} f_{\g} + \frac{\alpha_\gamma(t)}{2\pi}\sum_f Q_{f}^2P_{Vf}^f\otimes \left(f_{f}+f_{\bar{f}}\right)~,\\ 
\frac{df_g}{dt} =&  P^v_{g} f_{g} +\frac{\alpha_{3}(t)}{2\pi}\left[C_A P_{VV}\otimes f_{g} + C_F P_{Vf}^f \otimes \sum_{q}\left(f_{q}+f_{\bar{q}}\right)\right]~, \label{eq:DGLAP_QED}
\end{split}
\ee
where $\ell = \{\ell_{\textrm{sea}}, \mu\}$, $\otimes$ indicates a convolution as in Eq.~\eqref{eq:DGLAP_massless} and we defined the evolution variable $t$ as
\be
    t \equiv \log\left(Q^2 / m_\mu^2\right)~,
    \label{eq:tdef}
\ee
such that we start the evolution with the initial condition in Eq.~\eqref{eq:initial_condition} at $t=0$.
The splitting functions are listed in App.~\ref{app:splitting} (we regulate the $(1-z)^{-1}$ poles using the standard $+$-distribution, see Eq.~\eqref{eq:def+}), while the values of the virtual coefficients $P^v_B$ can be found in App.~\ref{app:radiative_QED}. Finally, the details for the RG evolution of QED and QCD couplings are reported in App.~\ref{app:SMcouplings}.
\begin{figure}[t]
\centering
\includegraphics[height=6cm]{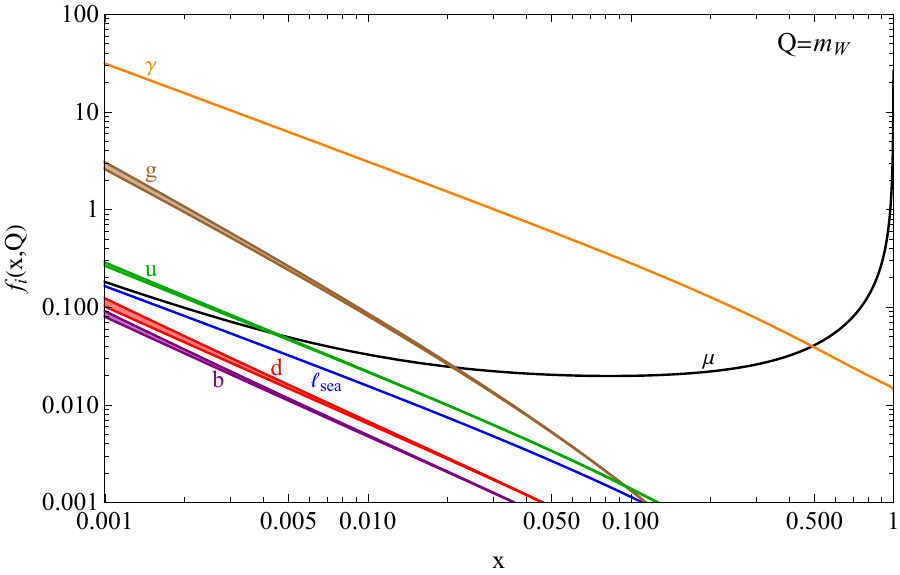} 
\caption{\label{fig:PDF_mZ} PDFs of a muon evaluated at the EW scale $Q_{EW} = m_W$. The colored regions corresponds to the variation obtained by modifying $Q_{\rm QCD}$ between 0.5 and 1 GeV (see Sec.~\ref{sec:uncertainties} for details). Here $\ell_{\rm sea}$ is defined as in Eq.~\eqref{eq:PDFs_QED_QCD}.}
\end{figure}
In Fig.~\ref{fig:PDF_mZ} we show the result of our numerical solution of DGLAP equations for a muon, evolved from the muon mass up to the EW scale $Q_{\rm EW} = m_W$ (solid lines). As can be seen, at small $x$ the gluon PDF becomes rather important, while quark PDFs have a size similar to the muon itself or sea leptons, as already showed in Refs.~\cite{Han:2020uid,Han:2021kes,Azatov:2022itm}.

\newpage
\subsection{Iterative solution for QED}

By solving iteratively the DGLAP equations order by order in $(\alpha_\gamma t) = (\alpha_\gamma \log Q^2 / m_Q^2)$, one gets approximate solutions for the PDFs. The LO contribution for the photon PDF is given by Eq.~\eqref{eq:epa} with the crucial substitution $E \to Q$.
By including terms up to $\mathcal{O}(\alpha^2 t^2)$ we get
\be
\begin{split}
	f_\mu^{(\alpha^2)}(x, t) &= \delta(1-x) + \frac{\alpha_\gamma}{2\pi}\, t \, \left(\frac{3}{2}\delta(1-x)+P_{ff}^V(x)\right) \\
	&+ \frac{1}{2} \left( \frac{\alpha_\gamma}{2\pi}\, t \right)^2 \left[\frac{9}{4}\delta(1-x)+ 3 P_{ff}^V(x)+ I_{fVVf}(x) + I_{ffff}(x) \right] ~, \\
	f_{\ell_{\rm sea}}^{(\alpha^2)}(x, t) &= \frac{1}{2} \left(  \frac{\alpha_\gamma}{2\pi}\, t \right)^2 I_{fVVf}(x) ~, \\
	f_{\gamma}^{(\alpha^2)}(x, t) &=  \frac{\alpha_\gamma}{2\pi}\, t\, P_{Vf}^f(x) + \frac{1}{2} \left(  \frac{\alpha_\gamma}{2\pi}\, t \right)^2 \left[ \left(\frac{3}{2}- \frac{2}{3}N_f^{\rm QED}\right) P_{Vf}^f(x)  + I_{Vfff}(x) \right] ~,
	\label{eq:appr_QED_sol}
\end{split}
\ee
where $t$ is defined in Eq. \ref{eq:tdef} and the $I_{ABBC}(x) \equiv \int_x^1 \frac{dz}{z} P^X_{AB}(z)P_{BC}^Y(x/z)$ integrals are collected in Eq.~\eqref{eq:appr_QED_integrals}.
Going up to $\mathcal{O}(\alpha^2 t^2)$ is required in order to describe the low-$x$ behavior of valence and sea lepton PDFs, that is dominated by the $\gamma \to \ell^+ \ell^-$ splitting (i.e. the $I_{fVVf}(x)$ term above).
We find good agreement between these expressions and our numerical results.

The gluon PDF starts formally at $\mathcal{O}(\alpha_\gamma^2\alpha_s t^3)$. However, due to the large value of $\alpha_s$ at low energies and the importance of a careful treatment of the matching at the $Q_{\rm QCD}$, a fixed-order solution would not be a good approximation for the correct result. For the same reason we do not report here also the approximated expression for quark PDFs, even if they formally arise already at $\mathcal{O}(\alpha_\gamma^2 t^2)$, since they receive large a QCD contribution from gluon splitting into $q\bar{q}$.
A correct description of the quark and gluon PDFs in a lepton therefore motivates a complete numerical solution of the DGLAP equations~\cite{Han:2021kes}.

\section{DGLAP evolution in the SM}
\label{sec:SM}

For energies above the EW scale, the splitting processes in the initial states can involve all SM interactions and fields, which must then be included in the DGLAP equations.
The chiral nature of EW interactions induces polarisation effects on PDFs, so in this region all gauge bosons and fermions polarisations are treated separately. Splitting functions for SM interactions in the unbroken phase have been computed in Refs.~\cite{Ciafaloni:2001mu,Ciafaloni:2005fm,Chen:2016wkt,Bauer:2017isx,Bauer:2017bnh,Bauer:2018arx}, with which we agree.
Another well known effect in EW PDFs is the interference between photon and transverse $Z_T$, which must be described with a $Z/\gamma$ mixed PDF, as well as between the Higgs boson and the longitudinal $Z_L$, which induces a $h/Z_L$ mixed PDF \cite{Ciafaloni:2000gm,Ciafaloni:2005fm,Chen:2016wkt}.
Also, the non-abelian nature of EW interactions induces Sudakov double logarithms, which must be resummed when one is interested at high energies \cite{Ciafaloni:1998xg,Ciafaloni:2000df,Ciafaloni:2001vt}.

Since we neglect all fermion masses except than the top and bottom quarks, several PDFs are related:
\be\begin{split}
	f_{e_L} &= f_{\tau_L} ~, \quad f_{\bar \ell_L} = f_{\bar e_L} = f_{\bar \mu_L} = f_{\bar\tau_L} ~,\\
	f_{e_R} &= f_{\tau_R} ~, \quad f_{\bar \ell_R} = f_{\bar e_R} = f_{\bar\mu_R} = f_{\bar\tau_R} ~,\\
	f_{\nu_e} &= f_{\nu_\tau} ~, \quad f_{\bar \nu_\ell} = f_{\bar \nu_e} = f_{\bar \nu_\mu} = f_{\bar \nu_\tau} ~,\\
	f_{u_L} &= f_{c_L} ~, \quad f_{\bar u_L} = f_{\bar c_L} ~, \quad
	f_{u_R} = f_{c_R} ~, \quad f_{\bar u_R} = f_{\bar c_R} ~, \\
	f_{d_L} &= f_{s_L} ~, \quad f_{\bar d_L} = f_{\bar s_L} ~, \quad
	f_{d_R} = f_{s_R} ~, \quad f_{\bar d_R} = f_{\bar s_R} ~.
\end{split}\ee
The fact that PDFs for right-handed fermions and their conjugate are different is due to the induced polarisation effects in the gluon, photon, and $Z$ boson PDFs.
The independent degrees of freedom are given in Table~\ref{tab:SM_DOF} in the case of PDFs of a muon beam, the final count is 42 independent PDFs.
Regarding the mass thresholds, we include each degrees of freedom right at the corresponding mass scale.

\begin{table}
\centering
\begin{tabular}{|c | c c c c c c c c c |}\hline
    Leptons         & $\mu_L$ & $\mu_R$ & $e_L$ & $e_R$ & $\nu_\mu$ & $\nu_e$ & $\bar{\ell}_L$ & $\bar{\ell}_R$ & $\bar{\nu}_\ell$ \\
    Quarks          & $u_L$ & $d_L$ & $u_R$ & $d_R$ & $t_L$ & $t_R$ & $b_L$ & $b_R$  & + h.c. \\
    Gauge Bosons    & $\gamma_\pm$ & $Z_\pm$ & $Z/\gamma_\pm$ & $W^\pm_\pm$ & $g_\pm$ &&&& \\
    Scalars         & $h$ & $Z_L$ & $h/Z_L$ & $W^\pm_L$ &&&&& \\\hline
\end{tabular}
\caption{\label{tab:SM_DOF} Independent degrees of freedom for our DGLAP evolution above the EW scale.}
\end{table}

\subsection{Electroweak symmetry breaking effects}
\label{sec:EWbreaking}

EW symmetry breaking effects can be classified as either due to non-vanishing parton masses or as new contributions to the splitting functions that vanish in the  limit of unbroken symmetry \cite{Chen:2016wkt}.
The latter are the so-called \emph{ultra-collinear} splittings, which are particularly relevant for the longitudinal polarisations of EW gauge bosons \cite{Chen:2016wkt,Cuomo:2019siu}, providing the leading contributions for their PDFs even when $Q \gg Q_{\rm EW}$.
In order to easily keep track of all these effects near the EW scale it is convenient to work in the broken phase and with the mass eigenstates. Then, as the DGLAP equations remain the same also for higher energies, we remain in the broken phase for any $Q > Q_{\rm EW}$.
We compute the splitting functions in the Goldstone equivalence gauge, following Ref.~\cite{Chen:2016wkt}. However, we checked that the results are the same as the ones in \cite{Cuomo:2019siu}, in which the theory is formulated in a standard $R_\xi$-gauge, with a $5$-dimensional polarisation vector for the longitudinal component of gauge bosons to take into account the Goldstone contribution.

In case any of the particles involved in the splitting $A \to B + C$ is massive, the kinematics of the process and the ensuing splitting functions are modified.
The relation between the $p_T$ and the virtuality of the parton $B$ entering the hard scattering process becomes
\be
	\widetilde{p}_T^2 \equiv \zb (m_B^2 - p_B^2) = p_T^2 + z m_C^2 + \zb m_B^2 - z \zb m_A^2 + \mathcal{O}\left(\frac{m^2}{E^2}, \frac{p_T^2}{E^2}\right)~,
	\label{eq:pT_virtuality}
\ee
where $\bar{z} = 1-z$ and $m^2/E^2$ or $p_T^2/E^2$ terms can be neglected in the regime where factorization can be applied.
In the DGLAP equations, this modified propagator of the virtual parton $B$ effectively corresponds to a rescaling of the splitting functions as \cite{Chen:2016wkt}
\be
	P_{BA}^C(z) \quad \to \quad \widetilde{P}_{BA}^C(z, p_T^2) = \left( \frac{p_T^2}{\widetilde{p}_T^2} \right)^2 P_{BA}^C(z)~. \label{eq:splitting_tilde}
\ee
Mass effects in matrix elements that are also present in the unbroken (i.e. massless) theory are instead suppressed by powers of $m/E$, therefore negligible.

Regarding the ultra-collinear splitting functions, the main feature of these contributions is that they do not scale logarithmically with $Q^2$ but at large factorization scales are suppressed as $v^2 / Q^2$ . Nevertheless, they provide important contributions that accumulate during the DGLAP evolution in the region $Q \sim Q_{\rm EW}$ and then remain almost constant at higher scales, see for instance the $W_L$ PDF in the right panel of Fig.~\ref{fig:EWA}.
We report the full set of splitting functions in App.~\ref{app:splitting}, while the list of SM DGLAP equations can be found in App.~\ref{app:DGLAP}.

In principle, with massive partons we should care of kinematics bounds: the splittings described by DGLAP equations involve partons $A$, $B$ and $C$ with energies $zE$, $xE$ and $(z-x)E$ respectively, all fractions of the energy of the beam $E$. Since the particle $C$ is emitted on-shell, we need $E_C\geq m_C$, that is $z\geq x + \frac{m_C}{E}$.
This means that the lower extreme of integration in Eq.~\eqref{eq:DGLAP_massless} should be modified. However, as in Eq.~\eqref{eq:pT_virtuality}, $m/E$ terms can be neglected and we can safely start the integration from $x$.

\subsection{Electroweak double logarithms}
\label{sec:EWdoublelogs}

The fact that the initial and final states are EW non-singlets has important implications, even for inclusive processes. 
The Bloch-Nordsieck theorem \cite{Bloch:1937pw}, that guarantees cancellation of IR divergencies between real emission and virtual corrections in such processes, is violated for non-abelian symmetries, which implies the presence of Sudakov double logarithms. While in the QCD case this effect vanishes upon averaging over color of the initial states, for $SU(2)_L$ we do not take such average and the initial state breaks explicitly the symmetry, hence double logs do appear also in inclusive processes~\cite{Ciafaloni:1998xg,Ciafaloni:2000gm,Ciafaloni:2000df,Ciafaloni:2000rp,Ciafaloni:2001vt}.
In our context they can be seen appearing in the terms of the DGLAP equations containing a $1/(1-z)$ pole and can be made explicit by introducing an IR cutoff in the integral, which allows also to resum the Sudakov double logs related to ISR~\cite{Amati:1980ch,Ciafaloni:2001mu,Bauer:2017isx,Bauer:2017bnh,Manohar:2018kfx}.\footnote{A complete description of all double logarithms in a full process requires, however, the inclusion of further contributions (e.g. virtual corrections, soft radiation, fragmentation, etc.), see for instance \cite{Ciafaloni:1998xg,Denner:2000jv,Denner:2001mn,Manohar:2018kfx,Pagani:2021vyk,Chen:2022msz,Chay:2022qlc}.}
We implement this following Ref.~\cite{Bauer:2017isx}, by modifying the boundaries of the integral in Eq.~\eqref{eq:DGLAP_massless} as
\be
   \frac{\alpha_{ABC}(Q)}{2\pi} \int_x^1 \frac{dz}{z} P_{BA}^C(z) f_A\left(\frac{x}{z}, Q^2 \right) \quad \to \quad
   \frac{\alpha_{ABC}(Q)}{2\pi} \int_x^{z_{\rm max}^{ABC}(Q)} \frac{dz}{z} P_{BA}^C(z) f_A\left(\frac{x}{z}, Q^2\right)~,
\ee
where $z_{\rm max}^{ABC}(Q)$ plays the role of an explicit IR cutoff for the $1/(1-z)$ poles, and is set equal to 1 (in which case we use the $+$-distribution to regulate the divergence and perform the numerical evaluation) except for the cases where the soft divergence is not cancelled between real emission and virtual corrections, in which case we set $z_{\rm max}^{ABC}(Q) = 1 - Q_{\rm EW} / Q$.
This modifies the computation of the virtual corrections, that becomes (see App.~\ref{app:radiative})
\be
    P^v_A(Q) \supset - \sum_{B,C} \frac{\alpha_{ABC}(Q)}{2\pi} \int_0^{z_{\rm max}^{ABC}(Q)} dz \, z \, P^C_{BA}(z) + \text{ultra-collinear}~.
\ee

The mismatch of IR divergencies between real and virtual contributions takes place in processes where the emitted radiation (e.g. a $W^\pm$ boson) changes the $SU(2)_L$ component of the initial state, and is always proportional to the amount of explicit breaking of the $SU(2)_L$ symmetry.
In fact, the physical effect of these double logs is to restore $SU(2)_L$ invariance at high scales. The case of PDFs of a proton is discussed in Ref.~\cite{Bauer:2017bnh}.
As an explicit example for leptons, let us consider the DGLAP equations for the $\mu_L$ and the corresponding $SU(2)_L$ partner $\nu_\mu$
\be\begin{split}
    \frac{d f_{\mu_L}}{d \log Q^2} &= \frac{\alpha_2}{2\pi} \frac{1}{2} \int_0^{z_{\rm max}(Q)} dz P_{ff}^V(z) \left( \frac{1}{z} f_{\nu_\mu}\left(\frac{x}{z}, Q^2 \right) - z f_{\mu_L}(x,Q^2)\right) + \ldots ~,\\
    \frac{d f_{\nu_\mu}}{d \log Q^2} &= \frac{\alpha_2}{2\pi} \frac{1}{2} \int_0^{z_{\rm max}(Q)} dz P_{ff}^V(z) \left( \frac{1}{z} f_{\mu_L}\left(\frac{x}{z}, Q^2 \right) - z f_{\nu_\mu}(x,Q^2)\right) + \ldots ~,
    \label{eq:doublelogexample}
\end{split}\ee
where the ellipses include other finite contributions to the integral, as well as other interactions. In the two parentheses, the first term is due to real emission of a $W^{\pm}_T$ boson, while the second is the corresponding virtual correction. The fact that the muon and neutrino PDFs are different is an explicit breaking of $SU(2)_L$ and implies a non-cancellation of the pole for $z\to 1$ inside $P_{ff}^V(z)$, which in turn generates the double log. To see this explicitly, let us isolate on the right-hand side only the terms that are singular in $z\to 1$, fixing $z=1$ everywhere else:
\be\begin{split}
    \frac{d f_{\mu_L}}{d \log Q^2} &\approx - \frac{\alpha_2}{4\pi} \Delta f_{L_2}(x) \, \int_0^{z_{\rm max}(Q)} dz \frac{2}{1-z} + \ldots \approx - \frac{\alpha_{2}}{4\pi} \log \frac{Q^2}{Q_{\rm EW}^2} \Delta f_{L_2}(x) + \ldots ~, \\
    \frac{d f_{\nu_\mu}}{d \log Q^2} &\approx \frac{\alpha_2}{4\pi} \Delta f_{L_2}(x) \int_0^{z_{\rm max}(Q)} dz \frac{2}{1-z} + \ldots \approx \frac{\alpha_{2}}{4\pi} \log \frac{Q^2}{Q_{\rm EW}^2} \Delta f_{L_2}(x) + \ldots ~,
\end{split}\ee
where $\Delta f_{L_2} \equiv f_{\mu_L} - f_{\nu_\mu}$.
Upon integration in $Q^2$, a $\log^2 Q^2 / Q_{\rm EW}^2$ contribution is generated, that tends to deplete the muon PDF and enhance the neutrino one.
This example also clearly shows how no such double log is generated for photon or $Z_T$ emission from a fermion or $W$, since the real and virtual contribution would be proportional to the PDF of the same parton.
A Sudakov double log is instead expected for splittings such that the splitting function is divergent in the soft limit, $z \to 1$, and the $A$ and $B$ partons are different:
\be
    z_{\rm max}^{ABC}(Q) = 1 - \frac{Q_{\rm EW}}{Q} \qquad
    \text{if} \quad P_{BA}^C, ~ U_{BA}^C \propto \frac{1}{1-z} \text{ and } A \neq B~,
\ee
otherwise we put $z_{\rm max} = 1$ and employ the standard $+$-distribution (see Eq.~\eqref{eq:def+}) to regulate the $z \to 1$ divergence, for a more stable numerical evaluation of the DGLAP equations.

In practice, this happens for $W^\pm_T$ emission off any parton (in correspondence to the poles in the $P_{ff}^V$,  $P_{VV}^V$, and  $P_{hh}^V$ splittings) and for $Z_T$ boson emission from an initial longitudinal $Z_L$ or Higgs (due to  the $P_{hh}^V$ splitting), since $Z$ emission changes $Z_L \longleftrightarrow h$.
Analogously, for ultra-collinear splittings this takes place for any $W_L$ emission and for $Z_L$ emission off and initial Higgs or $Z_L$.

This procedure amounts to a double-logarithmic (DL) approximation to the Sudakov factor for initial-state radiation. This could be further improved to LL or NLL resummation by suitably modifying the scale at which the coupling constant $\alpha_2(Q)$ is evaluated, as discussed in Refs.~\cite{Bauer:2018xag,Bauer:2018arx}. Since we are interested in energies where $\alpha_2 \log^2 Q^2/Q_{\rm EW}^2 \sim \mathcal{O}(1)$ but $\alpha_2 \log Q^2/Q_{\rm EW}^2\ll 1 $, we limit ourselves to the DL approximation for electroweak corrections in the present work.

\subsection{Top quark as parton}
\label{sec:top}

When the energy of the hard process $Q$ is much larger than the top mass, processes with a collinearly emitted top quark can develop a logarithmic enhancement proportional to $\log Q^2 / m_t^2$. It can therefore become useful to resum these logarithms by including the top quark among the other partons \cite{Barnett:1987jw,Olness:1987ep}.
The question of whether or not one should include it as a parton depends on the process considered and on the optimal way to rearrange the perturbative series \cite{Dawson:2014pea,Han:2014nja}.
With this in mind, we provide two versions of our PDFs of leptons, one in the 5-flavour-scheme (5FS) and one in a 6FS, where the top quark is added in the DGLAP evolution for scales above $m_t$.
While codes that include the top quark in proton PDFs assume it is massless, in our approach we keep a finite top mass in the same spirit in which we keep finite $W$ and $Z$ masses for the weak bosons PDFs.
This is justified by the fact that in our case, contrary to proton PDFs, EW interactions and EW symmetry breaking effects are crucial, and the top mass is one of such effects. For a detailed discussion of different schemes for the top mass in the computation of hadron collider observables with a top quark PDF see \cite{Han:2014nja}. 

The DGLAP equations for the $t_{L}$, $\bar t_L$, $t_R$, $\bar t_R$ are reported explicitly in Eqs.~(\ref{eq:tL_DGLAP}, \ref{eq:tLb_DGLAP}, \ref{eq:tR_DGLAP}, \ref{eq:tRb_DGLAP}).
We checked numerically that the dominant contributions are those from initial transverse gauge bosons, with electroweak bosons, photon and gluon terms being approximately of similar size. Instead, ultra-collinear contributions are practically negligible. Nevertheless, in our numerical evaluation we keep all terms.

\newpage
\subsection{Effective Vector Boson Approximation}
\label{sec:EVA}

The EPA has been generalized to describe EW gauge bosons in high-energy collisions since the '80s \cite{Cahn:1983ip,Dawson:1984gx,Chanowitz:1984ne,Kane:1984bb}, in what is now known as the Effective Vector Boson Approximation (EVA) \cite{AlAli:2021let,Ruiz:2021tdt}. When the hard scattering energy is much larger than the EW gauge boson mass and the $p_T$ of the collinearly emitted gauge boson, then the cross section of the process can be factorized into an almost on-shell collinear emission and the subsequent hard scattering \cite{Kunszt:1987tk,Borel:2012by,Buttazzo:2018qqp,Costantini:2020stv}.

\begin{figure}[t]
\centering
\includegraphics[height=5.3cm]{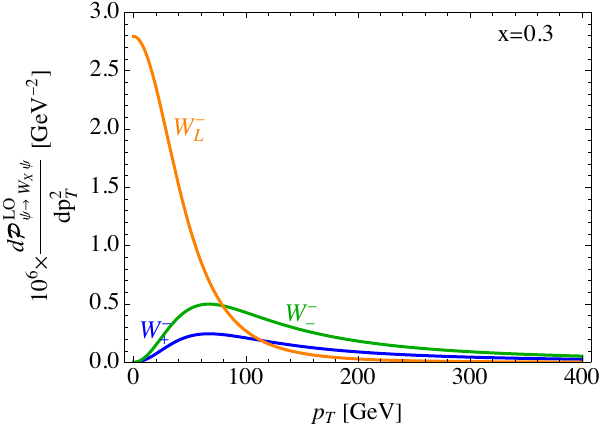} 
\includegraphics[height=5.3cm]{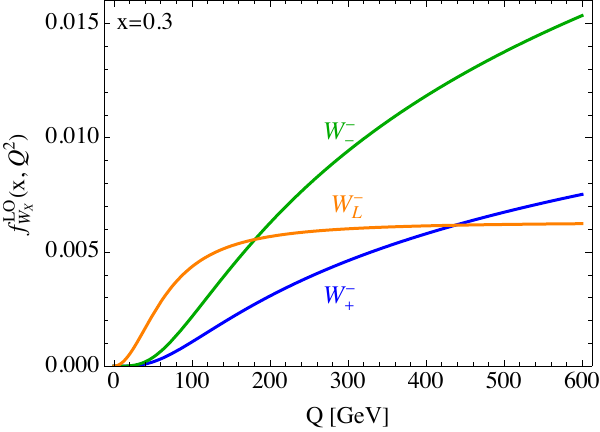} 
\caption{\label{fig:EWA} Integrand functions  (left) and resulting PDFs (right) for the computation of the EW PDFs for transverse and longitudinal $W$ boson of Eqs.~(\ref{eq:EWA_WT},\ref{eq:EWA_WL}).}
\end{figure}

One can compute the EW gauge bosons PDFs by evaluating the DGLAP equations at fixed order, similarly to what we did in Section~\ref{sec:QED} for QED, using $f_{\mu_L}^{(0)}(x) = f_{\mu_R}^{(0)}(x) = \frac{1}{2} \delta(1-x)$ for the muon PDF.
Taking the DGLAP equations for the transverse and longitudinal $W^-$ boson, Eqs.~(\ref{eq:Wmp_DGLAP},\ref{eq:Wmm_DGLAP},\ref{eq:WmL_DGLAP}) one gets the leading order results
\begin{eqnarray}
    f_{W^-_\pm}^{(\alpha)}(x, Q^2) &=& \int_{m_\mu^2}^{Q^2} d p_T^2 \frac{1}{2} \frac{d \mathcal{P}_{\psi \to W_T \psi}}{d p_T^2}(x, p_T^2) \, 
    = \int_{m_\mu^2}^{Q^2} d p_T^2 \, \frac{\alpha_2}{8 \pi} \frac{p_T^2}{(p_T^2 + (1-x) m_W^2)^2} P_{V_\pm f_L}^f(x) = \nonumber \\
    &=& \frac{\alpha_2}{8 \pi} P_{V_\pm f_L}^f(x) \left( \log \frac{Q^2 + (1-x) m_W^2}{m_{\mu}^2 + (1-x) m_W^2} 
    - \frac{Q^2}{Q^2 + (1-x) m_W^2} \right) = \nonumber \\
    &\approx& \frac{\alpha_2}{8 \pi} P_{V_\pm f_L}^f(x) \left( \log \frac{Q^2}{m_W^2} - \log(1-x) - 1\right) + \mathcal{O}\left( \frac{m_W^2}{Q^2} \right)~, \label{eq:EWA_WT} \\
    f_{W^-_L}^{(\alpha)}(x, Q^2) &=& \int_{0}^{Q^2} d p_T^2 \frac{1}{2} \frac{d \mathcal{P}_{\psi \to W_L \psi}}{d p_T^2} (x, p_T^2) \, 
    = \int_{0}^{Q^2} d p_T^2 \, \frac{\alpha_2}{4 \pi} \frac{m_W^2}{(p_T^2 + (1-x) m_W^2)^2} \frac{(1-x)^2}{x} = \nonumber \\
    &=& \frac{\alpha_2}{4 \pi} \frac{1-x}{x}
    \frac{Q^2}{Q^2 + (1-x) m_W^2}
    \approx \frac{\alpha_2}{4 \pi} \frac{1-x}{x} + \mathcal{O}\left( \frac{m_W^2}{Q^2} \right) ~,\label{eq:EWA_WL}
\end{eqnarray}
and analogously for the $Z$ and $Z/\gamma$ PDFs
\begin{eqnarray}
    f_{Z_\pm}^{(\alpha)}(x, Q^2) &=& 
    \frac{\alpha_2}{4 \pi c_W^2} \left(P_{V_\pm f_L}^f(x) (Q^Z_{\mu_L})^2 + P_{V_\pm f_R}^f(x) (Q^Z_{\mu_R})^2 \right) \nonumber \\
    && \quad \left( \log \frac{Q^2 + (1-x) m_Z^2}{m_{\mu}^2 + (1-x) m_Z^2} 
    - \frac{Q^2}{Q^2 + (1-x) m_Z^2} \right) ~, \label{eq:EWA_ZT} \\
    f_{Z/\gamma_\pm}^{(\alpha)}(x, Q^2) &=& 
    -\frac{\sqrt{\alpha_\gamma \alpha_2}}{2 \pi c_W} \left(P_{V_\pm f_L}^f(x) Q^Z_{\mu_L} + P_{V_\pm f_R}^f(x) Q^Z_{\mu_R} \right) \log \frac{Q^2 + (1-x) m_Z^2}{m_{\mu}^2 + (1-x) m_Z^2}  ~, \label{eq:EWA_ZgaT} \\
    f_{Z_L}^{(\alpha)}(x, Q^2) &=&
    \frac{\alpha_2}{2 \pi c_W^2} \frac{1-x}{x} \left( (Q^Z_{\mu_L})^2 + (Q^Z_{\mu_R})^2\right) 
    \frac{Q^2}{Q^2 + (1-x) m_Z^2} ~,\label{eq:EWA_ZL}
\end{eqnarray}
where $P_{V_+ f_L}^f(x) = P_{V_- f_R}^f(x) = (1-x)^2 / x$ and $P_{V_- f_L}^f(x) = P_{V_+ f_R}^f(x) = 1 / x$.
The muon mass here serves as an IR cutoff for the logarithm in the transverse case to cure the $x\to 1$ limit, while we neglect it in the other terms. Notably, the $W^+$ has no contribution at this order.

In Fig.~\ref{fig:EWA} we show the dependence in $\sqrt{p_T^2}$ of the integrands (left), and the resulting PDFs (right), fixing a value $x=0.3$ and showing separately the two polarisation of the transverse $W^-_{\pm}$. One can see that the integrands are peaked before the EW scale and, while in the case of $W_T$ it decreases as $\sim 1 / p_T^2$ inducing the logarithmic grow of the PDF, for the longitudinal $W$ polarisation the contribution to the integral is localised in $p_T^2$ before the EW scale and the PDF tends to a constant at large scales.

In our numerical integration of DGLAP equations, the effects due to EW interactions are introduced only above the $Q_{\rm EW}$ matching scale. Since we employ $p_T$ as factorization scale, this effectively corresponds to performing the integration in Eqs.~(\ref{eq:EWA_WT},\ref{eq:EWA_WL}) only for $p_T^2 > Q_{\rm EW}^2$, missing the region $0 < p_T^2 < Q^2_{\rm EW}$. To address this issue we match the gauge bosons PDFs at $Q_{\rm EW}$ to the analytically computed one for the same scale and use it as boundary conditions,
\be
    f_{A}(x, Q_{\rm EW}^2) \equiv f_{A}^{(\alpha)}(x, Q_{\rm EW}^2)~
    \quad\text{for}\quad
    A = W^-_{L,\pm},~Z_{L,\pm},~Z/\gamma_{\pm}~,
\ee
and then continue the integration numerically to higher scales.
The boundary conditions for the PDFs of other heavy states ($h$, $h/Z_L$, top quark) are instead set to zero at the corresponding mass scales.

\section{Results}
\label{sec:results}

\begin{figure}[t]
\centering
\includegraphics[width=12cm]{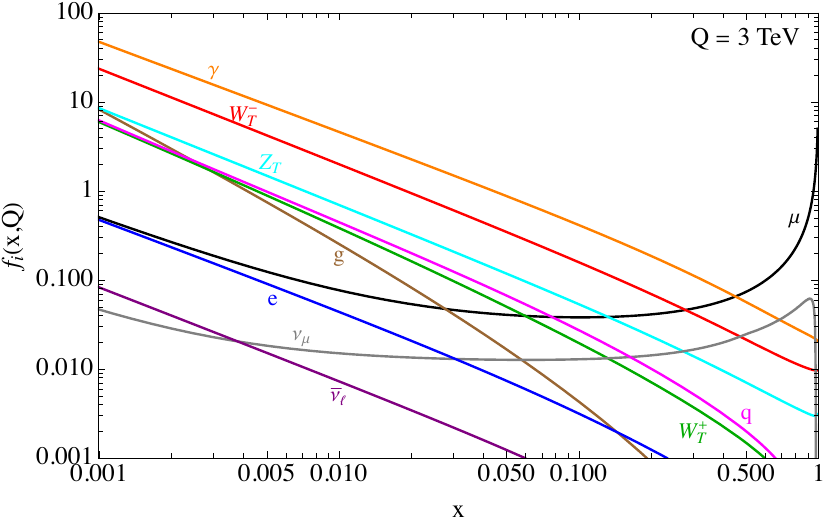}\\
\includegraphics[width=12cm]{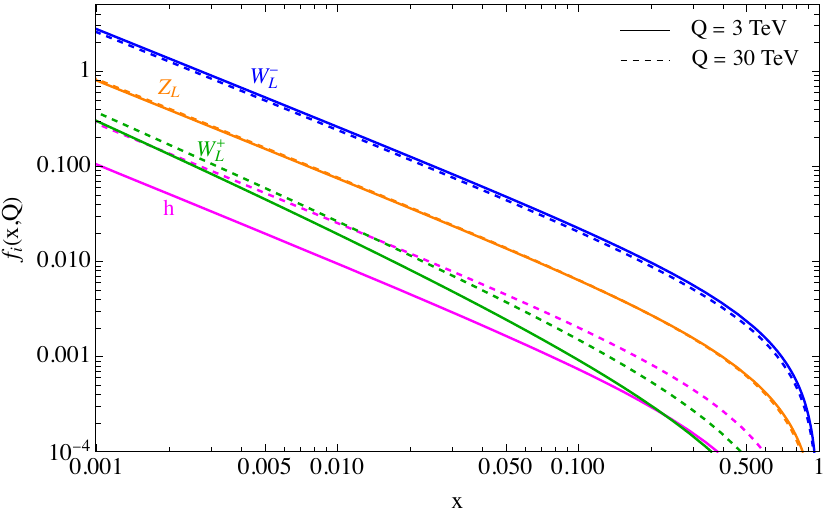}
\caption{\label{fig:summaryPDF} (Top panel): Sample of PDFs evaluated at a scale $Q = 3 ~\TeV$ for a muon. For this plot we sum over polarisations and $q$ represents the sum of all quark PDFs except for the top. (Bottom panel): PDFs for the scalar degrees of freedom in the SM. Solid (dashed) lines are evaluated at a scale of $Q= \, 3 \, (30) ~\TeV$.}
\end{figure}

\begin{table}
\centering
\begin{tabular}{ |c||c|c|c|c| }
\hline
 field & $Q = 3~\rm TeV$ & $Q = 10~\rm TeV$ & $Q = 30~\rm TeV$\\
 \hline
$\mu_L$   & $49.48 \%$ & $48.72 \%$    &  $47.76 \%$ \\
$\mu_R$ & $46.98 \%$ & $44.12 \%$     &  $41.12 \%$ \\
$\nu_\mu$ & $1.28 \%$ & $2.83 \%$    &  $4.85 \%$ \\
$\nu_\ell$ & $0.0004 \%$ & $0.0009 \%$    &  $0.001 \%$ \\
$\ell$ & $0.005 \%$ & $0.007 \%$    &  $0.01 \%$ \\
$q$ & $0.038 \%$ & $0.05 \%$    &  $0.07 \%$ \\
$\gamma$ & $1.3 \%$ & $1.4 \%$    &  $1.46 \%$ \\
$W_T^-$  & $0.52 \%$ & $0.64 \%$    &  $0.74 \%$ \\
$W_T^+$  & $0.03 \%$ & $0.06 \%$    &  $0.11 \%$ \\
$Z_T$ & $0.17 \%$ & $0.22 \%$    &  $0.28 \%$ \\
$g$   & $0.001 \%$ & $0.002 \%$    &  $0.003 \%$ \\
 \hline
\end{tabular}
\caption{Fraction of the momentum carried by each parton at $Q = 3 ,10,30~\rm TeV$.}
\label{tab:momemtum_frac}
\end{table}

Here we discuss several aspects of our LePDFs. The details of our numerical implementation of the DGLAP equations are collected in App.~\ref{app:numerical}.

Fig.~\ref{fig:summaryPDF} (top panel) collects, as an example, a set of PDFs evaluated at the scale $Q = 3~\TeV$.
One first thing to notice is that, as expected, for $x \gtrsim 0.5$ the muon PDF dominates, while for smaller $x$ the largest PDF is the photon one. However, the transverse negative $W^-_T$ PDF is only a factor $\sim 2$ smaller and the transverse $Z_T$ boson is another factor of $2$ smaller than that: they both receive contributions from the emission off an initial muon. Analogously, the muon neutrino $\nu_\mu$ has a large PDF at large $x$ values due to the emission off a $\mu_L^-$, which has also a Sudakov double-log enhancement. The positive transverse $W^+_T$ PDF is instead more suppressed because its leading contribution arises from the emission off the muon neutrino and off another gauge boson.
The importance of EW gauge bosons PDFs reflects the common lore that \emph{a high-energy lepton collider is also a weak boson collider}.

In the bottom panel of Fig.~\ref{fig:summaryPDF} we show the PDFs for the longitudinal polarisations of EW gauge bosons and the Higgs, evaluated at scales of 3~TeV (solid lines) and 30~TeV (dashed lines). The PDFs for $W^-_L$ and $Z_L$ are mostly scale independent, since they receive the dominant contribution from the ultra-collinear splitting off a muon. On the other hand, the ultra-collinear contribution to the $W^+_L$ PDF comes mostly from the muon neutrino, which has a PDF suppressed with respect to the muon one. Therefore, other contributions from standard splitting functions (e.g. from $P_{hV}^h$ and $P_{hh}^V$) are sizeable and induce a scale dependence.
In case of the Higgs boson there is no ultra-collinear contribution from massless fermions, so one does not expect ultra-collinear terms to dominate and indeed its PDF shows a large scale dependence.

\begin{figure}[t]
\centering
\includegraphics[height=5.5cm]{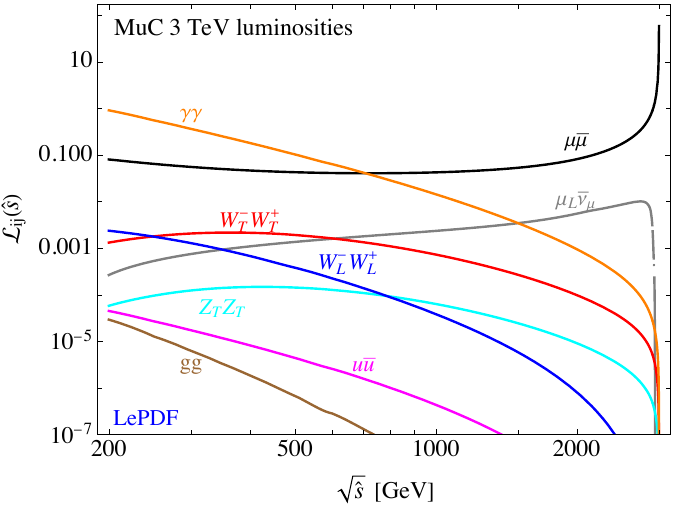}
\includegraphics[height=5.5cm]{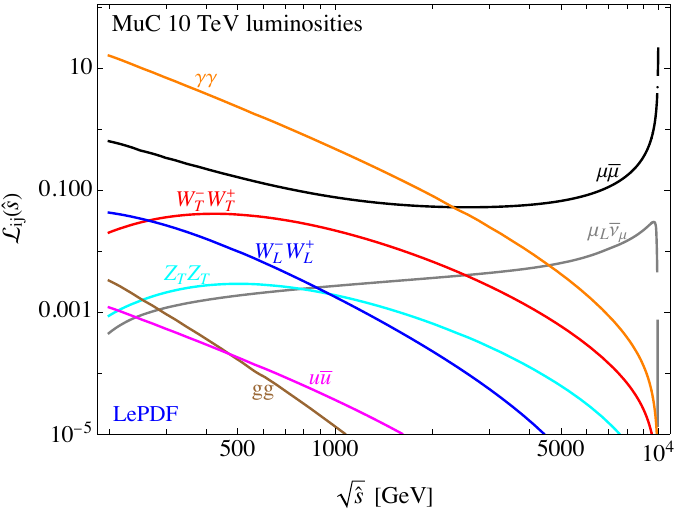}
\caption{\label{fig:LumiMuC}Examples of parton luminosities at a 3~TeV (left) and 10~TeV (right) MuC. Unless specified, for this plot we sum over polarizations.}
\end{figure}

The fraction of the momentum carried by each of the partonic components is given by the $n=2$ Mellin transform of the PDF (see App. \ref{app:radiative} for more details) and it can be used to evaluate the relevant role of the various individual components at different scales. To this end, in Table \ref{tab:momemtum_frac} we give three examples for the set of PDFs shown in Fig. \ref{fig:summaryPDF} at scales $3~\rm TeV$, $10~\rm TeV$, and $30~\rm TeV$. We observe that as the energy of the hard process is increased the percentages of both the left- and right-handed muon components are decreased and those of all other partons increased, which illustrates the importance of electroweak interactions at higher energies.

In Fig.~\ref{fig:LumiMuC} we show some examples of parton luminosities for a 3 and 10 TeV muon colliders where, unless specified, we sum over polarizations. Parton luminosities can be useful for computing cross sections integrated over angular variables. In case of a muon collider they are defined from the convolution of the PDFs of parton $i$ from the muon and parton $j$ from the anti-muon, as follows:
\be
     \mathcal{L}_{ij}(\hat{s}) \equiv \int_{\hat{s}/s_0}^1 \frac{dx}{x} f_i^{(\mu)}\left(x, \frac{\sqrt{\hat{s}}}{2} \right) \, f_j^{(\bar{\mu})}\left(\frac{\hat{s}}{x s_0}, \frac{\sqrt{\hat{s}}}{2} \right)~,
\ee
where $\sqrt{s_0}$ is the collider center of mass energy and $\sqrt{\hat{s}}$ is the invariant mass of the two-parton system.
From Fig.~\ref{fig:LumiMuC} we can notice that, even at small invariant masses, the $\mu\bar{\mu}$ luminosity is  much larger than the $W^+W^-$ luminosity (also the charged-current $\mu_L\bar{\nu}_\mu$ luminosity is sizeable). The impact of this channel in VBF studies at muon colliders should therefore be studied in more details.
It is also interesting to point out that the QCD-related luminosities (gluons and quarks) are very small, which is going to strongly suppress QCD-induced backgrounds in electroweak processes.

\subsection{Polarisation}
\label{sec:polarisation}

\begin{figure}[t]
\centering
\includegraphics[height=3.9cm]{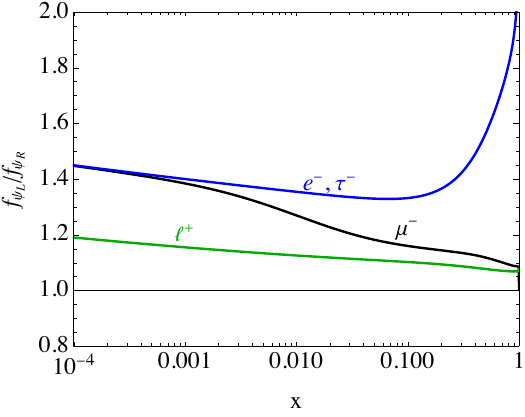} 
\includegraphics[height=3.9cm]{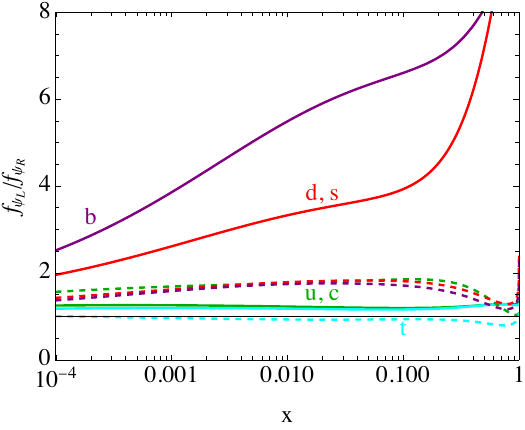} 
\includegraphics[height=3.8cm]{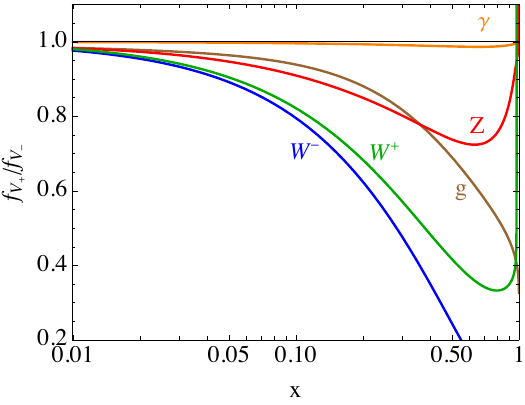} 
\caption{\label{fig:polarisation}Polarisation ratios for the PDFs of leptons (left), quarks (center), and gauge bosons (right) at a scale $Q = 3 ~\TeV$. Dashed lines in the central panel are for the corresponding anti-quarks.}
\end{figure}

The chiral structure of SM interactions above the EW scale induces polarisation effects for the PDFs \cite{Bauer:2018arx}. In Fig.~\ref{fig:polarisation} we show polarisation ratios for several PDFs at a scale $Q=3~\TeV$. The observed behavior can be easily understood as follows.
The $W^-_T$, $W^+_T$, and $Z_T$ PDFs receive the dominant contribution from the emission off an initial $\mu_{L,R}^-$ or $\nu_\mu$. Since the $P_{V_+ f_L}^f$ splitting function goes to zero for $z\to 1$, while $P_{V_- f_L}^f$ tends to a constant, the positive helicity of the EW gauge bosons will be suppressed for $x \to 1$. In case of the photon, the leading contribution comes from the muon splitting and it is vector-like at leading order, so the polarisation effect will be suppressed.

In case of fermions, left-handed chiralities (and their conjugate) receive contributions from $W$ bosons splitting to $\psi_L \bar \psi^\prime_L$, therefore their PDF is expected to be larger than the right-handed counterparts. The tendency increases with $x$ in the case of the leptons (except for the muon) and down-type quarks as opposed to antileptons and up-type quarks, since the left-handed parts of the former receive contributions from $W^-_T$, while the latter from $W^+_T$, and the $W^+_T$ PDF falls faster than the $W^-_T$ PDF at high $x$ (see Fig. \ref{fig:summaryPDF}) This effect can be of $\mathcal{O}(1)$, since the $W^-$ PDF is comparable in size to the photon one.
The $b_L$ PDF is further enhanced compared to $b_R$ due to the $W_L^- \to b_L \bar{t}_R$ splitting proportional to $y_t$.

\newpage
\subsection{Comparison with the Effective Vector Boson Approximation}

\begin{figure}[t]
\centering
\includegraphics[height=6.5cm]{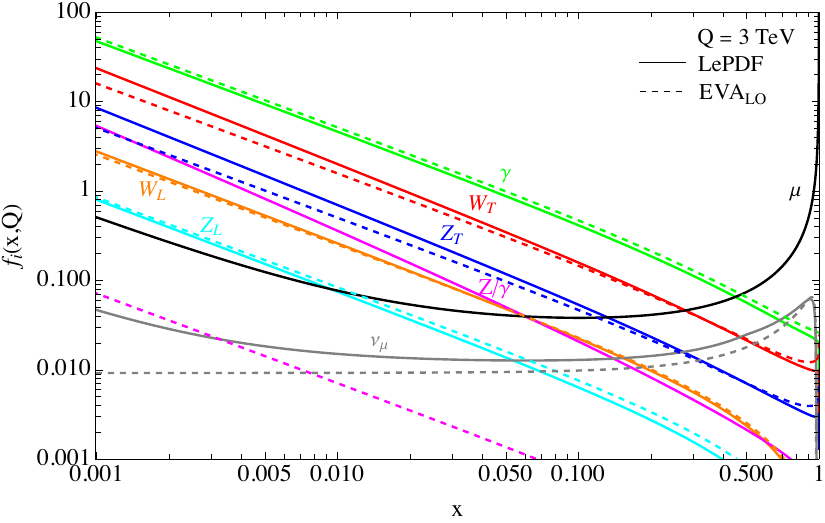}
\caption{\label{fig:EWgaugePDF} PDFs for EW gauge bosons (plus muon and muon neutrino) at a scale $Q=3~\TeV$. Solid lines are from the numerical solution of DGLAP equations while dashed ones are from the LO EVA expressions in Eqs.~\eqref{eq:EWA_WT}-\eqref{eq:EWA_ZL}.}
\end{figure}

In Fig.~\ref{fig:EWgaugePDF} we show our results for EW gauge bosons PDFs, compared with the LO EVA result discussed in Section~\ref{sec:EVA} (to reduce the number of lines plotted we show the sum of transverse polarisations).
Several things can be noticed.

For $Z/\gamma$, the EVA result is two orders of magnitude smaller than what we find with the numerical evolution.
This is due to the fact that $Q^Z_{\mu_L} + Q^Z_{\mu_R} = - \frac{1}{2} + 2 s_W^2 \ll 1$ is accidentally suppressed (indeed, it becomes zero when evolving the Weinberg angle at a scale of about 3.6TeV). This cancellation takes place because in EVA it is assumed that the initial-state muon is not polarised. However, in the evolution from the EW scale upward, electroweak interactions induce a polarisation of the muon PDF, which becomes up to $\sim 40\%$ at a scale of 3 TeV, as can be seen in Fig.~\ref{fig:polarisation} (left panel). Therefore, in the full numerical evolution there is no such tuned cancellation in the $Z/\gamma$ PDF. This clearly shows that the EVA result is not reliable for this PDF.

\begin{figure}[t]
\centering
\includegraphics[height=5.5cm]{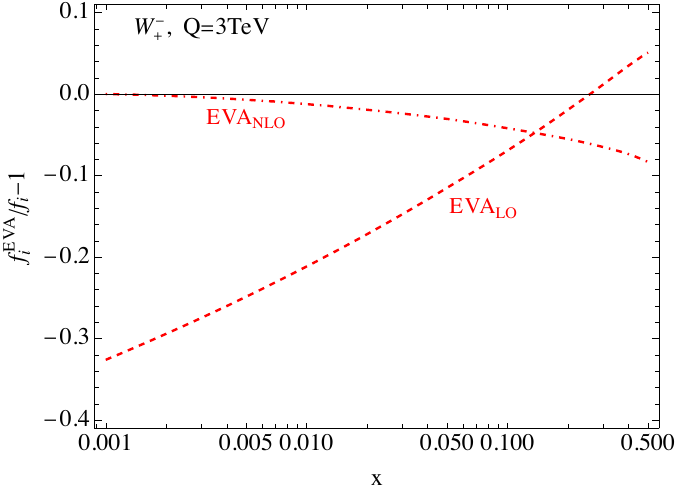} 
\includegraphics[height=5.5cm]{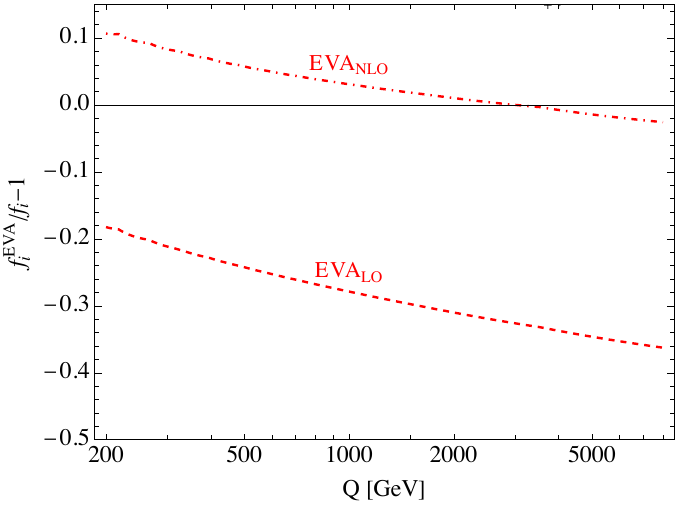} 
\caption{\label{fig:PDFvsEVA} Relative difference between the numerical result for the $W^-_+$ PDF and the LO (dashed) or NLO (dot-dashed) EVA expression for a fixed scale $Q = 3 ~\TeV$ (left) and for fixed $x=0.001$ (right).}
\end{figure}

\begin{figure}[t]
\centering
\includegraphics[height=8cm]{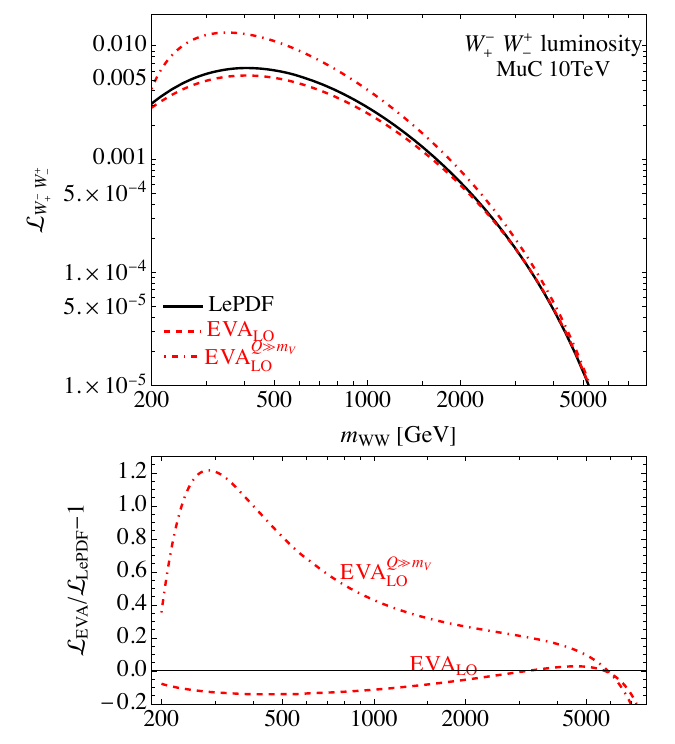} \includegraphics[height=8cm]
{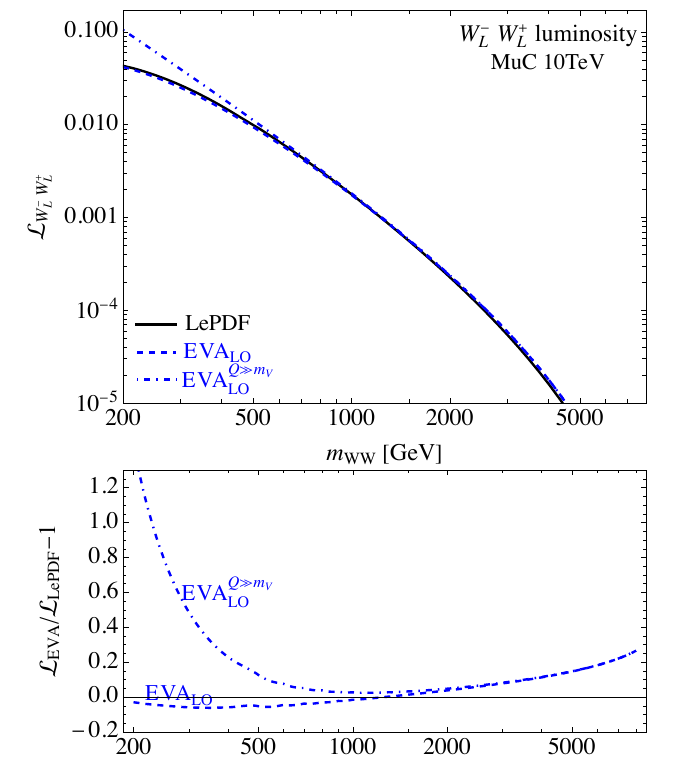} 
\caption{\label{fig:lumiWW}Parton luminosities for $W^-_+ W^+_-$ (left) and $W^-_L W^+_L$ (right) at a 10 TeV muon collider. We show a comparison with luminosities obtained with the LO EVA result in Eq.~\eqref{eq:EWA_WT} (dashed) and with the $Q\gg m_W$ approximation, implemented in \cite{Ruiz:2021tdt} (dot-dashed).}
\end{figure}

For the longitudinal polarisations, the EVA provides a good description of the PDFs, to within $\sim 10\%$ accuracy.
In case of the transverse $W$ and $Z$ polarisations, instead, there is a noticeable discrepancy which grows even up to $\mathcal{O}(50\%)$ at multi-TeV scales for small $x$ values.
This is dominantly due to the missing contributions from $V \to V  V$ splittings, that start at NLO. Such contributions become important due to two effects: the PDF of the initial-state gauge boson at small $x$ is much larger than the muon PDF and they include IR Sudakov corrections that induce a parametric dependence as $\alpha^2 \log^3 (Q^2 / m_W^2)$.
We perform two checks to verify this. First, we do a run of our numerical code  setting to zero the $P_{VV}$ splitting functions (both in real emission and radiative corrections): the resulting EW gauge bosons PDFs agree well with LO EVA. Second, focussing only on the $W^-_+$ PDF for simplicity, we compute iteratively the $\mathcal{O}(\alpha^2)$ contributions to the weak bosons PDFs, adding the real emissions from $P_{VV}^V$ and $P_{Vf}^f$ splittings, and the corresponding virtual corrections.
In practice, we take the DGLAP equations for transverse EW gauge bosons in Section~\ref{sec:DGLAP_VT} and use the $\mathcal{O}(\alpha)$ results for the gauge bosons PDFs (i.e. the LO EVA of Eqs.~(\ref{eq:EWA_WT}-\ref{eq:EWA_ZL})) and the muon.
For simplicity, at this step we use approximate expressions for the LO EVA by keeping only the $\log \frac{Q^2}{m_{W,Z}^2}$ term, which makes our NLO EVA result reliable only for $Q \gg m_W$ and $x \ll 1$. We also neglect contributions from longitudinal modes and ultra-collinear ones. We then perform analytically the convolution with the splitting functions and finally the integral from $m_W$ up to the factorization scale $Q$:
\be\begin{split}
f_{\mu_L}^{(\alpha)}(x, t) &\simeq \int_{t_{m_W}}^{t} dt' \left( 
    \frac{1}{2} P_{\mu_L}^v(t') \delta(1-x) 
    + \frac{\alpha_\gamma}{4\pi} P_{ff}^V(x) 
    + \frac{\alpha_2}{4\pi c_W^2} (Q_{\mu_L}^Z)^2 P_{ff}^V(x) 
    \right)~, \\
    f_{W^-_+}^{(\alpha^2)}(x, t) &\simeq \int_{t_{m_W}}^{t} dt' \left(
    P_{W^-_+}^{v} f_{W^-_+}^{(\alpha)} 
    + \frac{\alpha_2}{4\pi} P_{V_+ f_L}^f \otimes f_{\mu_L}^{(\alpha)} 
    + \frac{\alpha_2}{2\pi} c_W^2 P_{V_+ V_s} \otimes (f_{W^-_s}^{(\alpha)} + f_{Z_s}^{(\alpha)}) + \right.\\
    & \left.  
    + \frac{\alpha_\gamma}{2\pi} P_{V_+ V_s} \otimes (f_{W^-_s}^{(\alpha)} + f_{\gamma_s}^{(\alpha)})
    + \frac{\sqrt{\alpha_{\gamma} \alpha_{2}}}{2\pi} c_W P_{V_+ V_s} \otimes f_{Z/\gamma_s}^{(\alpha)}
    \right)~.
\end{split}\ee
It can be noted that Sudakov double logs appear here in the virtual contributions to $f_{\mu_L}^{(\alpha)}$ and to $f_{W^-_+}^{(\alpha^2)}$, as well as in the $P_{V_+ V_+}$ terms from the neutral gauge bosons.

In Fig.~\ref{fig:PDFvsEVA} we show the relative deviation of the LO (dashed) and NLO (dot-dashed) EVA results from the complete numerical PDF as function of $x$ (left panel) and as function of the scale (right panel).
We observe that the NLO EVA result improves substantially the agreement with the full numerical result, while the LO EVA has large deviations at small $x$. The missing terms in the LO EVA become more and more important with larger scales, confirming the argument made above.

In Fig.~\ref{fig:lumiWW}, instead, we plot the $W^-_+ W^+_-$ (left panel) and $W^-_L W^+_L$ (right panel) parton luminosities for a 10 TeV MuC. We show a comparison between the LePDF result (solid lines) and the LO EVA expression in the $Q \gg m_W$ approximation (dot-dashed), that is the one implemented in Ref.~\cite{Ruiz:2021tdt}, or with the complete $W$ mass dependence as in Eq.~\eqref{eq:EWA_WT} (dashed). We see that, at the level of luminosity, the LO EVA with the complete mass dependence provides a good approximation of the resummed LePDF result up to $\sim 15\%$ deviations for the transverse modes. This means that the much larger deviations we observe at the PDF level for small $x$ are diluted when the luminosities are calculated. On the other hand, the massless approximation deviates up to $\mathcal{O}(1)$ even at the luminosity level and in particular at partonic center of mass energies of few hundreds of GeV, where the weak-boson fusion process cross sections are the largest.

\subsection{Muon neutrino PDF}
\label{sec:neutrinoPDF}

The leading contribution to the muon neutrino PDF arises already at $\mathcal{O}(\alpha_2)$ from the $\mu_L \to W^- \nu_\mu$ splitting, which presents an IR soft divergence that is cutoff by the $W$ mass.
As already discussed in Sec.~\ref{sec:EWdoublelogs}, the missing counterpart of this divergence in the virtual correction is at the origin of the Sudakov double log. In the same spirit as done for the EVA, we can compute the neutrino PDF by iteratively solving the DGLAP equations up to $\mathcal{O}(\alpha)$, using the zeroth-order expression for the $\mu_L$ PDF:
\be\begin{split}
    \frac{d f_{\nu_\mu}}{d \log Q^2} &= \frac{\alpha_2}{4\pi} \int_x^{1-m_W/Q} \frac{dz}{z} \frac{Q^4}{(Q^2 + z m_W^2)^2} P_{ff}^V(z) \frac{1}{2} \delta\left(1- \frac{x}{z}\right) + \mathcal{O}(\alpha^2) = \\
        &= \frac{\alpha_2}{8\pi} \frac{Q^4}{(Q^2 + x m_W^2)^2} P_{ff}^V(x) \; \theta\!\left(1 - \frac{m_W}{Q} - x \right)  + \mathcal{O}(\alpha^2)~.
\end{split}\ee
The Heaviside theta is the result of the IR cutoff $z_{\rm max} = 1 - \frac{m_W}{Q}$ in the integral. Integrating this differential equation from $m_W^2$ up to $Q^2$ we get:
\be\begin{split}
    f_{\nu_\mu}^{(\alpha)}(x,Q^2) &= \frac{\alpha_2}{8\pi} \; \theta\!\left(Q^2 -\frac{m_W^2}{(1-x)^2}\right) P_{ff}^V(x) \left( \log \frac{Q^2 + x m_W^2}{m_W^2} +  \right. \\
    & \qquad \left. + \log \frac{(1-x)^2}{1 + x(1-x)^2} + \frac{x m_W^2}{Q^2 + x m_W^2} + \frac{1}{1+x (1-x)^2} - 1\right)~.
\end{split}\ee
We observe here the single logarithm due to the standard collinear divergence, while the Sudakov log is absent because the initial muon PDF at zeroth order is just a delta function. It will appear, however, in the computation of an inclusive cross section upon integration of the PDF, due to the $x\to 1$ divergence inside $P_{ff}^V(x)$, that is not cancelled by a virtual correction at the same order but is instead cut off at the $W$ mass by the theta function.

In Fig.~\ref{fig:EWgaugePDF} we show a comparison between this $\mathcal{O}(\alpha)$ approximation of the muon neutrino PDF (dashed gray line) with the one from LePDF (solid gray), at a scale $Q = 3 \TeV$. We obseve a very good agreement at large $x$ values, i.e. where the $\mu_L \to W^- \nu_\mu$ splitting dominates. At smaller $x$ values the $\mathcal{O}(\alpha^2)$ contribution from $Z \to \nu_\mu \bar{\nu}_\mu$ will instead dominate.
Comparing to the results of Refs.~\cite{Han:2020uid,Han:2021kes}, we observe a different behavior of the $\nu_\mu$ PDF for $x \to 1$: while both our analytic result described above and LePDF show a cutoff (due to the IR $m_W$ cutoff) at $x\lesssim 1$, in their result the $\nu_\mu$ PDF increases similarly to the muon PDF up to $x=1$.

\subsection{Mass effects}
\label{sec:masseffects}

\begin{figure}[t]
\centering
\includegraphics[height=5.7cm]{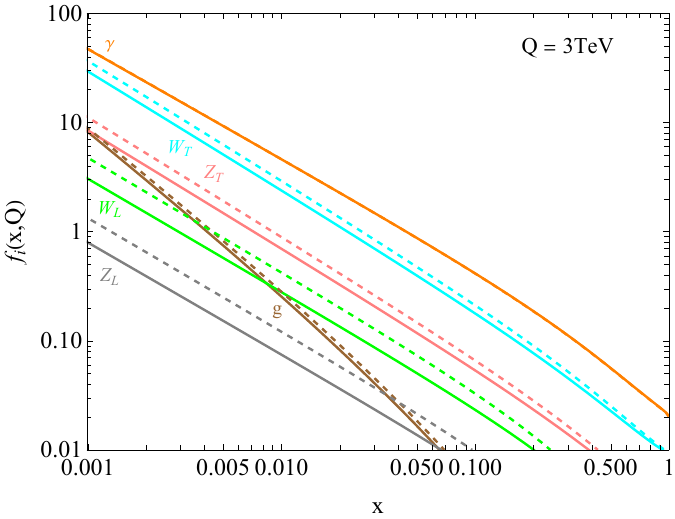} 
\includegraphics[height=5.7cm]{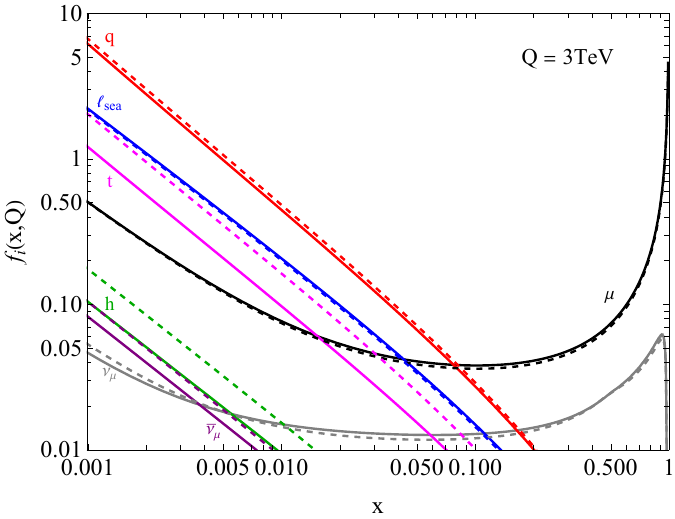} 
\caption{\label{fig:masseffectsPDF} PDFs for gauge bosons (left), fermions and Higgs (right) at a scale $Q=3~\TeV$. Solid lines are the full result while dashed ones are obtained neglecting the masses in the propagators. For simplicity in this plot we sum over polarisations.}
\end{figure}

We showed in Section~\ref{sec:EWbreaking} that massive particles modify the virtuality of the particle $B$ as in Eq.~\eqref{eq:pT_virtuality}. As already discussed in \cite{Chen:2016wkt,AlAli:2021let}, the impact of the latter effect is important since, due to the presence of the masses in the denominators of the DGLAP equations, the PDFs are lowered or enhanced when $m_{B,C}\neq 0$ and $m_A\neq 0$ respectively.
In Fig.~\ref{fig:masseffectsPDF} we show the PDFs computed keeping and neglecting the masses in $\widetilde{p}_T$, still starting the evolution of a given massive parton at the scale corresponding to its mass.

For instance, in case of EW gauge bosons (both transverse and longitudinal) one can expect that mass effects lower their PDFs, with the biggest effects at small $x$ due to the $(1-x)$ factor appearing in front of the mass corrections. This can also be verified from the LO EVA in Eqs.~(\ref{eq:EWA_WT}-\ref{eq:EWA_ZL}).
In case of the Higgs, its interactions have the form $A\to h +A$, with $A=W,Z,t$, which implies $\delta_{p_T}^2=x^2m_A^2+(1-x)m_h^2>0$, which explains why the Higgs PDF is bigger when masses are neglected.

\subsection{Top quark PDF}
\label{sec:top_results}

\begin{figure}[t]
\centering
\includegraphics[height=6.5cm]{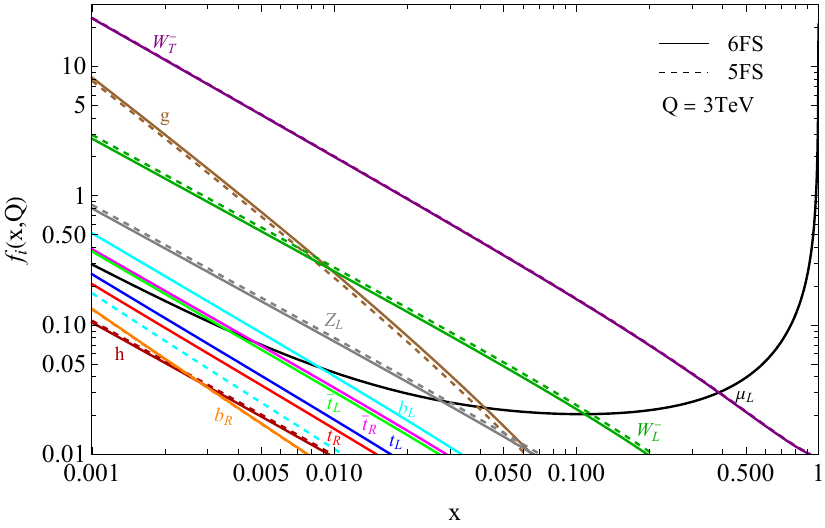} 
\caption{\label{fig:5vs6FS} Comparison between 5-flavour-scheme (dashed) and 6-flavour-schemes (solid) at a scale $Q = 3 ~\TeV$.}
\end{figure}

In Fig.~\ref{fig:5vs6FS} we compare the results between the 5FS and the 6FS showing the top PDFs together with the PDFs mostly affected by the inclusion of the top. As already mentioned in Section~\ref{sec:top} the top PDFs are mainly driven by the collinear emission off the transverse gauge bosons and the differences are more noticeable in high energies (we use the benchmark $Q = 3~\rm TeV$).
Since the PDFs of $W_T^-$ and $W_L^-$ are  large compared to the $Z$ and $W_T^+$, we expect a larger PDF for $\bar{t}_L$ and $\bar{t}_R$ than $t_L$ or $t_R$. Also, the same splittings $W_T^- \to b_L \bar{t}_L$ and $W_L^- \to b_L \bar{t}_R$ will induce a large enhancement of the $b_L$ PDF compared to $b_R$ in the 6FS.

Additionally, we observe that the PDFs of the EW transverse gauge bosons themselves are almost unaffected, since they are dominated by splitting off a muon. The gluon PDF, instead, receives a noticeable further contribution. Shifts of comparable size are also induced in the PDFs of the longitudinal gauge bosons (and even smaller for the Higgs), but in this case the PDFs are decreased due to a mass effect similar to the ones discussed in the previous Section~\ref{sec:masseffects}. 

\subsection{Uncertainties}
\label{sec:uncertainties}
Here we discuss several sources of uncertainties in our computation. Some are physical, such as the choice of $Q_{\rm QCD}$ or missing higher orders, while others are intrinsic in the numerical implementation of the DGLAP equations and can be improved simply by dedicating more computational time to the task.
\subsubsection*{QCD matching scale}
\begin{figure}[t]
\centering
\includegraphics[height=5.5cm]{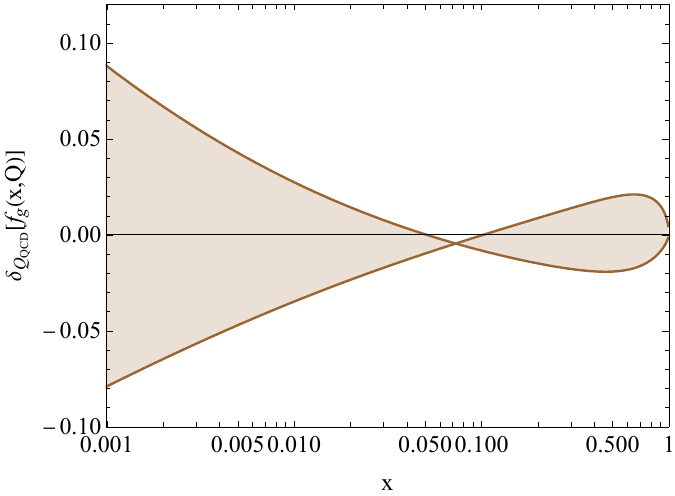} 
\includegraphics[height=5.5cm]{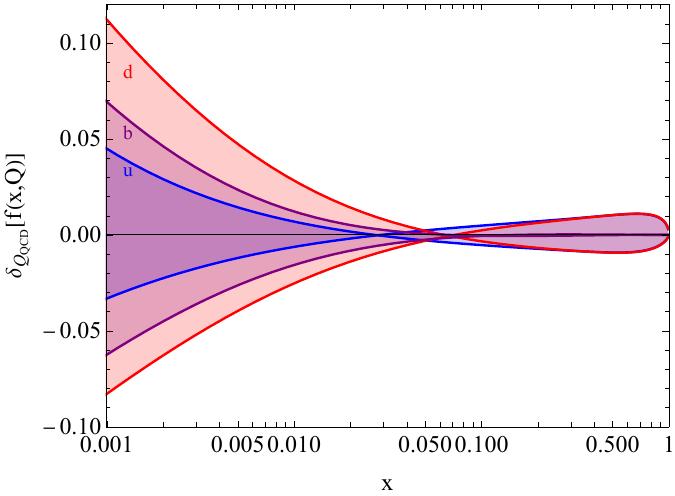} 
\caption{\label{fig:tQCDeffects} Effects of the choice of $Q_{\rm QCD}$ on the PDFs of gluon (left) and quarks (right) at the EW scale $Q = m_W$.}
\end{figure}
As already discussed in Section~\ref{sec:QED} the QCD scale $Q_{\rm QCD}$ is not clearly determined and different choices of this parameter can have a non negligible impact on the PDF, in particular for the colored particles. To study the dependence of our results on $Q_{\rm QCD}$ we repeat the evolution for $Q_{\rm QCD}=0.52~\GeV$ and $Q_{\rm QCD}=1~\GeV$ and we compute the relative differences with respect to the chosen value of $0.7 ~\GeV$
\be
\delta_{Q_{\rm QCD}}\left[f_A(x,Q)\right] = \frac{f_A(x,Q)|_{Q_{\rm QCD}}-f_A(x,Q)|_{0.7~\GeV}}{f_A(x,Q)|_{0.7~\GeV}}, \quad Q_{\rm QCD} = \left\{0.52~\GeV,1~\GeV\right\}.
\ee
In Fig.~\ref{fig:tQCDeffects} we show the results for the colored particles, which are the most affected by the choice of the QCD scale, while for the photon and the leptons the relative differences are smaller than $10^{-5}$. We report the results at $Q = m_W$ after the QED+QCD evolution, since this is the phase in which these effects are stronger.

\subsubsection*{Discretization}
\begin{figure}[t]
\centering
\includegraphics[height=5.5cm]{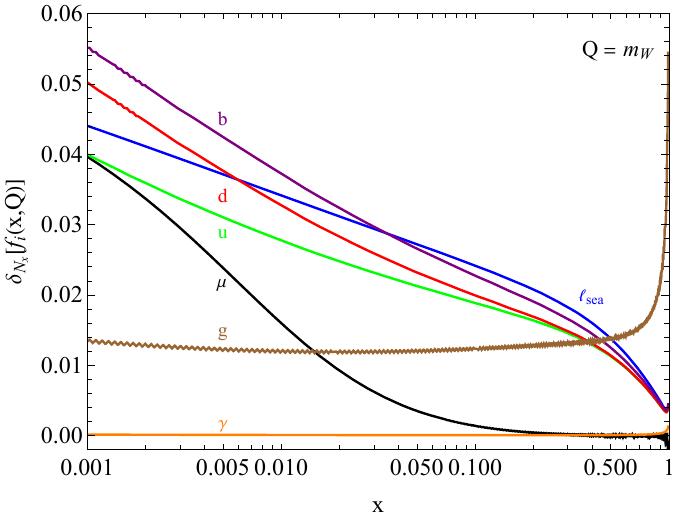} 
\includegraphics[height=5.5cm]{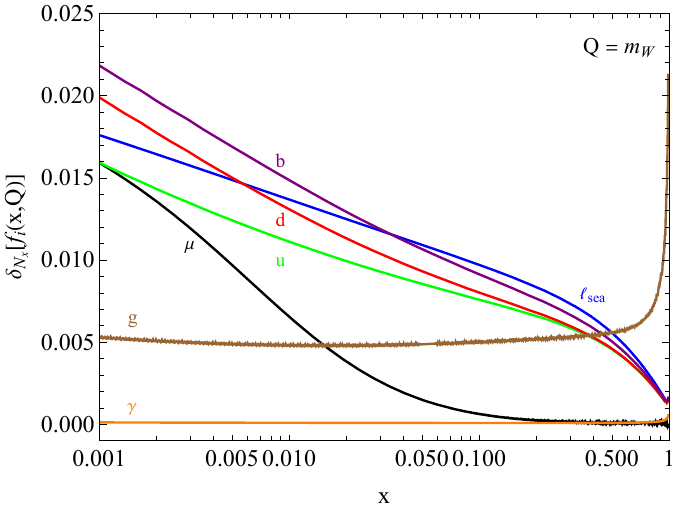} 
\caption{\label{fig:Nxeffects} Effects of $N_x$ on the PDFs at the EW scale $Q = m_W$. The relative differences correspond to $N_x = \left\{300; 600\right\}$ (left) and to $N_x = \left\{600; 1000\right\}$ (right).}
\end{figure}
The second source of uncertainty we take into account is the discretization, that is the number of grid points $N_x$. Again we focus for simplicity on the first phase of the evolution, since these effects do not change much with the energy scale\footnote{We checked that this is actually the case computing the PDFs in the full SM both with $N_x=600$ and $N_x=1000$: for instance at a scale $Q=3\TeV$ the relative differences are smaller then $2\%$.}. As for $Q_{\rm QCD}$ we repeat the evolution for different values of $N_x$ and compute the relative differences of the PDFs obtained. The results at the EW scale $m_W$, obtained varying $N_x$ from $300$ to $600$ and then to $1000$, are reported in Fig.~\ref{fig:Nxeffects}: it is clear that as we increase $N_x$ the relative differences are reduced, as expected since in this way we are approaching the continuum limit. Being the relative difference between $N_x=600$ and $N_x=1000$ already of $\mathcal{O}(10^{-2})$, we take the latter as reference value, since a further increase will introduce even smaller corrections. 

\subsubsection*{Integration step}

Numerical uncertainties also arise due to the discretization in $t$, depending on the choice of the integration step $dt$, as shown in Eq.~\eqref{eq:rungekutta}. As for the previous cases, we compute the relative differences of the PDFs obtained for two different values of $dt$, in particular we choose $dt = t(m_W)/N_t$, with $N_t = \left\{100;200\right\}$: this means that we consider $N_t$ steps in the first phase of the evolution. We do not report any plot, since we checked that the relative differences are at most of $\mathcal{O}(10^{-3})$, both at $Q=m_W$ and at higher scales.

\subsubsection*{Higher orders} 

The largest theoretical uncertainties in our results originate from neglecting higher order corrections. In particular, in the DL approximation terms of $\mathcal{O}(\alpha_2 \log (Q/Q_{\rm EW}))$ are not consistently resummed~\cite{Bauer:2018xag}. For example, at $Q=3~\rm TeV$, these terms already amount to $10 \%$, while at $Q=10~\rm TeV$ to $14\%$. We notice that promoting our approximation to the full LL result does not improve the situation~\cite{Bauer:2018xag}, since single-log terms of the same size from the NLL expansion are still present. However, performing the NLO matching as prescribed in Refs.~\cite{Bauer:2017bnh} can reduce the uncertainties to $4\%$ for $Q=3~\rm TeV$. Extending our formalism to NLL order would eventually correspond to $<3\%$ accuracy regardless of the energy scale.

\subsection{PDFs for electron beams}
\label{sec:electron}

\begin{figure}[t]
\centering
\includegraphics[height=7cm]{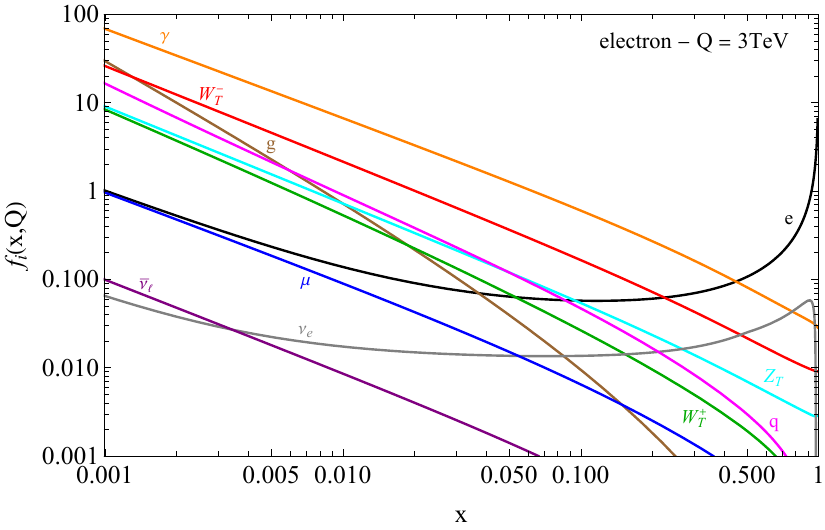} 
\caption{\label{fig:summary_electron}  In this plot we show some PDFs evaluated at a scale $Q = 3 ~\TeV$ for an electron beam. For this plot we sum over polarisations and $q$ represents the sum of all quark PDFs except for the top.}
\end{figure}

Our numerical code, with obvious substitutions, can also be used to derive LePDFs for electron beams. While most future projects for $e^+ e^-$ colliders are focussed on EW-scale energies to perform high-precision studies of EW gauge bosons, the Higgs, and top quark, linear collider projects also envisage later stages with TeV-scale center of mass energies. In this case our SM PDFs can provide a useful tool. 
We therefore provide public PDFs for electron/positron beams alongside those for muons and anti-muons. In Fig.~\ref{fig:summary_electron} we show only an example plot for some PDFs of an electron. As for the muon case, in case of EW gauge bosons PDFs we did a comparison with the EVA approximation at LO and NLO, obtaining similar results as shown above for the muon.
The main difference in the PDFs of an electron compared to those of a muon is that photon, charged leptons and quarks PDFs are larger. This is due to the longer QED evolution from $m_e$ to $m_\mu$. A consequence of larger quarks PDF is also a larger gluon one, even if $Q_{\rm QCD} > m_\mu$. On the other hand, EW gauge boson PDFs are very similar since, at first order, their PDF is insensitive on physics at scales below the EW one.

\section{Conclusions}
\label{sec:conclusions}

Obtaining precise predictions for multi-TeV lepton colliders is a topic of active studies. In fact, while QCD radiation plays a minor role, in comparison to hadron colliders, electroweak corrections in this energy regime can become very large due to single and double logarithmic enhancements. Furthermore, the electroweak sector presents several features that are absent when dealing with QED or QCD radiation. 
Within this context, the resummation of a subset of large logarithms related to the factorisable emission of initial-state radiation can be viewed as an ingredient for a complete description of collider phenomenology at such machines. 

In this paper we solved the set of DGLAP equations for an initial-state lepton, evolving the complete set of PDFs from the infrared up to multi-TeV scales. Our computation is performed with LO splitting functions, keeping into account all relevant mass thresholds, EW symmetry breaking terms, masses of all EW states, and resumming EW Sudakov logs at the double-logarithmic level. The residual uncertainty is dominated by the incomplete single-log EW resummation and can be estimated to be of $\mathcal{O}(10\%)$, while we show that other systematic uncertainties are fully under control. Improving our results to include single-log resummation is left for future work.

Using this result we discuss several notable features of LePDFs. Polarisation effects, due to the chiral nature of SM interactions, are shown to be of $\mathcal{O}(1)$ in Fig.~\ref{fig:polarisation}, or even larger than that in case of the $b$ quark. This last effect is due to the interaction with the top quark via the large top Yukawa coupling, and becomes much smaller in the 5FS, where the top is not included in the evolution.
As already observed in previous studies, we confirm that including EW states masses in the propagators gives sizeable correction to the PDFs, even for large factorisation scales.

Furthermore, we perform a detailed comparison of our results for the EW gauge bosons PDFs with the widely used Effective Vector Approximation.
This comparison, presented in Fig.~\ref{fig:EWgaugePDF}, illustrates how the EVA fails to describe correctly the transverse EW gauge bosons PDFs at small $x$ values, with deviations reaching even $\mathcal{O}(50\%)$ for $W^\pm_T$ and $Z_T$ at large scales, or even missing the target by more than one order of magnitude in case of the off-diagonal $Z/\gamma$ PDF. The cause of the latter large deviation is the well-known accidental cancellation in the vector-like coupling of a lepton to the $Z$ boson, which does not take place in the full result since the muon gains a strong polarisation. The smaller, but still substantial deviations in the PDFs of transverse EW gauge bosons are due to the fact that the EVA, treated at LO, does not include contributions from gauge boson splitting off an initial gauge boson. Such terms, while being formally of higher order, become enhanced due to the large $\gamma, W, Z$ PDFs and the fact that these splittings come with associated Sudakov double logs. In fact, by extending iteratively the EVA up to $\mathcal{O}(\alpha^2)$ we obtain a much better agreement with the complete numerical result, as shown in Fig.~\ref{fig:PDFvsEVA}.
In light of this, we recommend the use of LePDF to derive precision predictions for SM and BSM processes at high-energy electron or muon colliders (they could also be used for lepton-hadron collisions), when one is interested in being inclusive on radiation emitted at small $p_T$ compared to the typical energy of the hard scattering, $E \gg p_T^{\rm coll. rad.}$, and when $E \gg m_W$, which are the conditions for factorisation to be valid.

Our numerical results for the LePDFs are made public\footnote{They are available from GitHub at {\tt\href{https://github.com/DavidMarzocca/LePDF}{https://github.com/DavidMarzocca/LePDF}}.} in the LHAPDF6 format (with some modifications due to the necessity of describing independently all helicity states). We provide results for initial-state muons, anti-muons, electrons, and positrons, all in both the 5 and 6 flavour schemes.

\section*{Acknowledgments}

We would like to thank Thomas Gehrmann, Johannes Michel, Luca Vecchi, and Andrea Wulzer for fruitful discussions.
DM acknowledges partial support by MIUR grant PRIN 2017L5W2PT and the European Research Council (ERC) under the European Union’s Horizon 2020 research and innovation programme, grant agreement 833280 (FLAY).
ST is supported by the Swiss National Science Foundation - project n. P500PT\_203156, and by the Center of Theoretical Physics at MIT.

\appendix

\section{Inputs and Formalism}
\label{app:inputs}

\subsection{Standard Model inputs}
\label{app:SMcouplings}

In order to set our notation, we define the covariant derivatives as
\begin{equation}
D_{\mu} = \partial_{\mu} - ig_{3}G_{\mu}^A t_{A} - i g_{2} W_{\mu}^a T^a - i g_{1} Y B_{\mu}~,
\end{equation}
where $t^A$ and $T^a$ are $SU(3)$ and $SU(2)$ generators, respectively, while $Y$ is the hypercharge. The electric charge is given by $Q = Y + T^3$.
In our analysis we neglect the CKM matrix and work with diagonal Yukawas
\begin{equation}
    \mathcal{L}_{Y} = - y_u^i \bar{Q}_{L}^i u_R^{i} H^c - y_d^i \bar{Q}_{L}^i d_R^{i} H - y_e^i \bar{L}_{L}^i e_R^{i} H + h.c.~,  \label{eq:smyuk}
\end{equation}
where $H^c = i \sigma_2 H^*$. In particular, in our implementation we keep only $y_u^3\equiv y_t\neq 0$.
Finally, we define the Higgs potential as
\begin{equation}
    V(H) = \lambda_h \left( H^\dagger H- \frac{v^2}{2}\right)^2.
\end{equation}
To shorten the notation, we define the following quantities
\begin{align}
c_W^2 = \frac{g_2^2}{g_1^1 + g_2^2}~, \quad e &= g_1 c_W~,\quad \alpha_{g_x} \equiv \frac{g_x^2}{4 \pi}~, \quad \alpha_\gamma = \frac{e^2}{4 \pi}~, \quad 
Q^Z_{f}= T_3-Q_fs_W^2~, \notag \\
\alpha_{\g2}(t) &= \sqrt{\alpha_{\g}(t)\alpha_2(t)}~, \quad C_F = \frac{N_c^2-1}{2N_c}~, \quad T_F = 1/2~, 
\label{eq:defQvf}
\end{align}
where $N_c = 3$.

\begin{table}
\centering
\begin{tabular}{| c | c | c | c | c | c |}\hline
    Parameter & $\alpha_3$ & $\alpha_2$ & $\alpha_1$ & $\alpha_{y_t}$  \\
    Value  & 0.1057 & 0.03329 & 0.01025 & 0.0679 \\\hline
\end{tabular}
\caption{\label{tab:SM_coupl_num} Numerical inputs for the SM parameters at a scale $Q = 200\, ~\GeV$ from \cite{Alam:2022cdv}. }
\end{table}

We evaluate the RG evolution for the QCD coupling using the 3-loop result \cite{Workman:2022ynf} with mass thresholds for the charm, bottom, and top quarks.
For the QED and EW couplings, as well as for the top Yukawa, we employ the corresponding 1-loop RGE. The numerical boundary conditions are taken at $Q = 200 ~\GeV$ from \cite{Alam:2022cdv}, we report the relevant ones in Table~\ref{tab:SM_coupl_num} for convenience.

\subsection{Formalism for the DGLAP equations}
\label{app:formalism}

 We consider a process $A+X\rightarrow C+Y$ mediated by the particle $B$ at tree level, with $B$ carrying a fraction $z\in[0,1]$ of the energy of $A$, as in Fig. \ref{fig:AX>CY}. The kinematics, up to quadratic order in the transverse momentum $p_T$, is the following
\begin{equation}
p_A^\mu = \left(E_A,0,0,\sqrt{E_A^2-m_A^2}\right), \label{eq:p}
\end{equation}
\begin{equation}
p_B^\mu = \left(zE_A,p_T,0,\sqrt{E_A^2-m_A^2}-\sqrt{\bar{z}^2E_A^2-m_C^2}+\frac{p_T^2}{2\sqrt{\bar{z}^2E_A^2-m_C^2}}\right), \label{eq:q}
\end{equation}
\begin{equation}
p_C^\mu = \left(\bar{z}E_A,-p_T,0,\sqrt{\bar{z}^2E_A^2-m_C^2}-\frac{p_T^2}{2\sqrt{\bar{z}^2E_A^2-m_C^2}}\right), \label{eq:k}
\end{equation}
where we remind that $\bar{z} = 1-z$.
In this way the emitted particle $C$ is on-shell ($p_C^2=m_C^2$), while for the particle $B$, neglecting $\mathcal{O}(p_T^4)$ terms, we have
\beq
p_B^2 =  m_A^2 + m_C^2 -2\bar{z}E_A^2+2\sqrt{(E_A^2-m_A^2)(\bar{z}^2E_A^2-m_C^2)}-\sqrt{\frac{(E_A^2-m_A^2)}{(\bar{z}^2E_A^2-m_C^2)}}p_T^2. \label{eq:q2} 
\eeq
Since we are working with high-energy initial beams, we can expand $p_B^2$ neglecting terms of order $m/E$ and $p_T/E$ or higher, so that the virtuality of the particle B is given by
\begin{equation}
m_B^2-p_B^2 = \frac{1}{\bar{z}}(p_T^2+zm_C^2+\bar{z}m_B^2-z\bar{z}m_A^2) + \mathcal{O}\left(\frac{m^2}{E^2}, \frac{p_T^2}{E^2}\right) \equiv \frac{\widetilde{p}_T^2}{\bar{z}}~. \label{eq:virtuality}
\end{equation}

Invariance of the cross section on the factorization scale $Q$, that we choose to be the $p_T$ of the emitted parton, gives the DGLAP equations \cite{Cacciari:1992pz,Drees:1994eu}
\be
    Q^2 \frac{d f_B(x, Q^2)}{dQ^2} = P_B^v(x,Q^2) f_B(x,Q^2) + \sum_{A,C} \int_x^{z_{\rm max}^{ABC}} \frac{dz}{z}  Q^2 \frac{d \mathcal{P}_{A \to B + C}}{dz dp_T^2}\left(z, Q^2 \right) f_A\left(\frac{x}{z}, Q^2\right)~, \label{eq:DGLAP}
\ee
where $P_B^v$ represents the virtual corrections (see App.~\ref{app:radiative}) and  
\begin{equation}
\frac{d \mathcal{P}_{A \to B + C}}{dz dp_T^2}(z, p_T^2) = \frac{1}{16 \pi^2 \widetilde{p}_T^4} z \bar{z} \left|\mathcal{M}_{A \to B + C}\right|^2~,
	\label{eq:splitting_generic}
\end{equation}
describes the splitting process. The standard matrix elements, which are also present in the unbroken phase (i.e. in a massless theory), are typically parametrized as
\be
    |\mathcal{M}_{A \to B + C}|^2 \equiv 8\pi\alpha_{ABC} \frac{p_T^2}{z\bar{z}}P_{BA}^{C} (z)~,
    \label{eq:ampl_splitting_generic_massless}
\ee
where $P_{BA}^{C}(z)$ is the splitting function and $\alpha_{ABC}$ the corresponding coupling.
Ultra-collinear matrix elements are instead proportional to $v^2$ and they can be parametrized with new splitting functions $U_{BA}^C$ as
\be
	\left|\mathcal{M}_{A \to B + C}\right|^2 \supset \frac{v^2}{z\zb} U_{BA}^C(z)~. \label{eq:ucsplitting}
\ee
The general DGLAP equation for a parton $B$ is then given by
\begin{equation}
    Q^2\frac{df_B(x,Q^2)}{d Q^2} = P_B^v \, f_B(x,Q^2)+\sum_{A,C}\frac{\alpha_{ABC}}{2\pi} \widetilde{P}_{BA}^C \otimes f_A + \frac{v^2}{16\pi^2Q^2}\sum_{A,C}\widetilde{U}_{BA}^{C}\otimes f_A ~,
\end{equation}
where $\widetilde{P}$ and $\widetilde{U}$ are the splitting functions for massive partons and are obtained from those for massless ones with the redefinition in Eq.~\eqref{eq:splitting_tilde}
\be
\widetilde{P}_{BA}^C(z, p_T^2) = \left( \frac{p_T^2}{\widetilde{p}_T^2} \right)^2 P_{BA}^C(z)~,
\ee
with
\be
	\widetilde{p}_T^2 \equiv \zb (m_B^2 - p_B^2) = p_T^2 + z m_C^2 + \zb m_B^2 - z \zb m_A^2 + \mathcal{O}\left(\frac{m^2}{E^2}, \frac{p_T^2}{E^2}\right)~.
\ee

\section{Splitting Functions}
\label{app:splitting}

Here we list all the splitting functions, including ultra-collinear ones. Some of them have a $\zb$ pole and therefore they introduce divergences in the DGLAP equations when $z_{\rm max}=1$. To deal with such divergences, we use the $+$ distribution, defined as
\begin{equation}
\int_x^1 dz\frac{f(z)}{(1-z)_+} = \int_x^1 dz\frac{f(z)-f(1)}{1-z}-f(1)\int_0^x\frac{dz}{1-z} = \int_x^1 dz\frac{f(z)-f(1)}{1-z}+f(1)\log(1-x)~. \label{eq:def+}
\end{equation}
As already discussed, since the SM is a chiral theory we separate vector polarizations and fermion helicities in the splitting functions.

\subsection{Massless splitting functions}
We start with the splitting functions of the form in Eq.~\eqref{eq:ampl_splitting_generic_massless}. Here $f$ labels a fermion, $V$ a gauge boson and $h$ a scalar. We do not specify the polarization of the particle $C$, since in the computation we sum over it.

\be\begin{split}
    P_{ff}^V(z)  &\equiv P_{f_Lf_L}^V(z) = P_{f_Rf_R}^V(z) = \frac{1+z^2}{\zb_+} ~,\\
    P_{V_+f_L}^{f}(z) &= P_{V_-f_R}^{f}(z) = \frac{\zb^2}{z}~,\\
    P_{V_-f_L}^{f}(z) &= P_{V_+f_R}^{f}(z) = \frac{1}{z}~,\\
    P_{f_LV_+}^{f}(z) &= P_{f_RV_-}^{f}(z) = \zb^2~,\\
    P_{f_LV_-}^{f}(z) &= P_{f_RV_+}^{f}(z) = z^2~,\\
\end{split}\ee
\be\begin{split}
    P_{ff}^h(z) &\equiv P_{f_Lf_R}^h(z) = P_{f_Rf_L}^h(z) = \frac{\zb}{2}~,\\
    P_{hf}^f(z) &\equiv P_{hf_L}^{f}(z) = P_{hf_R}^{f}(z) = \frac{z}{2}~,\\
    P_{fh}^f(z) &\equiv P_{f_Lh}^{f}(z) = P_{f_Rh}^{f}(z) = \frac{1}{2}~,\\
\end{split}\ee
\be\begin{split}
    P_{V_\pm h}^h(z) &= \frac{\zb}{z}~,\\
    P_{Vh}^h(z) &\equiv P_{V_+ h}^h(z)+P_{V_- h}^h(z) = \frac{2\zb}{z}~,\\
    P_{hh}^V(z) &= \frac{2z}{\zb_+}~,\\
    P_{hV}^h(z) &\equiv P_{hV_+}^h(z) = P_{hV_-}^h(z) = z\zb~,\\
\end{split}\ee
\be\begin{split}
    P_{V_+V_+}^V(z) &= P_{V_-V_-}^V(z) = \frac{1+z^4}{z\zb_+}~,\\
    P_{V_+V_-}^V(z) &= P_{V_-V_+}^V(z) =\frac{\zb^3}{z}~.\\
\end{split}
\ee
Since QED and QCD are vectorlike theories, we also define the splitting functions properly summed over the vector polarizations and fermion helicities:
\be\begin{split}
    P_{Vf}^f(z) &\equiv \frac{P_{V_+f_L}^{f}(z)+P_{V_+f_R}^{f}(z)+P_{V_-f_L}^{f}(z)+P_{V_-f_R}^{f}(z)}{2}= \frac{1+\zb^2}{z}~,\\
    P_{fV}^f(z) &\equiv \frac{P_{f_LV_+}^{f}(z)+P_{f_RV_+}^{f}(z)+P_{f_LV_-}^{f}(z)+P_{f_RV_-}^{f}(z)}{2} = z^2 + \zb^2~,\\
    P_{VV}^V(z) &\equiv \frac{P_{V_+V_+}^V(z)+P_{V_+V_-}^V(z)+P_{V_-V_+}^V(z)+P_{V_-V_-}^V(z)}{2} = 2\frac{(1-z\zb)^2}{z\zb_+}~.
\end{split}\ee
Finally, we report here the integrals appearing in Eq.~\eqref{eq:appr_QED_sol}:
\be\begin{split}
	I_{fVVf}(x) &= \int_x^1 \frac{dz}{z} P_{fV}^f(z) P_{Vf}^f\left(\frac{x}{z}\right) = \frac{4+3x-3x^2-4x^3}{3x}+2(1+x) \log x~,\\
	I_{Vfff}(x) &= \int_x^1 \frac{dz}{z} P_{Vf}^f\left(\frac{x}{z}\right) P_{ff}^V(z) = \\
	&= 2\log(1-x) P_{Vf}^f(x)+\frac{(1-x)(2x-3)}{x}+(2-x) \log x~,\\
	I_{ffff}(x) &= \int_x^1 \frac{dz}{z} P_{ff}^V\left(\frac{x}{z}\right) P_{ff}^V(z) = \\
	&= \frac{-2(1-x)^2+4(1+x^2)\log(1-x)-(1+3x^2)\log x}{1-x}~.
    \label{eq:appr_QED_integrals}
\end{split}\ee

\subsection{Ultra-collinear splitting functions}

 The top quark is explicitly written, while for other fermions we write $f$. $s=L,R$ is the helicity of the fermion, $T=\pm$ is a transverse polarization of the gauge bosons. If inside a splitting we write $f_sf_{-s}$ it means that the two fermions have opposite helicity (same for the gauge bosons). $N_c^f$ is $1$ for leptons and $N_c$ for quarks.\\

\noindent Splitting $f\to f+V_T$ ($f = t,b$):
\be\begin{split}
    U_{\tr\bl}^{\wm^-}(z)  &= U_{\trb\blb}^{\Wp^+}(z)  = \frac{1}{2}g_2^2y_t^2\zb ~,\\
    U_{\wm^-\bl}^{\tr}(z)  &= U_{\Wp^+\blb}^{\trb}(z)  = \frac{1}{2}g_2^2y_t^2z ~,\\
    U_{\bl\tr}^{\Wp^+}(z)  &= U_{\blb\trb}^{\wm^-}(z)  = \frac{1}{2}g_2^2y_t^2\zb z^2~,\\
    U_{\Wp^+\tr}^{\bl}(z)  &= U_{\wm^-\trb}^{\blb}(z)  = \frac{1}{2}g_2^2y_t^2z \zb^2 ~,
\end{split}\ee
\be\begin{split}
    U_{tt}^{g}(z)&\equiv U_{\tr\tl}^{g_-}(z)  = U_{\trb\tlb}^{g_+}(z)  = U_{\tl\tr}^{g_+}(z)  = U_{\tlb\trb}^{g_-}(z)  = C_Fg_3^2y_t^2\zb^3 ~,\\
    U_{g t}^{t}(z)&\equiv U_{g_-\tl}^{\tr}(z)  = U_{g_+\tlb}^{\trb}(z)  = U_{g_+\tr}^{\tl}(z)  = U_{g_-\trb}^{,\tlb}(z)  = C_Fg_3^2y_t^2z^3 ~,\\
    U_{tt}^{\g}(z)&\equiv U_{\tr\tl}^{\gm}(z)  = U_{\trb\tlb}^{,\gp}(z)  = U_{\tl\tr}^{\gp}(z)  = U_{\tlb\trb}^{\gm}(z)  = Q_u^2e^2y_t^2\zb^3 ~,\\
    U_{\g t}^{t}(z)&\equiv U_{\gm\tl}^{\tr}(z)  = U_{\gp\tlb}^{\trb}(z)  = U_{\gp\tr}^{\tl}(z)  = U_{\gm\trb}^{\tlb}(z)  = Q_u^2e^2y_t^2z^3 ~,
\end{split}\ee
\be\begin{split}
    U_{\tr\tl}^{\zm}(z)  &= U_{\trb\tlb}^{\zp}(z)  = \frac{g_2^2y_t^2}{c_W^2}\zb\left(\frac{1}{2}-Q_us_W^2\zb\right)^2 ~,\\
    U_{\zm\tl}^{\tr}(z)  &= U_{\zp\tlb}^{\trb}(z)  = \frac{g_2^2y_t^2}{c_W^2}z\left(\frac{1}{2}-Q_us_W^2z\right)^2 ~,\\
    U_{\tl\tr}^{\zp}(z)  &= U_{\tlb\trb}^{\zm}(z)  = \frac{g_2^2y_t^2}{c_W^2}\zb\left(\frac{z}{2}+Q_us_W^2\zb\right)^2 ~,\\
    U_{\zp\tr}^{\tl}(z)  &= U_{\zm\trb}^{\tlb}(z)  = \frac{g_2^2y_t^2}{c_W^2}z\left(\frac{\zb}{2}+Q_us_W^2z\right)^2 ~,\\
    U_{\zgm\tl}^{\tr}(z)  &= U_{\zgp\tlb}^{\trb}(z)  = 2Q_u\frac{eg_2}{c_W}y_t^2z^2\left(\frac{1}{2}-Q_us_W^2z\right) ~,\\
    U_{\zgp\tr}^{\tl}(z)  &= U_{\zgm\trb}^{\tlb}(z)  = -2Q_u\frac{eg_2}{c_W}y_t^2z^2\left(\frac{\zb}{2}+Q_us_W^2z\right) ~.\\
\end{split}
\ee
Splitting $V_T\to f+\bar{f}$ ($f = t,b$):
\be\begin{split}
    U_{\bl\wm^-}^{\trb}(z)  &= U_{\blb\Wp^+}^{\tr}(z)  = \frac{N_c}{2}g_2^2y_t^2z^2 ~,\\
    U_{\trb\wm^-}^{\bl}(z)  &= U_{\tr\Wp^+}^{\blb}(z)  = \frac{N_c}{2}g_2^2y_t^2\zb^2 ~,\\
\end{split}
\ee
\be\begin{split}
    U_{tg}^{t}(z)&\equiv U_{\tl g_-}^{\trb}(z)  = U_{\trb g_-}^{\tl}(z)  = U_{\tr g_+}^{\tlb}(z)  = U_{\tlb g_+}^{\tr}(z)  = T_Fg_3^2y_t^2 ~,\\
    U_{t\g}^{t}(z)&\equiv U_{\tl\gm}^{\trb}(z)  = U_{\trb\gm}^{\tl}(z)  = U_{\tr\gp}^{\tlb}(z)  = U_{\tlb\gp}^{\tr}(z)  = N_c Q_u^2e^2y_t^2 ~,\\
    \end{split}
\ee
\be\begin{split}
    U_{\tl\zm}^{\trb}(z)  &= U_{\tlb\zp}^{\tr}(z)  = N_c\frac{g_2^2y_t^2}{c_W^2}\left(\frac{z}{2}-Q_us_W^2\right)^2 ~,\\
    U_{\trb\zm}^{\tl}(z)  &= U_{\tr\zp}^{\tlb}(z)  = N_c\frac{g_2^2y_t^2}{c_W^2}\left(\frac{\zb}{2}-Q_us_W^2\right)^2 ~,\\
    U_{\tl\zgm}^{\trb}(z)  &= U_{\tlb\zgp}^{\tr}(z)  = N_cQ_ue\frac{g_2}{c_W}y_t^2\left(\frac{z}{2}-Q_us_W^2\right) ~,\\
    U_{\trb\zgm}^{\tl}(z)  &= U_{\tr\zgp}^{\tlb}(z)  = N_cQ_ue\frac{g_2}{c_W}y_t^2\left(\frac{\zb}{2}-Q_us_W^2\right) ~.\\
\end{split}\ee
Splitting $f\to f + V_L$:
\be\begin{split}
    U_{f_Lf_L}^{\zl}(z) &= U_{\fbar_L\fbar_L}^{\zl}(z)  = \left(T_3 y_f^2\zb^2+\frac{g_2^2}{c_W^2}Q^Z_{f_L}z\right)^2\frac{1}{\zb_+} ~,\\
    U_{Z_Lf_L}^{\fl}(z) &= U_{Z_L\fbar_L}^{\flb}(z) = \left(T_3 y_f^2z^2+\frac{g_2^2}{c_W^2}Q^Z_{f_L}\zb\right)^2\frac{1}{z} ~,\\
    U_{f_Rf_R}^{\zl}(z) &= U_{\fbar_R\fbar_R}^{\zl}(z)  = \left(T_3 y_f^2\zb^2-\frac{g_2^2}{c_W^2}Q^Z_{f_R}z\right)^2\frac{1}{\zb_+} ~,\\
    U_{Z_Lf_R}^{\fr}(z) &= U_{Z_L\fbar_R}^{\frb}(z) = \left(T_3 y_f^2z^2-\frac{g_2^2}{c_W^2}Q^Z_{f_R}\zb\right)^2\frac{1}{z} ~,\\
    U_{f^{(2)}_Lf^{(1)}_L}^{\wl}(z) &= U_{\fbar^{(2)}_L\fbar^{(1)}_L}^{\wl}(z)  = \left(y_{f_1}^2z\zb-y_{f_2}^2\zb-g_2^2z\right)^2\frac{1}{2\zb_+} ~,\\
    U_{W_Lf^{(1)}_L}^{f^{(2)}_L}(z) &= U_{W_L\fbar^{(1)}_L}^{\fbar^{(2)}_L}(z) = \left(y_{f_1}^2z\zb-y_{f_2}^2z-g_2^2\zb\right)^2\frac{1}{2z} ~.\\
\end{split}
\ee
Splitting $V_L\to f+\bar{f}$:
\be\begin{split}
    U_{f_LZ_L}^{\flb}(z) &= U_{\fbar_LZ_L}^{\fl}(z)  = N_c^f\left(T_3 y_f^2-\frac{g_2^2}{c_W^2}Q^Z_{f_L}z\zb\right)^2 ~,\\
    U_{f_RZ_L}^{\frb}(z) &= U_{\fbar_RZ_L}^{\fr}(z)  = N_c^f\left(T_3 y_f^2-\frac{g_2^2}{c_W^2}Q^Z_{f_R}z\zb\right)^2 ~,\\
    U_{f^{(1)}_LW_L}^{\fbar^{(2)}_L}(z) &= U_{\fbar^{(1)}_LW_L}^{f^{(2)}_L}(z)  = \frac{N_c^f}{2}\left(y_{f_1}^2\zb+y_{f_2}^2z-g_2^2z\zb\right)^2 ~.\\
\end{split}
\ee
Splitting $t\to t+h$:
\be\begin{split}
    U_{tt}^{h}(z)&\equiv U_{\tl\tl}^{h}(z)  = U_{\tr\tr}^{h}(z)  = U_{\trb\trb}^{h}(z)  = U_{\tlb\tlb}^{h}(z)  = \frac{y_t^4}{4}\zb(1+z)^2 ~,\\
    U_{ht}^{t}(z)&\equiv U_{h\tl}^{\tl}(z)  = U_{h\tr}^{\tr}(z)  = U_{h\trb}^{\trb}(z)  = U_{h\tlb}^{\tlb}(z)  = \frac{y_t^4}{4}z(1+\zb)^2 ~.\\
\end{split}
\ee
Splitting $h\to t+\bar{t}$:
\be
    U_{th}^{t}(z)\equiv  U_{\tl h}^{\tlb}(z)  = U_{\tr h}^{\trb}(z)  = U_{\trb h}^{\tr}(z)  = U_{\tlb h}^{\tl}(z)  = N_c\frac{y_t^4}{4}(\zb-z)^2.
\ee
Splitting $V_T\to V_L+V_T$:
\be\begin{split}
    U_{\wl\wt}^{\gt}(z)  &= e^2g_2^2\frac{\zb^3}{z} ~,\\
    U_{\gt\wt}^{\wl}(z)  &= e^2g_2^2\frac{z^3}{\zb_+} ~,\\
    U_{\wl\wt}^{\zt}(z)  &= \frac{1}{4}c_W^2g_2^4(1+\zb+t_W^2z)^2\frac{\zb}{z} ~,\\
    U_{\zt \wt}^{\wl}(z)  &= \frac{1}{4}c_W^2g_2^4(1+z+t_W^2\zb)^2\frac{z}{\zb_+}~,\\
    U_{\zgt \wt}^{\wl}(z)  &= c_Weg_2^3(1+z+t_W^2\zb)\frac{z^2}{\zb_+}~,\\
    U_{\zl\wt}^{\wt}(z)  &= \frac{1}{4}g_2^4(1+\zb)^2\frac{\zb}{z} ~,\\
    U_{\wt \wt}^{\zl}(z)  &= \frac{1}{4}g_2^4(1+z)^2\frac{z}{\zb_+}~,\\
    U_{\wl\gt}^{\wt}(z)  &= e^2g_2^2\frac{\zb}{z} ~,\\
    U_{\wt\gt}^{\wl}(z)  &= e^2g_2^2\frac{z}{\zb_+} ~,\\
    U_{\wl \zt}^{\wt}(z)  &= \frac{1}{4}c_W^2g_2^4(1+\zb-t_W^2z)^2\frac{\zb}{z}~,\\
    U_{\wt \zt}^{\wl}(z)  &= \frac{1}{4}c_W^2g_2^4(1+z-t_W^2\zb)^2\frac{z}{\zb_+}~,\\
    U_{\wl \zgt}^{\wt}(z)  &= \frac{1}{2}c_Weg_2^3(1+\zb-t_W^2z)\frac{\zb}{z}~,\\
    U_{\wt \zgt}^{\wl}(z)  &= \frac{1}{2}c_Weg_2^3(1+z-t_W^2\zb)\frac{z}{\zb_+}~.\\
\end{split}
\ee
Splitting $V_L\to V_T+V_{-T}$:

\be\begin{split}
    U_{\gt \wl}^{\wtm}(z)  &= e^2g_2^2z^3\zb~,\\
    U_{\wt \wl}^{\gtm}(z)  &= e^2g_2^2z\zb^3~,\\
    U_{\zt \wl}^{\wtm}(z)  &= \frac{1}{4}c_W^2g_2^4z\zb(\zb -z+t_W^2)^2~,\\
    U_{\wt \wl}^{\ztm}(z)  &= \frac{1}{4}c_W^2g_2^4z\zb(z -\zb+t_W^2)^2~,\\
    U_{\zgt \wl}^{\wtm}(z)  &= -c_Weg_2^3z^2\zb(\zb -z+t_W^2) ~,\\
    U_{\wt \zl}^{\wtm}(z)  &= \frac{1}{4}g_2^4z\zb(\zb -z)^2~.\\
\end{split}
\ee
Splitting $V_T\to h+V_T$:
\be\begin{split}
    U_{h\wt}^{\wt}(z) &= U_{\wt\wt}^{h}(z) = \frac{1}{4}g_2^4z\zb~,\\
    U_{h\zt}^{\zt}(z) &= U_{\zt\zt}^{h}(z) = \frac{1}{4}\frac{g_2^4}{c_W^4}z\zb~.\\
\end{split}
\ee
Splitting $h\to V_T+V_{-T}$:
\be\begin{split}
    U_{\wt h}^{\wtm}(z)  &= \frac{1}{4}g_2^4z\zb~,\\
    U_{\zt h}^{\ztm}(z)  &= \frac{1}{8}\frac{g_2^4}{c_W^4}z\zb~.\\
\end{split}
\ee
Splitting $V_L\to V_L+V_L$:
\be\begin{split}
    U_{\zl\wl}^{\wl}(z) &= \frac{1}{16}g_2^4[(\zb-z)(2+z\zb)-t_W^2\zb(1+\zb)]^2\frac{1}{z\zb_+}~,\\
    U_{\wl\wl}^{\zl}(z) &= \frac{1}{16}g_2^4[(z-\zb)(2+z\zb)-t_W^2z(1+z)]^2\frac{1}{z\zb_+}~,\\
    U_{\wl\zl}^{\wl}(z) &= \frac{1}{16}g_2^4(z-\zb)^2(2+z\zb-t_W^2z\zb)^2\frac{1}{z\zb_+}~.\\
\end{split}
\ee
Splitting $h\to V_L+V_L$:
\be\begin{split}
    U_{\zl h}^{\zl}(z) &= \frac{1}{8}\left[\frac{g_2^2}{c_W^2}(1-z\zb)-4\lambda_hz\zb\right]^2\frac{1}{z\zb_+}~,\\
    U_{\wl h}^{\wl}(z) &= \frac{1}{4}[g_2^2(1-z\zb)-4\lambda_hz\zb]^2\frac{1}{z\zb_+}~.\\
\end{split}
\ee
Splitting $V_L\to h+V_L$:
\be\begin{split}
    U_{h \zl}^{\zl}(z) &= \frac{1}{4}\left[\frac{g_2^2}{c_W^2}(1-z\zb)+4\lambda_h\zb\right]^2\frac{z}{\zb_+}~,\\
    U_{\zl \zl}^{h}(z) &= \frac{1}{4}\left[\frac{g_2^2}{c_W^2}(1-z\zb)+4\lambda_hz\right]^2\frac{\zb}{z}~,\\
    U_{h\wl}^{\wl}(z) &= \frac{1}{4}[g_2^2(1-z\zb)+4\lambda_h\zb]^2\frac{z}{\zb_+}~,\\
    U_{\wl\wl}^{h}(z) &= \frac{1}{4}[g_2^2(1-z\zb)+4\lambda_hz]^2\frac{\zb}{z}~.\\
\end{split}
\ee
Splitting $h\to h+h$:
\be
U_{h h}^{h}(z) = 18\lambda_h^2z\zb~.
\ee
Splitting involving the mixed state $hZ_L$:
\be\begin{split}
     U_{hZ_L\tl}^{\tl}(z) &= -U_{hZ_L\tlb}^{\tlb}(z) =y_t^2(1+\zb)\left(\frac{y_t^2}{2}z^2+g_Z^2 Q^Z_{\tl}\zb\right) ~,\\
     U_{hZ_L\tr}^{\tr}(z) &= -U_{hZ_L\trb}^{\trb}(z) =-y_t^2(1+\zb)\left(\frac{y_t^2}{2}z^2-g_Z^2Q^Z_{\tr}\zb\right) ~,\\
     U_{\tl hZ_L}^{\tlb}(z) &= U_{\tlb hZ_L}^{\tl}(z) =-\frac{y_t^2}{2}(\zb-z)\left(\frac{y_t^2}{2}-g_Z^2Q^Z_{\tl}z\zb\right) ~,\\
     U_{\tr hZ_L}^{\trb}(z) &= U_{\trb hZ_L}^{\tr}(z) =\frac{y_t^2}{2}(\zb-z)\left(\frac{y_t^2}{2}-g_Z^2Q^Z_{\tr}z\zb\right) ~,\\
     \end{split}
\ee
    \be\begin{split}
    U_{hZ_L\wl}^{\wl}(z) &\equiv U_{hZ_L\wl^+}^{\wl^+}(z) = -U_{hZ_L\wl^-}^{\wl^-}(z)\\
    &=\frac{g_2^2}{4}[g_2^2(1-z\zb)+4\lambda_h\zb][(\zb -z)(2+z\zb)-t_W^2\zb(1-\zb)]\frac{1}{\zb_+}~,\\
    U_{\wl hZ_L}^{\wl}(z) &\equiv U_{\wl^+ hZ_L}^{\wl^-}(z) = -U_{\wl^- hZ_L}^{\wl^+}(z) \\
    &= \frac{g_2^2}{8}[g_2^2(1-z\zb)-4\lambda_h z\zb](\zb-z)[(2+(1-t_W^2)z\zb]\frac{1}{z\zb_+}~,\\
    \end{split}
\ee
    \be\begin{split}
    U_{hZ_L\wt}^{\wt}(z) &\equiv U_{hZ_L\wt^+}^{\wt^+}(z) = U_{hZ_L\wt^-}^{\wt^-}(z)= \frac{g_2^4}{2}\zb(1+\zb)~,\\
    U_{\wt hZ_L}^{\wt}(z) &\equiv U_{\wt^+ hZ_L}^{\wt^-}(z) = -U_{\wt^- hZ_L}^{\wt^+}(z)= -\frac{g_2^4}{4}z\zb(\zb-z)~.\\
\end{split}
\ee
%

\section{Radiative corrections}
\label{app:radiative}

To cancel the IR divergences appearing in the splitting functions in the soft limit $z\to1$ we need to add virtual corrections. Instead of explicitly computing the corresponding Feynman diagrams, we use that virtual corrections correspond to a process in which the particles $B$ and $A$ are the same and $C$ is absent, so in the splitting formalism they can be treated introducing a new splitting function for each particle of the form $P_{BB}^v(x,t)=P_B^v(t)\delta(1-x)$. Inserting the new term in the DGLAP equations we get
\be
    \frac{df_{B}(x,t)}{dt} \supset P_{BB}^v\otimes f_B =  P_B^v(t)f_B(x,t)~. \label{eq:virtual_dglap}
\ee
The coefficients $P_B^v(t)$ can be computed using momentum conservation
\be
    \sum_{i}\int_0^1 dx \, x f_i(x,t) = \sum_i f_i^{(2)}(t) = 1 \quad \forall t~, \label{eq:momcons}
\ee
where $f_i^{(2)}$ represents the $n=2$ Mellin transform of the PDF
\be
    f^{(n)} = \int_0^1\frac{dx}{x}x^nf(x)~.
\ee
Deriving in $t$ and using Eqs.~(\ref{eq:DGLAP}, \ref{eq:virtual_dglap}) we obtain for each particle $A$
\be
    P_A^v(t)+\sum_{B,C}\left(\frac{d \mathcal{P}_{A \to B + C}}{dz dp_T^2}\right)^{(2)} = 0~.
\ee
From the definitions in Eqs.~(\ref{eq:splitting_generic}, \ref{eq:ampl_splitting_generic_massless}, \ref{eq:splitting_tilde}, \ref{eq:ucsplitting}), we get
\be
    P_A^v(t) = -\sum_{B,C}\left(\frac{\alpha_{ABC}}{2\pi}\widetilde{P}_{BA}^{C(2)}+\frac{v^2}{16\pi^2Q^2(t)}\widetilde{U}_{BA}^{C(2)}\right)~.
\ee
In order to reproduce the non cancellation of IR divergences in $SU(2)_L$ interactions, the computation of virtual corrections is modified by changing the boundary of the integral as discussed in Section~\ref{sec:EWdoublelogs}
\be
    P_A^v(t) = -\sum_{B,C} \int_0^{z_{\rm max}^{ABC}(t)} dz z  \left(\frac{\alpha_{ABC}}{2\pi}\widetilde{P}_{BA}^{C}(z)+\frac{v^2}{16\pi^2Q^2(t)}\widetilde{U}_{BA}^{C}(z)\right)~. \label{eq:virtual_coefficients}
\ee

\label{app:radiative_QED}

We show explicitly the virtual corrections for the QED+QCD phase, with $N_\ell$ charged leptons, $N_u$ up and $N_d$ down quarks. All the particles are treated as massless, so momentum conservation equations become just relations between the $n=2$ Mellin transform of the splitting functions and the virtual coefficients. Applying Eq.~\eqref{eq:virtual_coefficients} respectively to the fermions, the photon and the gluon we get, using Eq.~\eqref{eq:DGLAP_QED}
\be \begin{split}
    P_f^v &= \frac{3}{2} \left( \frac{\alpha_\gamma}{2\pi} Q_f^2 + \frac{\alpha_3}{2\pi} C_F \delta_{f,q}  \right)~,\\
    P_{\gamma}^v&= -\frac{\alpha_\gamma}{2\pi} \frac{2}{3}N_f^{\rm QED}~,\\
    P_{g}^v&= \frac{\alpha_3}{2\pi} \left(\frac{11}{6} C_A -\frac{2}{3} T_F N_q\right)~,
\end{split}
\ee
where $N_f^{\rm QED} = N_\ell + N_c (N_uQ_u^2+N_dQ_d^2)$ is the effective number of fermions, $N_q = N_u+N_d$ is the number of quarks and $\delta_{f,q}=1$ for quarks and $0$ for leptons. In our specific setup, $N_\ell = N_d = 3$ and $N_u = 2$, so $N_q = 5$ and $N_f^{\rm QED} = \frac{20}{3}$.
We omit writing the explicit expressions for virtual corrections in the full SM phase, as they are many and too lengthy.

\section{DGLAP evolution equations in SM}
\label{app:DGLAP}

Here we list the full set of DGLAP equations we used above the EW scale. We use the following notation for transverse vector polarization:
\begin{equation}
P_{BV_s}^C\otimes f_{V_s} = P_{BV_+}^C\otimes f_{V_+} + P_{BV_-}^C\otimes f_{V_-}~,
\end{equation}
\begin{equation}
P_{BV_s}^C\otimes f_{V_{-s}} = P_{BV_+}^C\otimes f_{V_-} + P_{BV_-}^C\otimes f_{V_+}~.
\end{equation}
In the ultra-collinear terms the factor $v^2/(16\pi^2Q(t)^2)$ appearing in front of each splitting function is omitted to shorten the notation.

\subsection{Leptons}

\begin{equation}
\begin{split}
\frac{df_{\nu_i}}{dt} &= P_{\nu_i}^v f_{\nu_i}+\frac{\alpha_{2}(t)}{2\pi c_W^2(t)}\frac{1}{4}\Big[\widetilde{P}_{ff}^V\otimes f_{\nu_i}
+ \widetilde{P}_{f_LV_s}^{f}\otimes f_{Z_s}\Big] \\
&+ \frac{\alpha_{2}(t)}{2\pi }\frac{1}{2}\Big[\widetilde{P}_{ff}^V\otimes f_{\ell_{L,i}} + \widetilde{P}_{f_LV_s}^{f}\otimes f_{W^+_s}\Big] \\
    &+ \widetilde{U}_{\nu \ell_L}^{W^-_L}\otimes f_{\ell_{L,i}} + \widetilde{U}_{\nu W_L^+}^{\bar{\ell}_L}\otimes f_{W^+_L}+\widetilde{U}_{\nu \nu}^{Z_L}\otimes f_{\nu_i} + \widetilde{U}_{\nu Z_L}^{\bar{\nu}}\otimes f_{Z_L}~,
\end{split}
\end{equation}
\begin{equation}
\begin{split}
\frac{df_{\bar{\nu}_i}}{dt} &= P_{\bar{\nu}_i}^v f_{\bar{\nu}_i}+\frac{\alpha_{2}(t)}{2\pi c_W^2(t)}\frac{1}{4}\Big[\widetilde{P}_{ff}^V\otimes f_{\bar{\nu}_i} + \widetilde{P}_{f_LV_s}^{f}\otimes f_{Z_{-s}}\Big] \\
&+ \frac{\alpha_{2}(t)}{2\pi }\frac{1}{2}\Big[\widetilde{P}_{ff}^V\otimes f_{\bar{\ell}_{L,i}} + \widetilde{P}_{f_LV_s}^{f}\otimes f_{W^-_{-s}}\Big] \\
&+ \widetilde{U}_{\bar{\nu}\bar{\ell}_L}^{W^+_L}\otimes f_{\bar{\ell}_{L,i}} + \widetilde{U}_{\bar{\nu}W_L^-}^{\ell_L}\otimes f_{W^-_L}+\widetilde{U}_{\bar{\nu}\bar{\nu}}^{Z_L}\otimes f_{\bar{\nu}_i} + \widetilde{U}_{\bar{\nu}Z_L}^{\nu}\otimes f_{Z_L}~,
\end{split}
\end{equation}
\begin{equation}
\begin{split}
\frac{df_{\ell_{L,i}}}{dt} &= P_{\ell_{L,i}}^v f_{\ell_{L,i}}+ \frac{\alpha_{\g}(t)}{2\pi}Q_{\ell}^2\Big[\widetilde{P}_{ff}^V\otimes f_{\ell_{L,i}} + \widetilde{P}_{f_LV_s}^{f}\otimes f_{\gamma_s}\Big] \\
&+ \frac{\alpha_{2}(t)}{2\pi c_W^2(t)}\left(\frac{1}{2}+Q_\ell s_W^2(t)\right)^2\Big[\widetilde{P}_{ff}^V\otimes f_{\ell_{L,i}} + \widetilde{P}_{f_LV_s}^{f}\otimes f_{Z_s}\Big] \\ 
&+ \frac{\alpha_{2}(t)}{2\pi }\frac{1}{2}\Big[\widetilde{P}_{ff}^V\otimes f_{\nu_i} + \widetilde{P}_{f_LV_s}^{f}\otimes f_{W^-_s}\Big]\\
&-\frac{\alpha_{\g2}(t)}{2\pi c_W(t)}Q_\ell\left(\frac{1}{2}+Q_\ell s_W^2(t)\right)\widetilde{P}_{f_LV_s}^{f}\otimes f_{Z\gamma_s} \\
&+\widetilde{U}_{\ell_L\nu}^{\Wpl}\otimes f_{\nu_i} + \widetilde{U}_{\ell_LW_L^-}^{\nbar}\otimes f_{W^-_L}+\widetilde{U}_{\ell_L\ell_L}^{\zl}\otimes f_{\ell_{L,i}} + \widetilde{U}_{\ell_LZ_L}^{\bar{\ell}_L}\otimes f_{Z_L}~,
\end{split}
\end{equation}
\begin{equation}
\begin{split}
\frac{df_{\bar{\ell}_{L,i}}}{dt} &= P_{\bar{\ell}_{L,i}}^v f_{\bar{\ell}_{L,i}}+ \frac{\alpha_{\g}(t)}{2\pi}Q_{\ell}^2\Big[\widetilde{P}_{ff}^V\otimes f_{\bar{\ell}_{L,i}} + \widetilde{P}_{f_LV_s}^{f}\otimes f_{\gamma_{-s}}\Big] \\
&+ \frac{\alpha_{2}(t)}{2\pi c_W^2(t)}\left(\frac{1}{2}+Q_\ell s_W^2(t)\right)^2\Big[\widetilde{P}_{ff}^V\otimes f_{\bar{\ell}_{L,i}} + \widetilde{P}_{f_LV_s}^{f}\otimes f_{Z_{-s}}\Big] \\
&+ \frac{\alpha_{2}(t)}{2\pi }\frac{1}{2}\Big[\widetilde{P}_{ff}^V\otimes f_{\bar{\nu}_i} + \widetilde{P}_{f_LV_s}^{f}\otimes f_{W^+_{-s}}\Big]\\
&-\frac{\alpha_{\g2}(t)}{2\pi c_W(t)}Q_\ell\left(\frac{1}{2}+Q_\ell s_W^2(t)\right)\widetilde{P}_{f_LV_s}^{f}\otimes f_{Z\gamma_{-s}} \\
&+\widetilde{U}_{\bar{\ell}_L\bar{\nu}}^{\Wml}\otimes f_{\bar{\nu}_i} + \widetilde{U}_{\bar{\ell}_LW_L^+}^{\nu}\otimes f_{W^+_L}+\widetilde{U}_{\bar{\ell}_L\bar{\ell}_L}^{\zl}\otimes f_{\bar{\ell}_{L,i}} + \widetilde{U}_{\bar{\ell}_LZ_L}^{\ell_L}\otimes f_{Z_L}~,
\end{split}
\end{equation}
\begin{equation}
\begin{split}
\frac{df_{\ell_{R,i}}}{dt} &= P_{\ell_{R,i}}^v f_{\ell_{R,i}}+ \frac{\alpha_{\g}(t)}{2\pi}Q_{\ell}^2\Big[\widetilde{P}_{ff}^V\otimes f_{\ell_{R,i}} + \widetilde{P}_{f_LV_s}^{f}\otimes f_{\gamma_{-s}}\Big] \\
&+ \frac{\alpha_{2}(t)}{2\pi c_W^2(t)}Q_\ell^2s_W^4(t)\Big[\widetilde{P}_{ff}^V\otimes f_{\ell_{R,i}} + \widetilde{P}_{f_LV_s}^{f}\otimes f_{Z_{-s}}\Big] \\
&- \frac{\alpha_{\g2}(t)}{2\pi c_W(t)}Q_\ell^2s_W^2(t)\widetilde{P}_{f_LV_s}^{f}\otimes f_{Z\gamma_{-s}} \\
&+\widetilde{U}_{\ell_R\ell_R}^{\zl}\otimes f_{\ell_{R,i}} + \widetilde{U}_{\ell_RZ_L}^{\bar{\ell}_R}\otimes f_{Z_L}~,
\end{split}
\end{equation}
\begin{equation}
\begin{split}
\frac{df_{\bar{\ell}_{R,i}}}{dt} &= P_{\bar{\ell}_{R,i}}^v f_{\bar{\ell}_{R,i}}+ \frac{\alpha_{\g}(t)}{2\pi}Q_{\ell}^2\Big[\widetilde{P}_{ff}^V\otimes f_{\bar{\ell}_{R,i}} + \widetilde{P}_{f_LV_s}^{f}\otimes f_{\gamma_s}\Big] \\
&+ \frac{\alpha_{2}(t)}{2\pi c_W^2(t)}Q_\ell^2s_W^4(t)\Big[\widetilde{P}_{ff}^V\otimes f_{\bar{\ell}_{R,i}} + \widetilde{P}_{f_LV_s}^{f}\otimes f_{Z_s}\Big] \\
&- \frac{\alpha_{\g2}(t)}{2\pi c_W(t)}Q_\ell^2s_W^2(t)\widetilde{P}_{f_LV_s}^{f}\otimes f_{Z\gamma_s} \\
&+\widetilde{U}_{\bar{\ell}_R\bar{\ell}_R}^{\zl}\otimes f_{\bar{\ell}_{R,i}} + \widetilde{U}_{\bar{\ell}_RZ_L}^{\ell_R}\otimes f_{Z_L}~.
\end{split}
\end{equation}
%

\subsection{Quarks}

\begin{equation}
\begin{split}
\frac{df_{u_{L,i}}}{dt} &= P_{u_{L,i}}^v f_{u_{L,i}}+ \frac{\alpha_{\g}(t)}{2\pi}Q_{u}^2\Big[\widetilde{P}_{ff}^V\otimes f_{u_{L,i}} + N_c\widetilde{P}_{f_LV_s}^{f}\otimes f_{\gamma_s}\Big]\\
&+ \frac{\alpha_{2}(t)}{2\pi c_W^2(t)}\left(\frac{1}{2}-Q_us_W^2(t)\right)^2\Big[\widetilde{P}_{ff}^V\otimes f_{u_{L,i}} + N_c\widetilde{P}_{f_LV_s}^{f}\otimes f_{Z_s}\Big] \\
&+ \frac{\alpha_{2}(t)}{2\pi }\frac{1}{2}\Big[\widetilde{P}_{ff}^V\otimes f_{d_{L,i}} + N_c\widetilde{P}_{f_LV_s}^{f}\otimes f_{W^+_s}\Big]\\
&+\frac{\alpha_{\g2}(t)}{2\pi}Q_uN_c\left(\frac{1}{2}-Q_us_W^2(t)\right)\widetilde{P}_{f_LV_s}^{f}\otimes f_{Z\gamma_s} \\
&+ \frac{\alpha_{3}(t)}{2\pi}\Big[C_F\widetilde{P}_{ff}^V\otimes f_{u_{L,i}} + T_F\widetilde{P}_{f_LV_s}^{f}\otimes f_{g_s}\Big] \\
&+\widetilde{U}_{u_Ld_L}^{\Wml}\otimes f_{d_{L,i}} + \widetilde{U}_{u_LW_L^+}^{\dlbar}\otimes f_{W^+_L}+\widetilde{U}_{u_Lu_L}^{\zl}\otimes f_{u_{L,i}} + \widetilde{U}_{u_LZ_L}^{\ulbar}\otimes f_{Z_L}~,
\end{split}
\end{equation}
\begin{equation}
\begin{split}
\frac{df_{\bar{u}_{L,i}}}{dt} &= P_{\bar{u}_{L,i}}^v f_{\bar{u}_{L,i}}+ \frac{\alpha_{\g}(t)}{2\pi}Q_{u}^2\Big[\widetilde{P}_{ff}^V\otimes f_{\bar{u}_{L,i}} + N_c\widetilde{P}_{f_LV_s}^{f}\otimes f_{\gamma_{-s}}\Big] \\
&+ \frac{\alpha_{2}(t)}{2\pi c_W^2(t)}\left(\frac{1}{2}-Q_us_W^2(t)\right)^2\Big[\widetilde{P}_{ff}^V\otimes f_{\bar{u}_{L,i}} + N_c\widetilde{P}_{f_LV_s}^{f}\otimes f_{Z_{-s}}\Big] \\ 
&+ \frac{\alpha_{2}(t)}{2\pi }\frac{1}{2}\Big[\widetilde{P}_{ff}^V\otimes f_{\bar{d}_{L,i}} + N_c\widetilde{P}_{f_LV_s}^{f}\otimes f_{W^-_{-s}}\Big]\\
&+\frac{\alpha_{\g2}(t)}{2\pi}Q_uN_c\left(\frac{1}{2}-Q_us_W^2(t)\right)\widetilde{P}_{f_LV_s}^{f}\otimes f_{Z\gamma_{-s}} \\
&+ \frac{\alpha_{3}(t)}{2\pi}\Big[C_F\widetilde{P}_{ff}^V\otimes f_{\bar{u}_{L,i}} + T_F\widetilde{P}_{f_LV_s}^{f}\otimes f_{g_{-s}}\Big] \\
&+\widetilde{U}_{\bar{u}_L\bar{d}_L}^{\Wpl}\otimes f_{\bar{d}_{L,i}} + \widetilde{U}_{\bar{u}_LW_L^-}^{d_L}\otimes f_{W^-_L}+\widetilde{U}_{\bar{u}_L\bar{u}_L}^{\zl}\otimes f_{\bar{u}_{L,i}} + \widetilde{U}_{\bar{u}_LZ_L}^{u_L}\otimes f_{Z_L}~,
\end{split}
\end{equation}
\begin{equation}
\begin{split}
\frac{df_{t_L}}{dt} &= P_{t_{L}}^v f_{t_{L}}+ \frac{\alpha_{\g}(t)}{2\pi}Q_{u}^2\Big[\widetilde{P}_{ff}^V\otimes f_{t_L} + N_c\widetilde{P}_{f_LV_s}^{f}\otimes f_{\gamma_s}\Big]\\
&+ \frac{\alpha_{2}(t)}{2\pi c_W^2(t)}\left(\frac{1}{2}-Q_us_W^2(t)\right)^2\Big[\widetilde{P}_{ff}^V\otimes f_{t_L} + N_c\widetilde{P}_{f_LV_s}^{f}\otimes f_{Z_s}\Big] \\ 
&+ \frac{\alpha_{2}(t)}{2\pi }\frac{1}{2}\Big[\widetilde{P}_{ff}^V\otimes f_{b_L} + N_c\widetilde{P}_{f_LV_s}^{f}\otimes f_{W^+_s}\Big]\\
&+\frac{\alpha_{\g2}(t)}{2\pi}Q_uN_c\left(\frac{1}{2}-Q_us_W^2(t)\right)\widetilde{P}_{f_LV_s}^{f}\otimes f_{Z\gamma_s} \\
&+ \frac{\alpha_{3}(t)}{2\pi}\Big[C_F\widetilde{P}_{ff}^V\otimes f_{t_L} + T_F\widetilde{P}_{f_LV_s}^{f}\otimes f_{g_s}\Big] \\
&+\frac{\alpha_{y}(t)}{2\pi}\frac{1}{2}\Big[\widetilde{P}_{ff}^{h}\otimes \left(f_{t_{R}}^{(h)}+f_{t_{R}}^{(Z_L)}\right) + N_c\widetilde{P}_{fh}^{f}\otimes \left(f_h+f_{Z_L}+f_{hZ_L}\right)\Big]\\
&+\widetilde{U}_{tt}^{g}\otimes f_{t_R} + \widetilde{U}_{tg}^{t}\otimes f_{g_-}\\
&+\widetilde{U}_{tt}^{\g}\otimes f_{t_R} + \widetilde{U}_{t\g}^{t}\otimes f_{\gm}+\widetilde{U}_{t_Lt_R}^{\zp}\otimes f_{t_R} + \widetilde{U}_{t_LZ_-}^{\trb}\otimes f_{Z_-} + \widetilde{U}_{t_L\zgm}^{\trb}\otimes f_{\zgm}\\
&+\widetilde{U}_{t_Lb_L}^{\Wml}\otimes f_{b_L} + \widetilde{U}_{t_LW_L^+}^{\blb}\otimes f_{W^+_L}\\
&+\widetilde{U}_{t_Lt_L}^{\zl}\otimes f_{t_L} + \widetilde{U}_{t_LZ_L}^{\tlb}\otimes f_{Z_L}+\widetilde{U}_{tt}^{h}\otimes f_{t_L} + \widetilde{U}_{th}^{t}\otimes f_{h}+ \widetilde{U}_{t_LhZ_L}^{\tlb}\otimes f_{hZ_L}~,
\label{eq:tL_DGLAP}
\end{split}
\end{equation}
\begin{equation}
\begin{split}
\frac{df_{\bar{t}_L}}{dt} &= P_{\bar{t}_{L}}^v f_{\bar{t}_{L}}+ \frac{\alpha_{\g}(t)}{2\pi}Q_{u}^2\Big[\widetilde{P}_{ff}^V\otimes f_{\bar{t}_L} + N_c\widetilde{P}_{f_LV_s}^{f}\otimes f_{\gamma_{-s}}\Big] \\
&+ \frac{\alpha_{2}(t)}{2\pi c_W^2(t)}\left(\frac{1}{2}-Q_us_W^2(t)\right)^2\Big[\widetilde{P}_{ff}^V\otimes f_{\bar{t}_L} + N_c\widetilde{P}_{f_LV_s}^{f}\otimes f_{Z_{-s}}\Big] \\ 
&+ \frac{\alpha_{2}(t)}{2\pi }\frac{1}{2}\Big[\widetilde{P}_{ff}^V\otimes f_{\bar{b}_L} + N_c\widetilde{P}_{f_LV_s}^{f}\otimes f_{W^-_{-s}}\Big]\\
&+\frac{\alpha_{\g2}(t)}{2\pi}Q_uN_c\left(\frac{1}{2}-Q_us_W^2(t)\right)\widetilde{P}_{f_LV_s}^{f}\otimes f_{Z\gamma_{-s}} \\
&+ \frac{\alpha_{3}(t)}{2\pi}\Big[C_F\widetilde{P}_{ff}^V\otimes f_{\bar{t}_L} + T_F\widetilde{P}_{f_LV_s}^{f}\otimes f_{g_{-s}}\Big] \\
&+\frac{\alpha_{y}(t)}{2\pi}\frac{1}{2}\Big[\widetilde{P}_{ff}^{h}\otimes \left(f_{\bar{t}_{R}}^{(h)}+f_{\bar{t}_{R}}^{(Z_L)}\right) + N_c\widetilde{P}_{fh}^{f}\otimes \left(f_h+f_{Z_L}-f_{hZ_L}\right)\Big]\\
&+\widetilde{U}_{tt}^{g}\otimes f_{\bar{t}_R} + \widetilde{U}_{tg}^{t}\otimes f_{g_+}\\
&+\widetilde{U}_{tt}^{\g}\otimes f_{\bar{t}_R} + \widetilde{U}_{t\g}^{t}\otimes f_{\gp}+\widetilde{U}_{\bar{t}_L\bar{t}_R}^{\zm}\otimes f_{\bar{t}_R} + \widetilde{U}_{\bar{t}_LZ_+}^{t_R}\otimes f_{Z_+} + \widetilde{U}_{\bar{t}_L\zgp}^{t_R}\otimes f_{\zgp}\\
&+\widetilde{U}_{\bar{t}_L\bar{b}_L}^{\Wpl}\otimes f_{\bar{b}_L} + \widetilde{U}_{\bar{t}_LW_L^-}^{b_L}\otimes f_{W^-_L}\\
&+\widetilde{U}_{\bar{t}_L\bar{t}_L}^{\zl}\otimes f_{\bar{t}_L} + \widetilde{U}_{\bar{t}_LZ_L}^{t_L}\otimes f_{Z_L}+\widetilde{U}_{tt}^{h}\otimes f_{\bar{t}_L} + \widetilde{U}_{th}^{t}\otimes f_{h}+ \widetilde{U}_{\bar{t}_LhZ_L}^{t_L}\otimes f_{hZ_L}~,
\label{eq:tLb_DGLAP}
\end{split}
\end{equation}
\begin{equation}
\begin{split}
\frac{df_{d_{L,i}}}{dt} &= P_{d_{L,i}}^v f_{d_{L,i}}+ \frac{\alpha_{\g}(t)}{2\pi}Q_{d}^2\Big[\widetilde{P}_{ff}^V\otimes f_{d_{L,i}} + N_c\widetilde{P}_{f_LV_s}^{f}\otimes f_{\gamma_s}\Big]\\
&+ \frac{\alpha_{2}(t)}{2\pi c_W^2(t)}\left(\frac{1}{2}+Q_ds_W^2(t)\right)^2\Big[\widetilde{P}_{ff}^V\otimes f_{d_{L,i}} + N_c\widetilde{P}_{f_LV_s}^{f}\otimes f_{Z_s}\Big] \\
&+ \frac{\alpha_{2}(t)}{2\pi }\frac{1}{2}\Big[\widetilde{P}_{ff}^V\otimes f_{u_{L,i}} + N_c\widetilde{P}_{f_LV_s}^{f}\otimes f_{W^-_s}\Big]\\
&-\frac{\alpha_{\g2}(t)}{2\pi}Q_dN_c\left(\frac{1}{2}+Q_ds_W^2(t)\right)\widetilde{P}_{f_LV_s}^{f}\otimes f_{Z\gamma_s} \\
&+ \frac{\alpha_{3}(t)}{2\pi}\Big[C_F\widetilde{P}_{ff}^V\otimes f_{d_{L,i}} + T_F\widetilde{P}_{f_LV_s}^{f}\otimes f_{g_s}\Big]\\
&+\widetilde{U}_{d_Lu_L}^{\Wpl}\otimes f_{u_{L,i}} + \widetilde{U}_{d_LW_L^-}^{\ulbar}\otimes f_{W^-_L}+\widetilde{U}_{d_Ld_L}^{\zl}\otimes f_{d_{L,i}} + \widetilde{U}_{d_LZ_L}^{\dlbar}\otimes f_{Z_L}~,
\end{split}
\end{equation}
\begin{equation}
\begin{split}
\frac{df_{\bar{d}_{L,i}}}{dt} &= P_{\bar{d}_{L,i}}^v f_{\bar{d}_{L,i}}+ \frac{\alpha_{\g}(t)}{2\pi}Q_{d}^2\Big[\widetilde{P}_{ff}^V\otimes f_{\bar{d}_{L,i}} + N_c\widetilde{P}_{f_LV_s}^{f}\otimes f_{\gamma_{-s}}\Big] \\
&+ \frac{\alpha_{2}(t)}{2\pi c_W^2(t)}\left(\frac{1}{2}+Q_ds_W^2(t)\right)^2\Big[\widetilde{P}_{ff}^V\otimes f_{\bar{d}_{L,i}} + N_c\widetilde{P}_{f_LV_s}^{f}\otimes f_{Z_{-s}}\Big] \\ 
&+ \frac{\alpha_{2}(t)}{2\pi }\frac{1}{2}\Big[\widetilde{P}_{ff}^V\otimes f_{\bar{u}_{L,i}} + N_c\widetilde{P}_{f_LV_s}^{f}\otimes f_{W^+_{-s}}\Big]\\
&-\frac{\alpha_{\g2}(t)}{2\pi}Q_dN_c\left(\frac{1}{2}+Q_ds_W^2(t)\right)\widetilde{P}_{f_LV_s}^{f}\otimes f_{Z\gamma_{-s}} \\
&+ \frac{\alpha_{3}(t)}{2\pi}\Big[C_F\widetilde{P}_{ff}^V\otimes f_{\bar{d}_{L,i}} + T_F\widetilde{P}_{f_LV_s}^{f}\otimes f_{g_{-s}}\Big] \\
&+\widetilde{U}_{\bar{d}_L\bar{u}_L}^{\Wml}\otimes f_{\bar{u}_{L,i}} + \widetilde{U}_{\bar{d}_LW_L^+}^{u_L}\otimes f_{W^+_L}+\widetilde{U}_{\bar{d}_L\bar{d}_L}^{\zl}\otimes f_{\bar{d}_{L,i}} + \widetilde{U}_{\bar{d}_LZ_L}^{d_L}\otimes f_{Z_L}~,
\end{split}
\end{equation}
\begin{equation}
\begin{split}
\frac{df_{b_L}}{dt} &= P_{b_{L}}^v f_{b_{L}}+ \frac{\alpha_{\g}(t)}{2\pi}Q_{d}^2\Big[\widetilde{P}_{ff}^V\otimes f_{b_L} + N_c\widetilde{P}_{f_LV_s}^{f}\otimes f_{\gamma_s}\Big]\\
&+ \frac{\alpha_{2}(t)}{2\pi c_W^2(t)}\left(\frac{1}{2}+Q_ds_W^2(t)\right)^2\Big[\widetilde{P}_{ff}^V\otimes f_{b_L} + N_c\widetilde{P}_{f_LV_s}^{f}\otimes f_{Z_s}\Big] \\ 
&+ \frac{\alpha_{2}(t)}{2\pi }\frac{1}{2}\Big[\widetilde{P}_{ff}^V\otimes f_{t_L} + N_c\widetilde{P}_{f_LV_s}^{f}\otimes f_{W^-_s}\Big]\\
&-\frac{\alpha_{\g2}(t)}{2\pi}Q_dN_c\left(\frac{1}{2}+Q_ds_W^2(t)\right)\widetilde{P}_{f_LV_s}^{f}\otimes f_{Z\gamma_s} \\
&+ \frac{\alpha_{3}(t)}{2\pi}\Big[C_F\widetilde{P}_{ff}^V\otimes f_{b_L} + T_F\widetilde{P}_{f_LV_s}^{f}\otimes f_{g_s}\Big]\\
&+ \frac{\alpha_{y}(t)}{2\pi}\Big[\widetilde{P}_{ff}^{h}\otimes f_{t_R} + N_c\widetilde{P}_{fh}^{f}\otimes f_{W^-_L}\Big]\\
&+\widetilde{U}_{b_Lt_R}^{\Wpp}\otimes f_{t_R} + \widetilde{U}_{b_LW_-^-}^{\trb}\otimes f_{W^-_-}\\
&+\widetilde{U}_{b_Lt_L}^{\Wpl}\otimes f_{t_L} + \widetilde{U}_{b_LW_L^-}^{\tlb}\otimes f_{W^-_L}+\widetilde{U}_{b_Lb_L}^{\zl}\otimes f_{b_L} + \widetilde{U}_{b_LZ_L}^{\blb}\otimes f_{Z_L}~,
\end{split}
\end{equation}
\begin{equation}
\begin{split}
\frac{df_{\bar{b}_L}}{dt} &= P_{\bar{b}_{L}}^v f_{\bar{b}_{L}}+ \frac{\alpha_{\g}(t)}{2\pi}Q_{d}^2\Big[\widetilde{P}_{ff}^V\otimes f_{\bar{b}_{L}} + N_c\widetilde{P}_{f_LV_s}^{f}\otimes f_{\gamma_{-s}}\Big] \\
&+ \frac{\alpha_{2}(t)}{2\pi c_W^2(t)}\left(\frac{1}{2}+Q_ds_W^2(t)\right)^2\Big[\widetilde{P}_{ff}^V\otimes f_{\bar{b}_{L}} + N_c\widetilde{P}_{f_LV_s}^{f}\otimes f_{Z_{-s}}\Big] \\ 
&+ \frac{\alpha_{2}(t)}{2\pi }\frac{1}{2}\Big[\widetilde{P}_{ff}^V\otimes f_{\bar{t}_{L}} + N_c\widetilde{P}_{f_LV_s}^{f}\otimes f_{W^+_{-s}}\Big]\\
&-\frac{\alpha_{\g2}(t)}{2\pi}Q_dN_c\left(\frac{1}{2}+Q_ds_W^2(t)\right)\widetilde{P}_{f_LV_s}^{f}\otimes f_{Z\gamma_{-s}} \\
&+ \frac{\alpha_{3}(t)}{2\pi}\Big[C_F\widetilde{P}_{ff}^V\otimes f_{\bar{b}_{L}} + T_F\widetilde{P}_{f_LV_s}^{f}\otimes f_{g_{-s}}\Big] \\
&+ \frac{\alpha_{y}(t)}{2\pi}\Big[\widetilde{P}_{ff}^{h}\otimes f_{\bar{t}_R} + N_c\widetilde{P}_{fh}^{f}\otimes f_{W^+_L}\Big]\\
&+\widetilde{U}_{\bar{b}_L\bar{t}_R}^{\Wmm}\otimes f_{\bar{t}_R} + \widetilde{U}_{\bar{b}_LW_+^+}^{t_R}\otimes f_{W^+_+}\\
&+\widetilde{U}_{\bar{b}_L\bar{t}_L}^{\Wml}\otimes f_{\bar{t}_{L}} + \widetilde{U}_{\bar{b}_LW_L^+}^{t_L}\otimes f_{W^+_L}+\widetilde{U}_{\bar{b}_L\bar{b}_L}^{\zl}\otimes f_{\bar{b}_{L}} + \widetilde{U}_{\bar{b}_LZ_L}^{t_L}\otimes f_{Z_L}~,
\end{split}
\end{equation}
\begin{equation}
\begin{split}
\frac{df_{u_{R,i}}}{dt} &= P_{u_{R,i}}^v f_{u_{R,i}}+ \frac{\alpha_{\g}(t)}{2\pi}Q_{u}^2\Big[\widetilde{P}_{ff}^V\otimes f_{u_{R,i}} + N_c\widetilde{P}_{f_LV_s}^{f}\otimes f_{\gamma_{-s}}\Big] \\
&+ \frac{\alpha_{2}(t)}{2\pi c_W^2(t)}Q_u^2s_W^4(t)\Big[\widetilde{P}_{ff}^V\otimes f_{u_{R,i}} + N_c\widetilde{P}_{f_LV_s}^{f}\otimes f_{Z_{-s}}\Big] \\
&- N_c\frac{\alpha_{\g2}(t)}{2\pi c_W(t)}Q_u^2s_W^2(t)\widetilde{P}_{f_LV_s}^{f}\otimes f_{Z\gamma_{-s}} \\
&+ \frac{\alpha_{3}(t)}{2\pi}\Big[C_F\widetilde{P}_{ff}^V\otimes f_{u_{R,i}} + T_F\widetilde{P}_{f_LV_s}^{f}\otimes f_{g_{-s}}\Big]\\
&+\widetilde{U}_{u_Ru_R}^{\zl}\otimes f_{u_{R,i}} + \widetilde{U}_{u_RZ_L}^{\urbar}\otimes f_{Z_L},
\end{split}
\end{equation}
\begin{equation}
\begin{split}
\frac{df_{\bar{u}_{R,i}}}{dt} &= P_{\bar{u}_{R,i}}^v f_{\bar{u}_{R,i}}+ \frac{\alpha_{\g}(t)}{2\pi}Q_{u}^2\Big[\widetilde{P}_{ff}^V\otimes f_{\bar{u}_{R,i}} + N_c\widetilde{P}_{f_LV_s}^{f}\otimes f_{\gamma_s}\Big] \\
&+ \frac{\alpha_{2}(t)}{2\pi c_W^2(t)}Q_u^2s_W^4(t)\Big[\widetilde{P}_{ff}^V\otimes f_{\bar{u}_{R,i}} + N_c\widetilde{P}_{f_LV_s}^{f}\otimes f_{Z_s}\Big] \\
&- N_c\frac{\alpha_{\g2}(t)}{2\pi c_W(t)}Q_u^2s_W^2(t)\widetilde{P}_{f_LV_s}^{f}\otimes f_{Z\gamma_s} \\
&+\frac{\alpha_{3}(t)}{2\pi}\Big[C_F\widetilde{P}_{ff}^V\otimes f_{\bar{u}_{R,i}} + T_F\widetilde{P}_{f_LV_s}^{f}\otimes f_{g_s}\Big]\\ &+\widetilde{U}_{\bar{u}_R\bar{u}_R}^{\zl}\otimes f_{\bar{u}_{R,i}} + \widetilde{U}_{\bar{u}_RZ_L}^{u_R}\otimes f_{Z_L}~,
\end{split}
\end{equation}
\begin{equation}
\begin{split}
\frac{df_{t_R}}{dt} &= P_{t_{R}}^v f_{t_{R}}+ \frac{\alpha_{\g}(t)}{2\pi}Q_{u}^2\Big[\widetilde{P}_{ff}^V\otimes f_{t_R} + N_c\widetilde{P}_{f_LV_s}^{f}\otimes f_{\gamma_{-s}}\Big] \\
&+ \frac{\alpha_{2}(t)}{2\pi c_W^2(t)}Q_u^2s_W^4(t)\Big[\widetilde{P}_{ff}^V\otimes f_{t_R} + N_c\widetilde{P}_{f_LV_s}^{f}\otimes f_{Z_{-s}}\Big] \\
&- N_c\frac{\alpha_{\g2}(t)}{2\pi c_W(t)}Q_u^2s_W^2(t)\widetilde{P}_{f_LV_s}^{f}\otimes f_{Z\gamma_{-s}} \\
&+ \frac{\alpha_{3}(t)}{2\pi}\Big[C_F\widetilde{P}_{ff}^V\otimes f_{t_R} + T_F\widetilde{P}_{f_LV_s}^{f}\otimes f_{g_{-s}}\Big]\\
&+\frac{\alpha_{y}(t)}{2\pi}\left[\widetilde{P}_{ff}^{h}\otimes \left(\frac{f_{t_{L}}^{(h)}+f_{t_{L}}^{(Z_L)}}{2}+f_{b_L}\right) + N_c\widetilde{P}_{fh}^{f}\otimes \left(\frac{f_h+f_{Z_L}-f_{hZ_L}}{2}+f_{W^+_L}\right)\right]\\
&+\widetilde{U}_{tt}^{g}\otimes f_{t_L} + \widetilde{U}_{tg}^{t}\otimes f_{g_+}\\
&+\widetilde{U}_{tt}^{\g}\otimes f_{t_L} + \widetilde{U}_{t\g}^{t}\otimes f_{\gp}+\widetilde{U}_{t_Rt_L}^{\zm}\otimes f_{t_L} + \widetilde{U}_{t_RZ_+}^{\tlb}\otimes f_{Z_+} + \widetilde{U}_{t_R\zgp}^{\tlb}\otimes f_{\zgp}\\
&+\widetilde{U}_{t_Rb_L}^{\Wmm}\otimes f_{b_L} + \widetilde{U}_{t_R\Wpp}^{\blb}\otimes f_{\Wpp}\\
&+\widetilde{U}_{t_Rt_R}^{\zl}\otimes f_{t_R} + \widetilde{U}_{t_RZ_L}^{\trb}\otimes f_{Z_L}+\widetilde{U}_{tt}^{h}\otimes f_{t_R} + \widetilde{U}_{th}^{t}\otimes f_{h}+ \widetilde{U}_{t_RhZ_L}^{\trb}\otimes f_{hZ_L}~,
\label{eq:tR_DGLAP}
\end{split}
\end{equation}
\begin{equation}
\begin{split}
\frac{df_{\bar{t}_{R}}}{dt} &= P_{\bar{t}_{R}}^v f_{\bar{t}_{R}}+ \frac{\alpha_{\g}(t)}{2\pi}Q_{u}^2\Big[\widetilde{P}_{ff}^V\otimes f_{\bar{t}_{R}} + N_c\widetilde{P}_{f_LV_s}^{f}\otimes f_{\gamma_s}\Big] \\
&+ \frac{\alpha_{2}(t)}{2\pi c_W^2(t)}Q_u^2s_W^4(t)\Big[\widetilde{P}_{ff}^V\otimes f_{\bar{t}_{R}} + N_c\widetilde{P}_{f_LV_s}^{f}\otimes f_{Z_s}\Big] \\
&- N_c\frac{\alpha_{\g2}(t)}{2\pi c_W(t)}Q_u^2s_W^2(t)\widetilde{P}_{f_LV_s}^{f}\otimes f_{Z\gamma_s} \\
&+\frac{\alpha_{3}(t)}{2\pi}\Big[C_F\widetilde{P}_{ff}^V\otimes f_{\bar{t}_{R}} + T_F\widetilde{P}_{f_LV_s}^{f}\otimes f_{g_s}\Big]\\
&+\frac{\alpha_{y}(t)}{2\pi}\left[\widetilde{P}_{ff}^{h}\otimes \left(\frac{f_{\bar{t}_{L}}^{(h)}+f_{\bar{t}_{L}}^{(Z_L)}}{2}+f_{\bar{b}_L}\right) + N_c\widetilde{P}_{fh}^{f}\otimes \left(\frac{f_h+f_{Z_L}+f_{hZ_L}}{2}+f_{W^-_L}\right)\right]\\
&+\widetilde{U}_{tt}^{g}\otimes f_{\bar{t}_L} + \widetilde{U}_{tg}^{t}\otimes f_{g_-}\\
&+\widetilde{U}_{tt}^{\g}\otimes f_{\bar{t}_L} + \widetilde{U}_{t\g}^{t}\otimes f_{\gm}+\widetilde{U}_{\bar{t}_R\bar{t}_L}^{\zp}\otimes f_{\bar{t}_L} + \widetilde{U}_{\bar{t}_RZ_-}^{t_L}\otimes f_{Z_-} + \widetilde{U}_{\bar{t}_R\zgm}^{t_L}\otimes f_{\zgm}\\
&+\widetilde{U}_{\bar{t}_R\bar{b}_L}^{\Wpp}\otimes f_{\bar{b}_L} + \widetilde{U}_{\bar{t}_R\Wmm}^{b_L}\otimes f_{\Wmm}\\
&+\widetilde{U}_{\bar{t}_R\bar{t}_R}^{\zl}\otimes f_{\bar{t}_{R}} + \widetilde{U}_{\bar{t}_RZ_L}^{t_R}\otimes f_{Z_L}+\widetilde{U}_{tt}^{h}\otimes f_{\bar{t}_R} + \widetilde{U}_{th}^{t}\otimes f_{h}+ \widetilde{U}_{\bar{t}_RhZ_L}^{t_R}\otimes f_{hZ_L}~,
\label{eq:tRb_DGLAP}
\end{split}
\end{equation}
\begin{equation}
\begin{split}
\frac{df_{d_{R,i}}}{dt} &= P_{d_{R,i}}^v f_{d_{R,i}}+ \frac{\alpha_{\g}(t)}{2\pi}Q_{d}^2\Big[\widetilde{P}_{ff}^V\otimes f_{d_{R,i}} + N_c\widetilde{P}_{f_LV_s}^{f}\otimes f_{\gamma_{-s}}\Big] \\
&+ \frac{\alpha_{2}(t)}{2\pi c_W^2(t)}Q_d^2s_W^4(t)\Big[\widetilde{P}_{ff}^V\otimes f_{d_{R,i}} + N_c\widetilde{P}_{f_LV_s}^{f}\otimes f_{Z_{-s}}\Big] \\
&- N_c\frac{\alpha_{\g2}(t)}{2\pi c_W(t)}Q_d^2s_W^2(t)\widetilde{P}_{f_LV_s}^{f}\otimes f_{Z\gamma_{-s}} \\
&+ \frac{\alpha_{3}(t)}{2\pi}\Big[C_F\widetilde{P}_{ff}^V\otimes f_{d_{R,i}} + T_F\widetilde{P}_{f_LV_s}^{f}\otimes f_{g_{-s}}\Big] \\
&+\widetilde{U}_{d_Rd_R}^{\zl}\otimes f_{d_{R,i}} + \widetilde{U}_{d_RZ_L}^{\drbar}\otimes f_{Z_L}~,
\end{split}
\end{equation}
\begin{equation}
\begin{split}
\frac{df_{\bar{d}_{R,i}}}{dt} &= P_{\bar{d}_{R,i}}^v f_{\bar{d}_{R,i}}+ \frac{\alpha_{\g}(t)}{2\pi}Q_{d}^2\Big[\widetilde{P}_{ff}^V\otimes f_{\bar{d}_{R,i}} + N_c\widetilde{P}_{f_LV_s}^{f}\otimes f_{\gamma_s}\Big] \\
&+ \frac{\alpha_{2}(t)}{2\pi c_W^2(t)}Q_d^2s_W^4(t)\Big[\widetilde{P}_{ff}^V\otimes f_{\bar{d}_{R,i}} + N_c\widetilde{P}_{f_LV_s}^{f}\otimes f_{Z_s}\Big] \\
&- N_c\frac{\alpha_{\g2}(t)}{2\pi c_W(t)}Q_d^2s_W^2(t)\widetilde{P}_{f_LV_s}^{f}\otimes f_{Z\gamma_s} \\
&+\frac{\alpha_{3}(t)}{2\pi}\Big[C_F\widetilde{P}_{ff}^V\otimes f_{\bar{d}_{R,i}} + T_F\widetilde{P}_{f_LV_s}^{f}\otimes f_{g_s}\Big]\\ &+\widetilde{U}_{\bar{d}_R\bar{d}_R}^{\zl}\otimes f_{\bar{d}_{R,i}} + \widetilde{U}_{\bar{d}_RZ_L}^{d_R}\otimes f_{Z_L}~.
\end{split}
\end{equation}

\subsection{Transverse gauge bosons}
\label{sec:DGLAP_VT}

\be
\begin{split}
\frac{df_{g_+}}{dt} &= P_{g_+}^v f_{g_+}+ \frac{\alpha_{3}(t)}{2\pi}\Big[C_{A}^{(3)}P_{V_+ V_s}^V\otimes f_{g_s} + C_{F}^{(3)}\widetilde{P}_{V_+ f_L}^{f} \otimes \sum_i\left(f_{u_L,i}+f_{d_L,i}+f_{\bar{u}_R,i}+f_{\bar{d}_R,i}\right)\\
&+ C_{F}^{(3)}\widetilde{P}_{V_- f_L}^{f} \otimes \sum_i\left(f_{u_R,i}+f_{d_R,i}+f_{\bar{u}_L,i}+f_{\bar{d}_L,i}\right)\Big] +\widetilde{U}_{g t}^{t}\otimes (f_{t_R}+f_{\bar{t}_L})~, \\
\end{split}
\ee
\be
\begin{split}
\frac{df_{g_-}}{dt} &= P_{g_-}^v f_{g_-}+ \frac{\alpha_{3}(t)}{2\pi}\Big[C_{A}^{(3)}P_{V_- V_s}^V\otimes f_{g_s} + C_{F}^{(3)}\widetilde{P}_{V_- f_L}^{f} \otimes \sum_i\left(f_{u_L,i}+f_{d_L,i}+f_{\bar{u}_R,i}+f_{\bar{d}_R,i}\right)\\
&+ C_{F}^{(3)}\widetilde{P}_{V_+ f_L}^{f} \otimes \sum_i\left(f_{u_R,i}+f_{d_R,i}+f_{\bar{u}_L,i}+f_{\bar{d}_L,i}\right)\Big] +\widetilde{U}_{g t}^{t}\otimes (f_{t_L}+f_{\bar{t}_R})~, \\
\end{split}
\ee
\be
\begin{split}
\frac{df_{\gamma_+}}{dt} &= P_{\gp}^v f_{\gp}+ \frac{\alpha_{\g}(t)}{2\pi}\widetilde{P}_{V_+ V_s}^V\otimes (f_{W^+_s}+f_{W^-_s}) + \frac{\alpha_{\g}(t)}{2\pi}\widetilde{P}_{V_+ h}^h\otimes (f_{W^+_L}+f_{W^-_L}) \\
&+ \frac{\alpha_{\g}(t)}{2\pi}\sum_{f}Q_f^2\left[\widetilde{P}_{V_+ f_L}^{f} \otimes \left(f_{f_L}+f_{\bar{f}_R}\right)
+ \widetilde{P}_{V_- f_L}^{f} \otimes \left(f_{f_R}+f_{\bar{f}_L}\right)\right] \\
&+\widetilde{U}_{\g t}^{t}\otimes (f_{t_R}+f_{\bar{t}_L}) +\widetilde{U}_{\gamma_TW_T}^{W_L}\otimes (f_{W_+^+}+f_{W_+^-}) +\widetilde{U}_{\gamma_TW_L}^{W_T}\otimes (f_{W_L^+}+f_{W_L^-})~,
\end{split}
\ee
\be
\begin{split}
\frac{df_{\gamma_-}}{dt} &= P_{\gm}^v f_{\gm}+ \frac{\alpha_{\g}(t)}{2\pi}\widetilde{P}_{V_- V_s}^V\otimes (f_{W^+_s}+f_{W^-_s}) + \frac{\alpha_{\g}(t)}{2\pi}\widetilde{P}_{V_- h}^h\otimes (f_{W^+_L}+f_{W^-_L}) \\
&+ \frac{\alpha_{\g}(t)}{2\pi}\sum_{f}Q_f^2\left[\widetilde{P}_{V_- f_L}^{f} \otimes \left(f_{f_L}+f_{\bar{f}_R}\right)
+ \widetilde{P}_{V_+ f_L}^{f} \otimes \left(f_{f_R}+f_{\bar{f}_L}\right)\right] \\
&+\widetilde{U}_{\g t}^{t}\otimes (f_{t_L}+f_{\bar{t}_R}) +\widetilde{U}_{\gamma_TW_T}^{W_L}\otimes (f_{W_-^+}+f_{W_-^-}) +\widetilde{U}_{\gamma_TW_L}^{W_T}\otimes (f_{W_L^+}+f_{W_L^-})~,
\end{split}
\ee
\be
\begin{split}
\frac{df_{Z_+}}{dt} &= P_{\zp}^v f_{\zp}+\frac{\alpha_{2}(t)}{2\pi}c_W^2(t)\widetilde{P}_{V_+ V_s}^V\otimes (f_{W^+_s}+f_{W^-_s})\\
&+ \frac{\alpha_{2}(t)}{2\pi}\frac{c_{2W}^2}{4c_W^2}\widetilde{P}_{V_+ h}^h\otimes (f_{W^+_L}+f_{W^-_L})
+ \frac{\alpha_{2}(t)}{2\pi}\frac{1}{4c_W^2}\widetilde{P}_{V_+ h}^h\otimes (f_{h}+f_{Z_L})\\
&+ \frac{\alpha_{2}(t)}{2\pi c_W^2}\sum_{f}\left[\widetilde{P}_{V_+ f_L}^{f} \otimes \left(Z_{f_L}^2f_{f_L}+Z_{f_R}^2f_{\bar{f}_R}\right)
+ \widetilde{P}_{V_- f_L}^{f} \otimes \left(Z_{f_R}^2f_{f_R}+Z_{f_L}^2f_{\bar{f}_L}\right)\right]\\
&+\widetilde{U}_{Z_+t_R}^{t_L}\otimes f_{t_R}+\widetilde{U}_{Z_-t_L}^{t_R}\otimes f_{\bar{t}_L}\\
&+\widetilde{U}_{Z_TW_T}^{W_L}\otimes (f_{W_+^+}+f_{W_+^-}) +\widetilde{U}_{Z_TW_L}^{W_T}\otimes (f_{W_L^+}+f_{W_L^-})\\
&+ \widetilde{U}_{Z_TZ_T}^{h}\otimes f_{Z_+} +\widetilde{U}_{Z_Th}^{Z_T}\otimes f_{h}~,
\end{split}
\ee
\be
\begin{split}
\frac{df_{Z_-}}{dt} &= P_{\zm}^v f_{\zm}+\frac{\alpha_{2}(t)}{2\pi}c_W^2(t)\widetilde{P}_{V_- V_s}^V\otimes (f_{W^+_s}+f_{W^-_s})\\
&+ \frac{\alpha_{2}(t)}{2\pi}\frac{c_{2W}^2}{4c_W^2}\widetilde{P}_{V_- h}^h\otimes (f_{W^+_L}+f_{W^-_L})
+ \frac{\alpha_{2}(t)}{2\pi}\frac{1}{4c_W^2}\widetilde{P}_{V_- h}^h\otimes (f_{h}+f_{Z_L})\\
&+ \frac{\alpha_{2}(t)}{2\pi c_W^2}\sum_{f}\left[\widetilde{P}_{V_- f_L}^{f} \otimes \left(Z_{f_L}^2f_{f_L}+Z_{f_R}^2f_{\bar{f}_R}\right)
+ \widetilde{P}_{V_+ f_L}^{f} \otimes \left(Z_{f_R}^2f_{f_R}+Z_{f_L}^2f_{\bar{f}_L}\right)\right]\\
&+\widetilde{U}_{Z_-t_L}^{t_R}\otimes f_{t_L}+\widetilde{U}_{Z_+t_R}^{t_L}\otimes f_{\bar{t}_R}\\
&+\widetilde{U}_{Z_TW_T}^{W_L}\otimes (f_{W_-^+}+f_{W_-^-}) +\widetilde{U}_{Z_TW_L}^{W_T}\otimes (f_{W_L^+}+f_{W_L^-})\\
&+ \widetilde{U}_{Z_TZ_T}^{h}\otimes f_{Z_-} +\widetilde{U}_{Z_Th}^{Z_T}\otimes f_{h}~,
\end{split}
\ee
\be
\begin{split}
\frac{df_{Z\gamma_+}}{dt} &= \frac{\alpha_{\g2}(t)}{2\pi}2c_W(t)\widetilde{P}_{V_+ V_s}^V\otimes (f_{W^+_s}+f_{W^-_s})+\frac{\alpha_{\g2}(t)}{2\pi}\frac{c_{2W}(t)}{c_W(t)}\widetilde{P}_{V_+ h}^h\otimes (f_{W^+_L}+f_{W^-_L})\\
&+\frac{\alpha_{\g2}(t)}{2\pi}\frac{2}{c_W(t)}\sum_{f}Q_{f}\left[\widetilde{P}_{V_+ f_L}^{f}\otimes \left(Z_{f_L}f_{f_L}+Z_{f_R}f_{\bar{f}_R}\right) + \widetilde{P}_{V_- f_L}^{f}\otimes \left(Z_{f_R}f_{f_R}+Z_{f_L}f_{\bar{f}_L}\right)\right]\\
&+\widetilde{U}_{\zg_+t_R}^{t_L}\otimes f_{t_R}+\widetilde{U}_{\zg_-t_L}^{t_R}\otimes f_{\bar{t}_L}\\
&+\widetilde{U}_{\zgt W_T}^{W_L}\otimes (f_{W_+^+}+f_{W_+^-}) +\widetilde{U}_{\zgt W_L}^{W_T}\otimes (f_{W_L^+}+f_{W_L^-})~,\\
\end{split}
\ee
\be
\begin{split}
\frac{df_{Z\gamma_-}}{dt} &= \frac{\alpha_{\g2}(t)}{2\pi}2c_W(t)\widetilde{P}_{V_- V_s}^V\otimes (f_{W^+_s}+f_{W^-_s})+\frac{\alpha_{\g2}(t)}{2\pi}\frac{c_{2W}(t)}{c_W(t)}\widetilde{P}_{V_- h}^h\otimes (f_{W^+_L}+f_{W^-_L})\\ 
&+\frac{\alpha_{\g2}(t)}{2\pi}\frac{2}{c_W(t)}\sum_{f}Q_{f}\left[\widetilde{P}_{V_- f_L}^{f}\otimes \left(Z_{f_L}f_{f_L}+Z_{f_R}f_{\bar{f}_R}\right) + \widetilde{P}_{V_+ f_L}^{f}\otimes \left(Z_{f_R}f_{f_R}+Z_{f_L}f_{\bar{f}_L}\right)\right]\\
&+\widetilde{U}_{\zg_-t_L}^{t_R}\otimes f_{t_L}+\widetilde{U}_{\zg_+t_R}^{t_L}\otimes f_{\bar{t}_R}\\
&+\widetilde{U}_{\zgt W_T}^{W_L}\otimes (f_{W_-^+}+f_{W_-^-}) +\widetilde{U}_{\zgt W_L}^{W_T}\otimes (f_{W_L^+}+f_{W_L^-})~,\\
\end{split}
\ee
\be
\begin{split}
\frac{df_{W^+_+}}{dt} &= P_{\Wpp}^v f_{\Wpp}+ \frac{\alpha_{2}(t)}{2\pi}c_W^2(t)\widetilde{P}_{V_+ V_s}^V\otimes (f_{W^+_s}+f_{Z_s}) + \frac{\alpha_{\g}(t)}{2\pi}\widetilde{P}_{V_+ V_s}^V\otimes (f_{W^+_s}+f_{\gamma_s}) \\
&+ \frac{\alpha_{\g2}(t)}{2\pi}c_W(t)\widetilde{P}_{V_+ V_s}^V\otimes f_{Z\gamma_s} +
\frac{\alpha_{2}(t)}{2\pi}\frac{1}{4}\widetilde{P}_{V_+ h}^h\otimes (f_{h}+f_{Z_L}+f_{hZ_L}+f_{W^+_L}^{(h)}+f_{W^+_L}^{(Z_L)})\\
&+ \frac{\alpha_{2}(t)}{2\pi}\frac{1}{2}\sum_{i}\left[\widetilde{P}_{V_+ f_L}^{f} \otimes \left(f_{u_{L,i}}+f_{\nu_i}\right)+ \widetilde{P}_{V_- f_L}^{f} \otimes \left(f_{\bar{d}_{L,i}}+f_{\bar{\ell}_{L,i}}\right)\right]\\
&+\widetilde{U}_{\Wpp t_R}^{b_L}\otimes f_{t_R}+\widetilde{U}_{\Wmm b_L}^{t_R}\otimes f_{\bar{b}_L}\\
&+\left(\widetilde{U}_{W_TW_L}^{\g_T} +\widetilde{U}_{W_TW_L}^{Z_T}\right)\otimes f_{W_L^+}+\left(\widetilde{U}_{W_TW_T}^{h}+\widetilde{U}_{W_TW_T}^{\zl}\right)\otimes f_{W_+^+}\\
&+\widetilde{U}_{W_T\gamma_T}^{W_L}\otimes f_{\gamma_+}+\widetilde{U}_{W_TZ_T}^{W_L}\otimes f_{Z_+} +\widetilde{U}_{W_TZ\gamma_T}^{W_L}\otimes f_{Z\gamma_+} \\ 
&+\widetilde{U}_{W_Th}^{W_T}\otimes f_{h}+\widetilde{U}_{W_TZ_L}^{W_T}\otimes f_{Z_L} +\widetilde{U}_{W_ThZ_L}^{W_T}\otimes f_{hZ_L}~,
\label{eq:Wpp_DGLAP}
\end{split}
\ee
\be
\begin{split}
\frac{df_{W^+_-}}{dt} &= P_{\Wpm}^v f_{\Wpm}+ \frac{\alpha_{2}(t)}{2\pi}c_W^2(t)\widetilde{P}_{V_- V_s}^V\otimes (f_{W^+_s}+f_{Z_s}) + \frac{\alpha_{\g}(t)}{2\pi}\widetilde{P}_{V_- V_s}^V\otimes (f_{W^+_s}+f_{\gamma_s}) \\
&+ \frac{\alpha_{\g2}(t)}{2\pi}c_W(t)\widetilde{P}_{V_- V_s}^V\otimes f_{Z\gamma_s} +
\frac{\alpha_{2}(t)}{2\pi}\frac{1}{4}\widetilde{P}_{V_- h}^h\otimes (f_{h}+f_{Z_L}+f_{hZ_L}+f_{W^+_L}^{(h)}+f_{W^+_L}^{(Z_L)})\\
&+ \frac{\alpha_{2}(t)}{2\pi}\frac{1}{2}\sum_{i}\left[\widetilde{P}_{V_- f_L}^{f} \otimes \left(f_{u_{L,i}}+f_{\nu_i}\right)+ \widetilde{P}_{V_+ f_L}^{f} \otimes \left(f_{\bar{d}_{L,i}}+f_{\bar{\ell}_{L,i}}\right)\right]\\
&+\left(\widetilde{U}_{W_TW_L}^{\g_T} +\widetilde{U}_{W_TW_L}^{Z_T}\right)\otimes f_{W_L^+}+\left(\widetilde{U}_{W_TW_T}^{h}+\widetilde{U}_{W_TW_T}^{\zl}\right)\otimes f_{W_-^+}\\
&+\widetilde{U}_{W_T\gamma_T}^{W_L}\otimes f_{\gamma_-}+\widetilde{U}_{W_TZ_T}^{W_L}\otimes f_{Z_-} +\widetilde{U}_{W_TZ\gamma_T}^{W_L}\otimes f_{Z\gamma_-} \\ 
&+\widetilde{U}_{W_Th}^{W_T}\otimes f_{h}+\widetilde{U}_{W_TZ_L}^{W_T}\otimes f_{Z_L} +\widetilde{U}_{W_ThZ_L}^{W_T}\otimes f_{hZ_L}~,
\label{eq:Wpm_DGLAP}
\end{split}
\ee
\be
\begin{split}
\frac{df_{W^-_+}}{dt} &= P_{\Wmp}^v f_{\Wmp}+ \frac{\alpha_{2}(t)}{2\pi}c_W^2(t)\widetilde{P}_{V_+ V_s}^V\otimes (f_{W^-_s}+f_{Z_s}) + \frac{\alpha_{\g}(t)}{2\pi}\widetilde{P}_{V_+ V_s}^V\otimes (f_{W^-_s}+f_{\gamma_s}) \\
&+ \frac{\alpha_{\g2}(t)}{2\pi}c_W(t)\widetilde{P}_{V_+ V_s}^V\otimes f_{Z\gamma_s} + \frac{\alpha_{2}(t)}{2\pi}\frac{1}{4}\widetilde{P}_{V_+ h}^h\otimes (f_{h}+f_{Z_L}-f_{hZ_L}+f_{W^-_L}^{(h)}+f_{W^-_L}^{(Z_L)}) \\
&+ \frac{\alpha_{2}(t)}{2\pi}\frac{1}{2}\sum_{i}\left[\widetilde{P}_{V_+ f_L}^{f} \otimes \left(f_{d_{L,i}}+f_{\ell_{L,i}}\right)+ \widetilde{P}_{V_- f_L}^{f} \otimes \left(f_{\bar{u}_{L,i}}+f_{\bar{\nu}_i}\right)\right] \\
&+\left(\widetilde{U}_{W_TW_L}^{\g_T}+\widetilde{U}_{W_TW_L}^{Z_T}\right)\otimes f_{W_L^-}+\left(\widetilde{U}_{W_TW_T}^{h}+\widetilde{U}_{W_TW_T}^{\zl}\right) \otimes f_{W_+^-}\\
&+\widetilde{U}_{W_T\gamma_T}^{W_L}\otimes f_{\gamma_+}  +\widetilde{U}_{W_TZ_T}^{W_L}\otimes f_{Z_+} +\widetilde{U}_{W_TZ\gamma_T}^{W_L}\otimes f_{Z\gamma_+} \\ 
&+\widetilde{U}_{W_Th}^{W_T}\otimes f_{h} +\widetilde{U}_{W_TZ_L}^{W_T}\otimes f_{Z_L} -\widetilde{U}_{W_ThZ_L}^{W_T}\otimes f_{hZ_L}~,
\label{eq:Wmp_DGLAP}
\end{split}
\ee
\be
\begin{split}
\frac{df_{W^-_-}}{dt} &= P_{\Wmm}^v f_{\Wmm}+ \frac{\alpha_{2}(t)}{2\pi}c_W^2(t)\widetilde{P}_{V_- V_s}^V\otimes (f_{W^-_s}+f_{Z_s}) + \frac{\alpha_{\g}(t)}{2\pi}\widetilde{P}_{V_- V_s}^V\otimes (f_{W^-_s}+f_{\gamma_s}) \\
&+ \frac{\alpha_{\g2}(t)}{2\pi}c_W(t)\widetilde{P}_{V_- V_s}^V\otimes f_{Z\gamma_s} + \frac{\alpha_{2}(t)}{2\pi}\frac{1}{4}\widetilde{P}_{V_- h}^h\otimes (f_{h}+f_{Z_L}-f_{hZ_L}+f_{W^-_L}^{(h)}+f_{W^-_L}^{(Z_L)}) \\
&+ \frac{\alpha_{2}(t)}{2\pi}\frac{1}{2}\sum_{i}\left[\widetilde{P}_{V_- f_L}^{f} \otimes \left(f_{d_{L,i}}+f_{\ell_{L,i}}\right)+ \widetilde{P}_{V_+ f_L}^{f} \otimes \left(f_{\bar{u}_{L,i}}+f_{\bar{\nu}_i}\right)\right] \\
&+\widetilde{P}_{\Wmm b_L}^{t_R}\otimes f_{b_L}+\widetilde{P}_{\Wpp t_R}^{b_L}\otimes f_{\bar{t}_R}\\
&+\left(\widetilde{U}_{W_TW_L}^{\g_T}+\widetilde{U}_{W_TW_L}^{Z_T}\right)\otimes f_{W_L^-}+\left(\widetilde{U}_{W_TW_T}^{h}+\widetilde{U}_{W_TW_T}^{\zl}\right) \otimes f_{W_-^-}\\
&+\widetilde{U}_{W_T\gamma_T}^{W_L}\otimes f_{\gamma_-}  +\widetilde{U}_{W_TZ_T}^{W_L}\otimes f_{Z_-} +\widetilde{U}_{W_TZ\gamma_T}^{W_L}\otimes f_{Z\gamma_-} \\ 
&+\widetilde{U}_{W_Th}^{W_T}\otimes f_{h} +\widetilde{U}_{W_TZ_L}^{W_T}\otimes f_{Z_L} -\widetilde{U}_{W_ThZ_L}^{W_T}\otimes f_{hZ_L}~.
\label{eq:Wmm_DGLAP}
\end{split}
\ee

\subsection{Higgs and longitudinal gauge bosons}

\be
\begin{split}
\frac{df_h}{dt} &= P_h^vf_h+ \frac{\alpha_{2}(t)}{2\pi}\frac{1}{4}\Big[\widetilde{P}_{hh}^V\otimes \left(f_{W_L^+} + f_{W_L^-}\right) + \widetilde{P}_{hV}^h\otimes \left(f_{W_+^+} + f_{W_-^+} + f_{W_+^-} +f_{W_-^-} \right)\Big] \\
&+\frac{\alpha_{2}(t)}{2\pi}\frac{1}{4c_W^2}\Big[\widetilde{P}_{hh}^V\otimes f_{Z_L}+ \widetilde{P}_{hV}^h\otimes \left(f_{Z_+} +f_{Z_-} \right)\Big] \\
&+\frac{\alpha_{y}(t)}{2\pi}\frac{1}{2}\widetilde{P}_{hf}^f\otimes \left(f_{t_L}+f_{t_R}+f_{\bar{t}_L}+f_{\bar{t}_R}\right) \\
&+\widetilde{U}_{ht}^{t}\otimes (f_{t_L}+f_{t_R}+f_{\bar{t}_L}+f_{\bar{t}_R}) \\
&+ \widetilde{U}_{hW_T}^{W_T}\otimes (f_{W_+^+}+f_{W_-^+}+f_{W_+^-}+f_{W_-^-}) + \widetilde{U}_{hZ_T}^{Z_T}\otimes (f_{Z_+}+f_{Z_-}) \\
&+ \widetilde{U}_{hW_L}^{W_L}\otimes (f_{W_L^+}+ f_{W_L^-}) + \widetilde{U}_{hZ_L}^{\zl}\otimes f_{Z_L}+\widetilde{U}_{hh}^{h}\otimes f_{h}~,
\end{split}
\ee
\be
\begin{split}
\frac{df_{Z_L}}{dt} &= P_{Z_L}^v f_{Z_L}+  \frac{\alpha_{2}(t)}{2\pi}\frac{1}{4}\left[\widetilde{P}_{hh}^V\otimes (f_{W_L^+}+f_{W_L^-})+ \widetilde{P}_{hV}^h\otimes (f_{W_+^+}+f_{W_-^+}+f_{W_+^-}+f_{W_-^-})\right] \\ 
&+\frac{\alpha_{2}(t)}{2\pi}\frac{1}{4c_W^2}\left[\widetilde{P}_{hh}^V\otimes f_{h}+ \widetilde{P}_{hV}^h\otimes (f_{Z_+}+f_{Z_-})\right] \\
&+\frac{\alpha_{y}(t)}{2\pi}\frac{1}{2}\widetilde{P}_{hf}^f\otimes \left(f_{t_L}+f_{t_R}+f_{\bar{t}_L}+f_{\bar{t}_R}\right) \\
&+\sum_{f_L}\widetilde{U}_{Z_Lf_L}^{f_L}\otimes f_{f_L} +\sum_{f_R}\widetilde{U}_{Z_Lf_R}^{f_R}\otimes f_{f_R} \\
&+ \sum_{\bar{f}_L}\widetilde{U}_{Z_L\bar{f}_L}^{\flb}\otimes f_{\bar{f}_L} +\sum_{\bar{f}_R}\widetilde{U}_{Z_L\bar{f}_R}^{\frb}\otimes f_{\bar{f}_R}  \\
&+\widetilde{U}_{Z_LW_T}^{W_T}\otimes (f_{W_+^+}+f_{W_-^+}+f_{W_+^-}+f_{W_-^-}) \\
&+ \widetilde{U}_{Z_LZ_L}^{h}\otimes f_{Z_L} +\widetilde{U}_{Z_Lh}^{\zl}\otimes f_{h} + \widetilde{U}_{Z_LW_L}^{W_L}\otimes (f_{W_L^+}+f_{W_L^-})~,
\end{split}
\ee
\be
\begin{split}
\frac{df_{W^+_L}}{dt} &= P_{W^+_L}^v f_{W^+_L}+  \frac{\alpha_{2}(t)}{2\pi}\frac{1}{4}\left[\widetilde{P}_{hh}^V\otimes (f_{h}+f_{Z_L}-f_{hZ_L})+ \widetilde{P}_{hV}^h\otimes \left(f_{W^+_+}^{(h)}+f_{W^+_+}^{(Z_L)}+f_{W^+_-}^{(h)}+f_{W^+_-}^{(Z_L)}\right)\right] \\
&+\frac{\alpha_{2}(t)}{2\pi}\frac{c_{2W}^2}{4c_W^2}\left[\widetilde{P}_{hh}^V\otimes f_{W_L^+}+ \widetilde{P}_{hV}^h\otimes (f_{Z_+}+f_{Z_-})\right]\\
&+\frac{\alpha_{\g}(t)}{2\pi}\left[\widetilde{P}_{hh}^V\otimes f_{W_L^+}+ \widetilde{P}_{hV}^h\otimes (f_{\gamma_+}+f_{\gamma_-})\right]  
+\frac{\alpha_{\g2}(t)}{2\pi}\frac{c_{2W}}{2c_W}\widetilde{P}_{hV}^h\otimes (f_{Z\gamma_+}+f_{Z\gamma_-}) \\
&+\frac{\alpha_{y}(t)}{2\pi}\widetilde{P}_{hf}^f\otimes \left(f_{t_R}+f_{\bar{b}_L}\right) \\
&+\sum_{f^{(1)}_Lf^{(2)}_L}\widetilde{U}_{W_L^+f^{(1)}_L}^{f^{(2)}_L}\otimes f_{f^{(1)}_L} +\sum_{\bar{f}^{(1)}_L\bar{f}^{(2)}_L}\widetilde{U}_{W_L^+\bar{f}^{(1)}_L}^{\bar{f}^{(2)}_L}\otimes f_{\bar{f}^{(1)}_L}\\ 
&+\widetilde{U}_{W_LW_T}^{\g_T}\otimes (f_{W_+^+}+f_{W_-^+}) + \widetilde{U}_{W_LW_T}^{Z_T}\otimes (f_{W_+^+}+f_{W_-^+}) \\
&+ \widetilde{U}_{W_L\gamma_T}^{W_T}\otimes (f_{\gamma_+}+f_{\gamma_-})+ \widetilde{U}_{W_LZ_T}^{W_T}\otimes (f_{Z_+}+f_{Z_-}) \\
&+ \widetilde{U}_{W_LZ\gamma_T}^{W_T}\otimes (f_{Z\gamma_+}+f_{Z\gamma_-}) \\
&+\widetilde{U}_{W_Lh}^{W_L}\otimes f_h +\widetilde{U}_{W_LZ_L}^{W_L}\otimes f_{Z_L} +\widetilde{U}_{W_LhZ_L}^{W_L}\otimes f_{hZ_L}\\
&+\widetilde{U}_{W_LW_L}^{h}\otimes f_{W^+_L} +\widetilde{U}_{W_LW_L}^{\zl}\otimes f_{W^+_L}~,
\label{eq:WpL_DGLAP}
\end{split}
\ee
\be
\begin{split}
\frac{df_{W^-_L}}{dt} &= P_{W^-_L}^v f_{W^-_L}+  \frac{\alpha_{2}(t)}{2\pi}\frac{1}{4}\left[\widetilde{P}_{hh}^V\otimes (f_{h}+f_{Z_L}+f_{hZ_L})+ \widetilde{P}_{hV}^h\otimes \left(f_{W^-_+}^{(h)}+f_{W^-_+}^{(Z_L)}+f_{W^-_-}^{(h)}+f_{W^-_-}^{(Z_L)}\right)\right] \\ 
&+\frac{\alpha_{2}(t)}{2\pi}\frac{c_{2W}^2}{4c_W^2}\left[\widetilde{P}_{hh}^V\otimes f_{W_L^-}+ \widetilde{P}_{hV}^h\otimes (f_{Z_+}+f_{Z_-})\right]\\
&+\frac{\alpha_{\g}(t)}{2\pi}\left[\widetilde{P}_{hh}^V\otimes f_{W_L^-}+ \widetilde{P}_{hV}^h\otimes (f_{\gamma_+}+f_{\gamma_-})\right] 
+\frac{\alpha_{\g2}(t)}{2\pi}\frac{c_{2W}}{2c_W}\widetilde{P}_{hV}^h\otimes (f_{Z\gamma_+}+f_{Z\gamma_-}) \\
&+\frac{\alpha_{y}(t)}{2\pi}\widetilde{P}_{hf}^f\otimes \left(f_{b_L}+f_{\bar{t}_R}\right) \\
&+\sum_{f^{(1)}_Lf^{(2)}_L}\widetilde{U}_{W_L^-f^{(1)}_L}^{f^{(2)}_L}\otimes f_{f^{(1)}_L} +\sum_{\bar{f}^{(1)}_L\bar{f}^{(2)}_L}\widetilde{U}_{W_L^-\bar{f}^{(1)}_L}^{\bar{f}^{(2)}_L}\otimes f_{\bar{f}^{(1)}_L}\\  &+\widetilde{U}_{W_LW_T}^{\g_T}\otimes (f_{W_+^-}+f_{W_-^-}) + \widetilde{U}_{W_LW_T}^{Z_T}\otimes (f_{W_+^-}+f_{W_-^-}) \\
&+ \widetilde{U}_{W_L\gamma_T}^{W_T}\otimes (f_{\gamma_+}+f_{\gamma_-}) + \widetilde{U}_{W_LZ_T}^{W_T}\otimes (f_{Z_+}+f_{Z_-}) \\
&+ \widetilde{U}_{W_LZ\gamma_T}^{W_T}\otimes (f_{Z\gamma_+}+f_{Z\gamma_-}) \\
&+\widetilde{U}_{W_Lh}^{W_L}\otimes f_h +\widetilde{U}_{W_LZ_L}^{W_L}\otimes f_{Z_L} -\widetilde{U}_{W_LhZ_L}^{W_L}\otimes f_{hZ_L} \\
&+\widetilde{U}_{W_LW_L}^{h}\otimes f_{W^-_L} +\widetilde{U}_{W_LW_L}^{\zl}\otimes f_{W^-_L}~,
\label{eq:WmL_DGLAP}
\end{split}
\ee
\be
\begin{split}
\frac{df_{hZ_L}}{dt} &= \frac{\alpha_{y}(t)}{2\pi}\widetilde{P}_{hf}^f\otimes (f_{t_L+}+f_{\tbar_R}-f_{t_R}-f_{\tbar_L})\\ 
&+\frac{\alpha_{2}(t)}{2\pi}\frac{1}{4}\left[\widetilde{P}_{hh}^V\otimes (f_{W_L^-}-f_{W_L^+})+ \widetilde{P}_{hV}^h\otimes (f_{W_+^+}+f_{W_-^+}-f_{W_+^-}-f_{W_-^-})\right] \\
&+\widetilde{U}_{hZ_L\tl}^{\tl}\otimes (f_{\tl}-f_{\tlb}) +\widetilde{U}_{hZ_L\tr}^{\tr}\otimes (f_{\tr}-f_{\trb})\\
&+\widetilde{U}_{hZ_LW_L}^{W_L}\otimes (f_{W^+_L}-f_{W^-_L}) +\widetilde{U}_{hZ_LW_T}^{W_T}\otimes (f_{W_+^+}+f_{W_-^+}-f_{W_+^-}-f_{W_-^-})~.
\end{split}
\ee
%

\section{Numerical implementation}
\label{app:numerical}

Here we show the details of our numerical implementation of the DGLAP equations. 
We discretize the $x$-space from a minimum value $x_0$ up to 1 in $N_x$ bins, $x_\alpha=\{x_0, x_1,\dots,x_{N_x}\equiv 1\}$, with spacing $\delta x_{\alpha} = x_{\alpha}-x_{\alpha-1}$. We choose a spacing that is denser near $x=1$ and sparser at small values, in practice we set $x_\alpha = 10^{-6 \left((N_x - \alpha)/N_x \right)^{2.5}}$ for $\alpha = 0, 1, \ldots, N_x$ to get values from $x_0=10^{-6}$ to $x_{N_x}=1$.
This allows us to obtain a set of ODEs, where the integrals are computed using the rectangles method.\footnote{Due to the use of the rectangles method, we note that special care should be taken when interpolating the LePDFs near the region of $x=1$, where the muon PDF changes very steeply. In this case we recommend using zeroth order interpolation for consistency.}

For the non-divergent terms or when $\zmax\neq 1$ we obtain
\be
\frac{df_B(x_{\beta},t)}{dt} \supset \int_{x_{\beta}}^{\zmax} \frac{dz}{z}\widetilde{P}_{BA}^{C}\left(\frac{x_{\beta}}{z}\right)f_A(z,t)= \sum_{\alpha=\beta+1}^{\Nmax} \frac{\delta x_{\alpha}}{x_\alpha}\widetilde{P}_{BA}^{C}\left(\frac{x_\beta}{x_\alpha}\right)f_A(x_\alpha,t) \label{eq:DGLAPdiscr1},
\ee
where $\Nmax$ is the index of the greatest point of the grid which is smaller than $\zmax$.
The remaining case is that of the splitting functions with the $+$ distribution, which we can write as 
\begin{equation}
\widetilde{P}_{BA}^{C}(z) = \frac{\widetilde{D}_{BA}^{C}(z)}{(1-z)_+}~.
\end{equation}
Using the definition in Eq.~\eqref{eq:def+} the corresponding terms in the DGLAP equations are:
\be
\begin{split}
\frac{df_B(x_{\beta},t)}{dt} & \supset \int_{x_{\beta}}^1 \frac{dy}{y(1-y)_+}\widetilde{D}_{BA}^{C}(y)f_A\left(\frac{x_{\beta}}{y},t\right) = \widetilde{D}_{BA}^{C}(1)\log(1-x_{\beta})f_A(x_{\beta},t) \\
&+\int_{x_{\beta}}^{\zmax} \frac{dy}{(1-y)}\left(\frac{\widetilde{D}_{BA}^{C}(y)f_A(\frac{x_{\beta}}{y},t)}{y}-\widetilde{D}_{BA}^{C}(1)f_A(x_{\beta},t)\right) \label{eq:DGLAP2}~.
\end{split}
\ee
With our discretization, the last integral becomes
\be
\begin{split}
&\int_{x_\beta}^1 \frac{dy}{(1-y)}\left(\frac{\widetilde{D}_{BA}^{C}(y)f_A(\frac{x_\beta}{y},t)}{y}-\widetilde{D}_{BA}^{C}(1)f_A(x_\beta,t)\right) \\
&= \int_{x_\beta}^1 dz\frac{x_\beta}{z^2}\frac{1}{1-\frac{x_\beta}{z}}\left[\frac{z}{x_\beta}\widetilde{D}_{BA}^{C}\left(\frac{x_\beta}{z}\right)f_A(z,t)-\widetilde{D}_{BA}^{C}(1)f_A(x_\beta,t)\right] \\
&= \int_{x_\beta}^1 dz\left[\frac{1}{z\left(1-\frac{x_\beta}{z}\right)}\widetilde{D}_{BA}^{C}\left(\frac{x_\beta}{z}\right)f_A(z,t)-\frac{x_\beta}{z^2}\frac{1}{1-\frac{x_\beta}{z}}\widetilde{D}_{BA}^{C}(1)f_A(x_\beta,t)\right]  \\
&= \sum_{\alpha=\beta+1}^{N_x} \frac{\delta x_{\alpha}}{x_\alpha}\widetilde{P}_{BA}^{C}\left(\frac{x_\beta}{x_\alpha}\right)f_A(x_\alpha,t)-\widetilde{D}_{BA}^{C}(1)f_A(x_\beta,t)x_\beta\sum_{\alpha=\beta+1}^{N_x} \frac{\delta x_{\alpha}}{x_\alpha^2}\frac{1}{1-\frac{x_\beta}{x_\alpha}}\\
&\equiv \sum_{\alpha=\beta+1}^{N_x} \frac{\delta x_{\alpha}}{x_\alpha}\widetilde{P}_{BA}^{C}\left(\frac{x_\beta}{x_\alpha}\right)f_A(x_\alpha,t)-\widetilde{D}_{BA}^{C}(1)X_\beta f_A(x_\beta,t)~,
\end{split}
\ee
where $X_\beta$ is given by
\begin{equation}
X_\beta \equiv x_\beta\sum_{\alpha=\beta+1}^{N_x} \frac{\delta x_{\alpha}}{x_\alpha^2}\frac{1}{1-\frac{x_\beta}{x_\alpha}}~.
\end{equation}
With this discretization, starting with $N_f$ PDFs we get a set of $(N_x+1)N_f$ equations for the variables $f_{B\beta}(t)\equiv f_B(x_\beta,t)$, with  $\beta=\{0,\dots,N_x\}$ and $B=\{1,\dots,N_f\}$
\begin{equation}
\begin{split}
\frac{df_{B\beta}(t)}{dt} &= P_B^v(t)f_{B\beta}(t)+ \sum_{A,C}\frac{\alpha_{ABC}(t)}{2\pi}\left(\log(1-x_\beta)-X_\beta\right)\widetilde{D}_{BA}^{C}(1)f_{A\beta}\\
&+\sum_{A,C}\frac{\alpha_{ABC}(t)}{2\pi}\sum_{\alpha=\beta+1}^{\Nmax} \frac{\delta x_{\alpha}}{x_\alpha}\widetilde{P}_{BA}^{C}\left(\frac{x_\beta}{x_\alpha}\right)f_{A\alpha}(t)~. \label{eq:DGLAP_discr}
\end{split}
\end{equation}
Once we have the equations with the proper initial conditions, we solve them numerically using a fourth order Runge-Kutta with integration step $dt$. For a set of equations of the form $y_i'(t) = F_i(t,y)$, starting with the solution $y_{i,n}$ at $t=t_n$ the step of the algorithm is
\be
\begin{split}
k_{1,i} &= dt F_i(t_n,y_{i,n})\\
k_{2,i} &= dt F_i(t_n+\frac{dt}{2},y_{i,n}+\frac{k_1}{2})\\
k_{3,i} &= dt F_i(t_n+\frac{dt}{2},y_{i,n}+\frac{k_2}{2})\\
k_{4,i} &= dt F_i(t_n+dt,y_{i,n}+k_3)\\
y_{i,n+1} &= y_{i,n}+\frac{k_{1,i}}{6}+\frac{k_{2,i}}{3}+\frac{k_{3,i}}{3}+\frac{k_{4,i}}{6} + \mathcal{O}(dt^5)~.
\end{split} \label{eq:rungekutta}
\ee
We then reduce the number of variables imposing momentum conservation after every step of the evolution, since performing a numerical integration it will be more and more violated, as done in \cite{Han:2021kes}. With our discretization Eq.~\eqref{eq:momcons} becomes
\begin{equation}
\sum_{i=1}^{N_f}\sum_{\alpha=1}^{N_x}\delta x_{\alpha} x_\alpha f_{i\alpha}(t) = 1 \label{eq:momconsdiscr}~.
\end{equation}
Since the initial conditions on PDFs are given by
\begin{equation}
f_\mu(0,x_\alpha) = \delta(1-x_\alpha) = \frac{1}{\delta x_{N_x}}\delta_{\alpha {N_x}}~,\qquad f_{i\neq \mu}(0,x) = 0~,
\end{equation}
only $f_{\mu\alpha}$ will be nonzero for $\alpha = N_x$ throughout the evolution. Then we fix 
\begin{equation}
f_{i{N_x}}(t) = \left\{ \begin{array}{ll}
\frac{L(t)}{\delta x_{N_x}} & i = \mu\\
0 & i \neq \mu\\
\end{array}\right. ~,\label{eq:fixmomentum}
\end{equation}
reducing by $N_f$ the number of variables and solving the remaining equations. The factor $L(t)$ is computed using momentum conservation
\begin{equation}
1 = \sum_{i=1}^{N_f}\sum_{\alpha=1}^{N_x-1}\delta x_{\alpha} x_\alpha f_{i\alpha}(t) + L(t)~,
\end{equation}
that is
\begin{equation}
L(t) = 1-\sum_{i=1}^{N_f}\sum_{\alpha=1}^{N_x-1}\delta x_{\alpha} x_\alpha f_{i\alpha}(t)~. \label{eq:Lt}
\end{equation}

The uncertainties due to the discretizations in $x$ and $t$ spaces are discussed in Section~\ref{sec:uncertainties}.

\section{LHAPDF format}
\label{app:publicPDF}

\begin{table}[t]
\centering
\begin{tabular}{|c c c c|}\hline
    $e_L$ & eL & 11 & -  \\
    $e_R$ & eR & 11 & + \\
    $\nu_e$ & nue & 12 & - \\
    $\mu_L$ & muL & 13 & -  \\
    $\mu_R$ & muR & 13 & + \\
    $\nu_{\mu}$ & numu & 14 & - \\
    $\tau_L$ & taL & 15 & -  \\
    $\tau_R$ & taR & 15 & + \\
    $\nu_{\tau}$ & nuta & 16 & - \\
    $\bar{e}_L$ & eLb & -11 & +  \\
    $\bar{e}_R$ & eRb & -11 & - \\
    $\bar{\nu}_e$ & nueb & -12 & + \\
    $\bar{\mu}_L$ & muLb & -13 & +  \\
    $\bar{\mu}_R$ & muRb & -13 & - \\
    $\bar{\nu}_{\mu}$ & numub & -14 & + \\
    $\bar{\tau}_L$ & taLb & -15 & +  \\
    $\bar{\tau}_R$ & taRb & -15 & - \\
    $\bar{\nu}_{\tau}$ & nutab & -16 & + \\
    \hline
\end{tabular}
\begin{tabular}{|c c c c|}\hline
    $d_L$ & dL & 1 & -  \\
    $d_R$ & dR & 1 & + \\
    $u_L$ & uL & 2 & -  \\
    $u_R$ & uR & 2 & + \\
    $s_L$ & sL & 3 & -  \\
    $s_R$ & sR & 3 & + \\
    $c_L$ & cL & 4 & -  \\
    $c_R$ & cR & 4 & + \\
    $b_L$ & bL & 5 & -  \\
    $b_R$ & bR & 5 & + \\
    $t_L$ & tL & 6 & -  \\
    $t_R$ & tR & 6 & + \\
    $\bar{d}_L$ & dLb & -1 & +  \\
    $\bar{d}_R$ & dRb & -1 & - \\
    $\bar{u}_L$ & uLb & -2 & +  \\
    $\bar{u}_R$ & uRb & -2 & - \\
    $\bar{s}_L$ & sLb & -3 & +  \\
    $\bar{s}_R$ & sRb & -3 & - \\
    $\bar{c}_L$ & cLb & -4 & +  \\
    $\bar{c}_R$ & cRb & -4 & - \\
    $\bar{b}_L$ & bLb & -5 & +  \\
    $\bar{b}_R$ & bRb & -5 & - \\
    $\bar{t}_L$ & tLb & -6 & +  \\
    $\bar{t}_R$ & tRb & -6 & - \\
    \hline
\end{tabular}
\begin{tabular}{|c c c c|}\hline
    $g_+$ & gp & 21 & +  \\
    $g_-$ & gm & 21 & - \\
    $\gp$ & gap & 22 & +  \\
    $\gm$ & gam & 22 & - \\
    $\zp$ & Zp & 23 & +  \\
    $\zm$ & Zm & 23 & - \\
    $\zl$ & ZL & 23 & 0 \\
    $\zgp$ & Zgap & 2223 & +  \\
    $\zgm$ & Zgam & 2223 & - \\
    $\Wpp$ & Wpp & 24 & +  \\
    $\Wpm$ & Wpm & 24 & - \\
    $\Wpl$ & WpL & 24 & 0 \\
    $\Wmp$ & Wmp & -24 & + \\
    $\Wmm$ & Wmm & -24 & - \\
    $\Wml$ & WmL & -24 & 0 \\
    $h$ & h & 25 & 0 \\
    $h/\zl$ & hZL & 2523 & 0 \\
    \hline
\end{tabular}
\caption{\label{tab:partlabels} Names, PDG ID and polarisations of the particles. In particular, the second, third and fourth columns of the tables correspond to the sixth, seventh and eighth lines of the output file. }
\end{table}

We publish our numerical results in a format inspired by the LHAPDF6 \cite{Buckley:2014ana} one used for proton PDFs. Some changes are required due to the polarisation of PDFs. The structure of the output is then as follows:
\begin{itemize}
    \item the first three lines just specify the format;
    \item in the fourth and fifth line are reported respectively the grids in $x$ and in $Q$ (the grid in $Q$ is a subset of the grid used in the numerical solution of the DGLAP equations);
    \item the next three lines report the particles' list as in Table \ref{tab:partlabels}: name, PDG ID (for the $Z/\gamma$ and $h/Z_L$ interference we join the PDG ID of the two states) and the additional label specifying the helicity (it is understood that for fermions or vectors it will be $\pm 1/2$ or $\pm 1$, respectively);
    \item in all the remaining lines we report the quantities $xf(x,Q)$: each column corresponds to a particle, following the order of the previous lines. We start at $x=x_0$ increasing $Q$ at each row and repeating for each $x$, so that the data have the form reported in Table \ref{tab:LHAPDFformat}.
\end{itemize}

\begin{table}
\centering
\begin{tabular}{c c c c}
    $xf_{e_L}(x_0,Q_0)$ & $\dots$ & $\dots$ & $xf_{h/Z_L}(x_0,Q_0)$  \\
    $xf_{e_L}(x_0,Q_1)$ & $\dots$ & $\dots$ & $xf_{h/Z_L}(x_0,Q_1)$ \\
    \vdots & \vdots & \vdots & \vdots \\
    $xf_{e_L}(x_0,Q_{N_{Q}})$ & $\dots$ & $\dots$ & $xf_{h/Z_L}(x_0,Q_{N_{Q}})$\\
    $xf_{e_L}(x_1,Q_0)$ & $\dots$ & $\dots$ & $xf_{h/Z_L}(x_1,Q_0)$  \\
    $xf_{e_L}(x_1,Q_1)$ & $\dots$ & $\dots$ & $xf_{h/Z_L}(x_1,Q_1)$ \\
    \vdots & \vdots & \vdots & \vdots \\
    $xf_{e_L}(x_1,Q_{N_{Q}})$ & $\dots$ & $\dots$ & $xf_{h/Z_L}(x_1,Q_{N_{Q}})$\\
    \vdots & \vdots & \vdots & \vdots \\
    \vdots & \vdots & \vdots & \vdots \\
    $xf_{e_L}(x_{N_{x}},Q_0)$ & $\dots$ & $\dots$ & $xf_{h/Z_L}(x_1,Q_0)$  \\
    $xf_{e_L}(x_{N_{x}},Q_1)$ & $\dots$ & $\dots$ & $xf_{h/Z_L}(x_1,Q_1)$ \\
    \vdots & \vdots & \vdots & \vdots \\
    $xf_{e_L}(x_{N_{x}},Q_{N_{Q}})$ & $\dots$ & $\dots$ & $xf_{h/Z_L}(x_{N_{x}},Q_{N_{Q}})$\\
\end{tabular}
\caption{\label{tab:LHAPDFformat} Structure of our PDF data in the LHAPDF6 format.}
\end{table}

\bibliographystyle{JHEP}
\bibliography{Biblio}

\end{document}